\documentclass[11pt]{article}

\usepackage[final]{acl}

\usepackage{times}
\usepackage{latexsym}

\usepackage[T1]{fontenc}
\usepackage[utf8]{inputenc}

\usepackage{microtype}
\usepackage{inconsolata}
\usepackage{graphicx}

\usepackage{amsmath}
\usepackage{amssymb}
\usepackage{amsthm}
\usepackage{mathtools}
\usepackage{bbm}

\usepackage{booktabs}
\usepackage{multirow}
\usepackage{subcaption}
\usepackage{colortbl}
\usepackage{xcolor}
\usepackage{arydshln}
\usepackage{placeins}
\usepackage{wrapfig}
\usepackage{siunitx}

\usepackage{algorithm}
\usepackage{algorithmic}

\usepackage[capitalize,noabbrev]{cleveref}
\usepackage{etoc}

\usepackage{listings}
\usepackage{upquote}
\lstdefinestyle{promptstyle}{
  basicstyle=\ttfamily\scriptsize,
  breaklines=true,
  breakatwhitespace=true,
  breakindent=1em,
  upquote=true,
  columns=fullflexible,
  keepspaces=true,
  showstringspaces=false,
  frame=single,
  framesep=4pt,
  xleftmargin=5pt,
  xrightmargin=1pt,
  aboveskip=6pt,
  belowskip=6pt
}

\usepackage{tikz}

\usepackage[disable,textsize=tiny]{todonotes}

\definecolor{winnerblue}{RGB}{220,235,255}

\newcommand{\ours}{\texttt{SpIDER}}
\newcommand{\oursbench}{\texttt{SpIDER-Bench}}
\newcommand{\swerankszs}{\texttt{SweRankEmbed-Small ZS}}
\newcommand{\codesage}{\texttt{CodeSAGE}}

\newcommand{\codesageszs}{\texttt{CodeSAGE-Small ZS}}
\newcommand{\codesagespeft}{\texttt{CodeSAGE-Small-PEFT}}
\newcommand{\bm}{\texttt{BM25}}
\newcommand{\locagent}{\texttt{LocAgent}}
\newcommand{\swepoly}{\texttt{SWE-PolyBench}}
\newcommand{\multiswe}{\texttt{Multi-SWEBench}}
\newcommand{\sweverified}{\texttt{SWEBench-Verified}}
\newcommand{\swelite}{\texttt{SWEBench-Lite}}
\newcommand{\swebench}{\texttt{SWEBench}}
\newcommand{\locbench}{\texttt{LocBench}}
\newcommand{\der}{\texttt{DER}}
\newcommand{\sr}{\texttt{SR}}
\newcommand{\spisr}{\texttt{SpISR}}

\newcommand{\gV}{\mathcal{V}}

\newcommand{\gS}{\mathcal{S}}

\newcommand{\oursfull}{Spatially Informed Dense Embedding Retrieval}
\newcommand{\acck}{Acc@$K$}
\newcommand{\recallk}{Recall@$K$}
\newcommand{\gP}{\mathcal{P}}
\newcommand{\gA}{\mathcal{A}}
\newcommand{\gD}{\mathcal{D}}
\newcommand{\gR}{\mathcal{R}}
\newcommand{\gC}{\mathcal{C}}
\newcommand{\gU}{\mathcal{U}}
\newcommand{\gL}{\mathcal{L}}
\newcommand{\gM}{\mathcal{M}}
\newcommand{\E}{\mathbb{E}}
\newcommand{\Reck}{\mathrm{Rec}@K}
\newcommand{\Acck}{\mathrm{Acc}@K}
\newif\ifcompacttheory \compacttheorytrue
\newif\ifdepthappendix \depthappendixtrue

\theoremstyle{plain}
\newtheorem{theorem}{Theorem}[section]
\newtheorem{proposition}[theorem]{Proposition}
\newtheorem{lemma}[theorem]{Lemma}
\newtheorem{corollary}[theorem]{Corollary}
\theoremstyle{definition}

\newtheorem{assumption}[theorem]{Assumption}
\theoremstyle{remark}
\newtheorem{remark}[theorem]{Remark}

\crefname{proposition}{Proposition}{Propositions}
\Crefname{proposition}{Proposition}{Propositions}
\crefname{lemma}{Lemma}{Lemmas}
\Crefname{lemma}{Lemma}{Lemmas}
\crefname{corollary}{Corollary}{Corollaries}
\Crefname{corollary}{Corollary}{Corollaries}
\crefname{definition}{Definition}{Definitions}
\Crefname{definition}{Definition}{Definitions}
\crefname{assumption}{Assumption}{Assumptions}
\Crefname{assumption}{Assumption}{Assumptions}
\crefname{remark}{Remark}{Remarks}
\Crefname{remark}{Remark}{Remarks}

\title{\ours{}: Spatially Informed Dense Embedding Retrieval for Software Issue Localization}

\author{%
  Shravan Chaudhari\textsuperscript{1,2}
  \And
  Rahul Thomas Jacob\textsuperscript{3}
  \AND
  Jiajun Cao\textsuperscript{3}
  \And
  Shihab Rashid\textsuperscript{3}
  \And
  Mononito Goswami\textsuperscript{3}
  \AND
  Christian Bock\textsuperscript{3}
}

\begin{document}
\maketitle

\footnotetext[1]{Department of Computer Science, Johns Hopkins University, Baltimore, MD, USA}
\footnotetext[2]{Work done while interning at AWS AI Labs}
\footnotetext[3]{AWS AI Labs, Seattle, WA, USA}
\setcounter{footnote}{3}

\begin{abstract}
Retrieving code functions, classes or files relevant to a user query, bug report or feature request from large codebases is a fundamental challenge for Large Language Model (LLM)-based coding agents. Agentic approaches typically employ sparse methods like \bm{} or dense embedding strategies to identify semantically relevant units. While dense embedding approaches can outperform \bm{} by large margins, both ignore the graph-structured characteristics of the codebase. To address this, we propose \ours{} (\oursfull{}), a dense retrieval approach that integrates LLM-based reasoning with graph-based exploration of the codebase. We further introduce \oursbench{}\footnote{Dataset is available at \url{https://huggingface.co/datasets/AmazonScience/SpIDER-Bench}, and code at \url{https://github.com/amazon-science/SpIDER}.}, a graph-structured benchmark curated from \swepoly{}, \sweverified{} and \multiswe{}, spanning Python, Java, JavaScript and TypeScript repositories. \ours{}'s graph-based candidate expansion attaches a structural reason for inclusion to each surfaced function (its seed and the edge type linking them), making the candidate set auditable while keeping the retrieval budget fixed. The graph is built from per-repository syntax trees, so it can be constructed on-demand at the start of a developer session rather than precomputed offline. Empirical results show that \ours{} consistently improves dense retrieval Recall@20 across every language and benchmark in \oursbench{}: by at least $13\%$ relative ($+0.05$ to $+0.12$ absolute) along containment edges, rising to at least $27\%$ relative ($+0.11$ to $+0.19$ absolute) once call edges are explored.
\end{abstract}

\section{Introduction}

\begin{figure*}[t]
  \centering
  \includegraphics[width=0.9\linewidth]{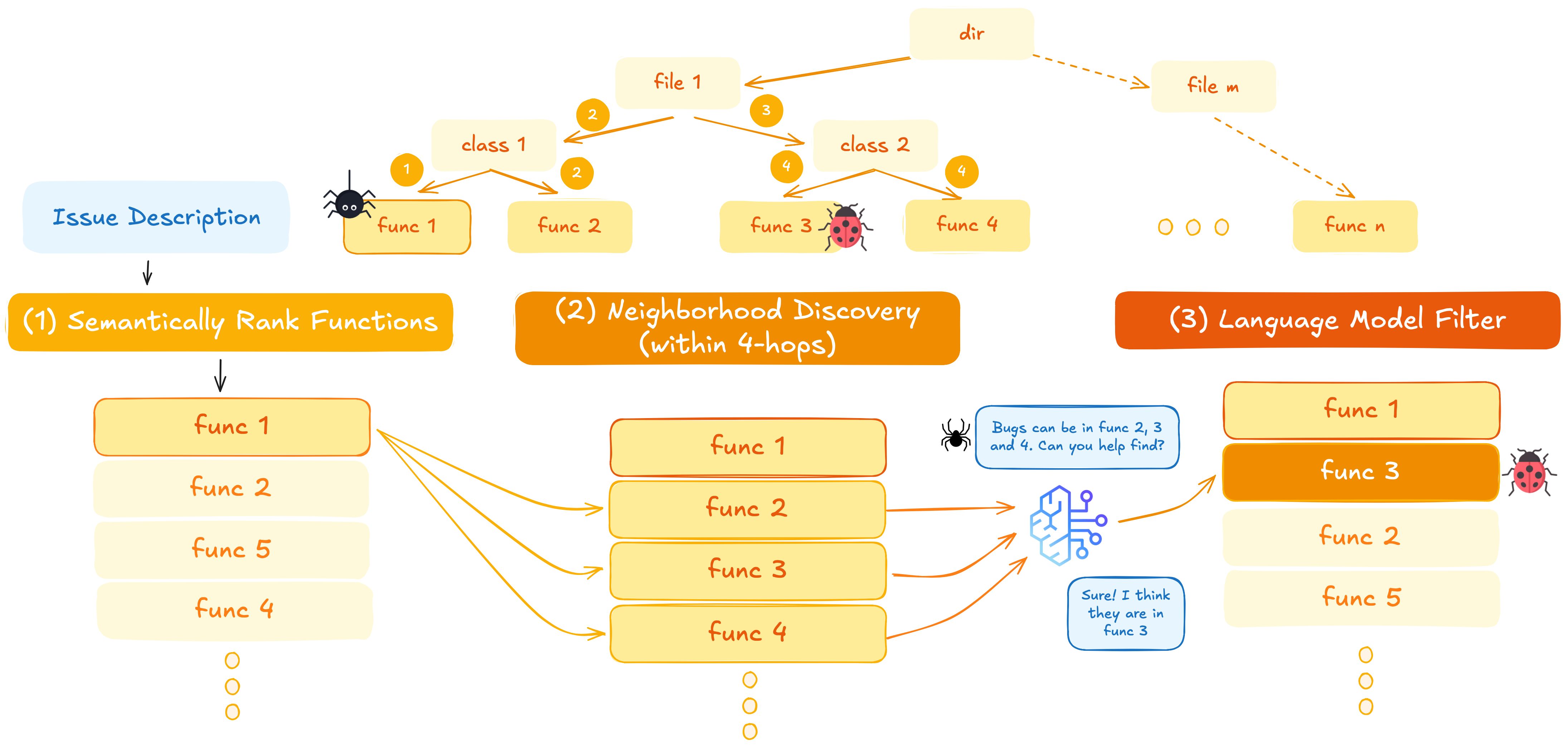}
    \caption{\textbf{\ours{} workflow.}
    SpIDER augments standard dense embedding retrieval with graph-aware spatial exploration and LLM-based reasoning to improve function-level code localization under a fixed retrieval (K). Given an issue description, all functions are first ranked by semantic similarity, and the top-K candidates are retrieved. The top-C (C<K) functions serve as centers (func 1)  for structured neighborhood exploration along contains edges in the code graph, where neighboring functions within d hops are considered if they also rank within the top-N semantically (funcs 2,3,4). An LLM then filters these candidates to identify functions that are likely relevant but under-ranked by semantic similarity alone (func3). The selected neighbors are inserted immediately below their corresponding centers while discarding an equal number of bottom ranked functions in the initial top-K list, thereby increasing coverage of structurally related buggy functions without increasing K. Further algorithmic details are provided in \cref{sec:algorithm_app} (\cref{alg:spider}).}
  \label{fig:spider_framework}
\end{figure*}

Recent work has demonstrated the potential of agents powered by Large Language Models~(LLMs) for automated bug repair and feature implementation within large-scale software repositories~\citep{yang2024swe, aider, xia2024agentlessdemystifyingllmbasedsoftware}. A critical prerequisite for effective code generation is the precise identification of relevant code units such as functions, classes, or files. This code localization task is a foundational yet difficult challenge, particularly at the fine-grained function level~\citep{jimenez2024swebench, rashid2025swepolybenchmultilanguagebenchmarkrepository, zan2025multiswebenchmultilingualbenchmarkissue, chen2025locagent}. Poor retrieval directly compromises generated code, leading to incorrect or incomplete fixes and increased computational cost. It is therefore essential not only for autonomous patch generation but also for \emph{developer oversight}: when retrieval choices are unexplained, reviewers cannot verify what the agent considered, eroding trust.

Current approaches to code localization fall into two primary categories. Some methods exploit graph representations of code repositories alongside the reasoning capabilities of LLMs~\citep{ouyang2025repograph, chen2025locagent, jiang2025cosilsoftwareissuelocalization}, mostly using sparse retrieval such as \bm{}~\citep{Robertson1994OkapiATBM25} to identify relevant graph nodes or subgraphs for agent exploration. Others train bimodal encoders with a contrastive objective to align dense embeddings of code units with issue descriptions~\citep{fehr-etal-2025-coret, reddy2025sweranksoftwareissuelocalization}, ranking units by embedding similarity.

These existing approaches face three critical limitations. First, contrastive learning-based retrieval methods overlook the structural context of code modules within a repository when ranking them \citep{reddy2025sweranksoftwareissuelocalization, zhang2024codesage, suresh2025cornstack}. We observe that buggy code modules are often spatially proximate within the codebase (e.g., in the same file or in neighboring functions), and top-ranked entities retrieved via dense semantic similarity frequently reside \textit{near} the actual buggy modules; relying exclusively on semantic similarity may thus overlook relevant candidates with \emph{marginally} lower embedding scores. Second, existing methods have been primarily evaluated on Python~\citep{chen2025locagent, fehr-etal-2025-coret}, particularly \swelite{}, \sweverified{}, and \locbench{}~\citep{jimenez2024swebench, chen2025locagent}, leaving their performance on other languages largely underexplored.
Third, function-level retrieval, the most challenging granularity compared to class-level and file-level retrieval~\citep{rashid2025swepolybenchmultilanguagebenchmarkrepository, reddy2025sweranksoftwareissuelocalization, chen2025locagent}, remains largely understudied.

These limitations matter most in interactive, developer-in-the-loop settings, where retrieval must run cheaply on a developer's active repository at session start and expose its choices to the reviewer. Function-level granularity matches how developers reason about edits, and \ours{}'s graph-based candidate expansion attaches a structural reason for inclusion to each surfaced function (its position relative to a top-ranked seed) rather than only an embedding score. The code graph is built on-demand from per-repository syntax trees, avoiding the offline corpus-wide preprocessing of many graph-based pipelines.

To address these limitations, we introduce \ours{}, a simple graph-aware dense retrieval strategy for \textit{function-level} code localization. \ours{} incorporates spatial locality as a complementary signal to semantic similarity, jointly exploiting code content and repository structure within a fixed retrieval budget. Our focus is on retrieval methods that augment downstream tasks such as precise localization and automated patch generation for GitHub-style issues. For evaluation across multiple programming languages, we also introduce \oursbench{}, covering Python, Java, JavaScript, and TypeScript repositories from \sweverified{}, \swepoly{}, and \multiswe{}.

Our primary contributions are: \underline{(1)} We propose \ours{}, a simple graph-aware dense retrieval strategy combining semantic content and code structure to identify the most relevant functions under a fixed retrieval budget. \underline{(2)} We introduce \oursbench{}, a heterogeneous graph-structured benchmark for multi-language code localization, built from GitHub repositories in \sweverified{}, \swepoly{}, and \multiswe{} covering Python, Java, JavaScript, and TypeScript, with source code as node-level features. We validate \ours{} against existing dense retrievers and the sparse \bm{} baseline across all four languages.

\section{Related Work}

\noindent
\textbf{Code Retrieval and Software Issue Localization} \quad Code retrieval refers to identifying code locations responsible for a software issue. Classical approaches are rooted in information retrieval, leveraging lexical or semantic similarity to rank candidate code snippets. Consequently, many existing systems rely on sparse retrievers such as BM25~\citep{Robertson1994OkapiATBM25}. While BM25 indexing is cheaper than dense embedding generation and vector similarity search, it typically underperforms when rich semantic representations are available \citep{reddy2025sweranksoftwareissuelocalization, fehr-etal-2025-coret}.

Recent dense retrieval approaches propose bimodal encoder models to encode both code chunks as well as issue embeddings in an aligned latent space~\citep{zhang2024codesage, suresh2025cornstack, feng-etal-2020-codebert, guo2021graphcodebertpretrainingcoderepresentations}. However, these models are typically pretrained on generic natural language–to–code objectives, which leads to suboptimal performance for issue-driven code retrieval. To address this gap, \citet{fehr-etal-2025-coret} and \citet{reddy2025sweranksoftwareissuelocalization} introduce bimodal encoders fine-tuned on GitHub-style issue resolution datasets such as \texttt{SWEBench}~\citep{jimenez2024swebench}, resulting in improved alignment between issue and code representations. Despite these gains, \texttt{SWEBench} is limited to Python repositories, and no comparably large multilingual dataset currently exists to support training such encoders across diverse programming languages. Consequently, as real-world codebases span multiple languages, these fine-tuned dense retrievers exhibit degraded performance under domain shifts, reducing their robustness and reliability. \emph{We argue that combining LLM-based reasoning with structural signals from code graphs complements semantic similarity and reduces sensitivity to surface-level language artifacts}. Augmenting dense retrievers with these components helps mitigate spurious correlations and improves generalization. We empirically validate this hypothesis on multilingual benchmarks, including \swepoly{} \citep{rashid2025swepolybenchmultilanguagebenchmarkrepository}, \multiswe{} \citep{zan2025multiswebenchmultilingualbenchmarkissue}, and \sweverified{} \citep{jimenez2024swebench}.

\noindent
\textbf{LLM-based retrieval methods} \quad Recent work has shown that large language models (LLMs) are highly effective at localizing and resolving GitHub-style software issues, largely due to their strong reasoning capabilities \citep{Kang_2024_llm-based_faultlocalization, xia2024agentlessdemystifyingllmbasedsoftware, luo2024repoagent, orwall2024moatless}. In the context of code retrieval, \citet{xia2024agentlessdemystifyingllmbasedsoftware} introduce a simple hierarchical localization strategy driven directly by an LLM, whereas \citet{ouyang2025repograph} employ an intermediate sparse BM25 retrieval stage to identify relevant files prior to code generation. Similarly, \citet{chen2025locagent} leverage inverted BM25 indexing to match issue-description keywords with node identifiers or code chunks. Both \citet{ouyang2025repograph} and \citet{chen2025locagent} further enable LLM-guided traversal of local subgraphs around the retrieved files or nodes, using edges defined by their respective code graph representations. In contrast, \citet{reddy2025sweranksoftwareissuelocalization} focus on re-ranking densely retrieved code chunks using LLMs, with the goal of enhancing semantic retrieval through reasoning. Although this approach improves retrieval performance, it does not exploit the structural signals encoded in code graphs. \textit{We aim to bridge this gap by jointly leveraging LLM reasoning, semantic embeddings, and the additional contextual information captured by code graphs, thereby improving overall retrieval effectiveness.  }

\noindent
\textbf{Code Graph Construction} \quad Several recent approaches leverage explicit code graph representations to expose structural information that is otherwise difficult to capture with purely textual or embedding-based methods. The specific utility of such graphs depends critically on how nodes and relations are defined. For example, \citet{ouyang2025repograph} construct graphs at the line-of-code level, where nodes correspond to variable or module definitions and references, and edges encode \textit{invokes} and \textit{contains} relationships. While this design enables fine-grained structural reasoning, it scales poorly with repository size, leading to dense and complex graphs that are challenging for LLM-based agents to traverse at inference time.

To improve scalability, other works adopt coarser graph abstractions. \citet{jiang2025cosilsoftwareissuelocalization} propose two complementary graph types: a module call graph, connecting files via \textit{imports}, and a function call graph, linking functions and classes through \textit{invokes} and \textit{inherits} relations. More recently, \citet{chen2025locagent} introduce a unified graph formulation in which nodes represent functions, classes, files, or directories, and edges capture \textit{contains}', \textit{invokes}', \textit{imports}', and `\textit{inherits}' relationships. This design strikes a balance between expressivity and tractability, enabling structural reasoning at multiple hierarchical levels while remaining amenable to LLM-guided traversal. \textit{Following this line of work, we adopt their graph construction, as it naturally supports retrieval at the file, class, and function levels—granularities that are common across programming languages and well aligned with multilingual code retrieval settings.}

\section{Problem Definition and Notations}
\label{sec:notations}
We formalize the function-level code retrieval task and its evaluation, applicable to any retrieval method operating over structured code representations.

Given a codebase and an issue description $Q$, the goal is to retrieve the top-$K$ functions that are most likely to require edits to resolve the issue.
We represent the codebase as a graph $\mathcal{G} = (\mathcal{V}, \mathcal{E})$ with $n$ nodes, where $\mathcal{V} = \{v_i\}_{i=1}^{n}$ denotes the set of code entities and $\mathcal{E} \subseteq \mathcal{V} \times \mathcal{V}$ encodes structural relations. Each node $v \in \mathcal{V}$ corresponds to a code entity, such as a function, class, file, or directory; and edges represent relationships of types \textit{contains}, \textit{invokes}, \textit{imports}, and \textit{inherits}. Let $\mathcal{V}^* = \{v_i^*\}_{i=1}^{m} \subseteq \mathcal{V}$ denote the ground-truth set of $m$ relevant function nodes for issue $Q$.

\textbf{Scoring Function} \quad A bi-modal encoder $\mathcal{F}(\cdot)$ maps both code units and issue descriptions to a shared embedding space. The relevance score of node $v$ for issue $Q$ is: $$s_Q(v) = \cos\bigl(\mathcal{F}(v), \mathcal{F}(Q)\bigr)$$

\textbf{Baseline Dense Retrieval} \quad Standard dense embedding retrieval (DER) ranks all nodes by their relevance scores and returns the top-$K$ candidates: $$\mathcal{S}_K(Q) := \underset{v \in \mathcal{V}}{\mathrm{arg\,top}_{K}}\; s_Q(v)$$ where $\mathrm{arg\,top}_K$ returns $K$ nodes with top-$K$ scores $s_Q(\cdot)$. We use notation $\mathcal{S}_K$ and $\mathcal{S}_K(Q)$ interchangeably.

\section{Methodology}
\begin{table*}[!tb]
\centering
\caption{\textbf{\oursbench{} data statistics after graph construction (GT = ground truth).} Function-level edits dominate across benchmarks; edit frequency and localization difficulty vary substantially across languages.}
\resizebox{0.85\textwidth}{!}{
    \begin{tabular}{cr|cc|cc|cc|cc}
    \toprule
    \multirow{3}{*}{\textbf{GT Type}} & \multirow{3}{*}{\textbf{Original Dataset}} & \multicolumn{2}{c|}{\textbf{Python}} & \multicolumn{2}{c|}{\textbf{Java}} & \multicolumn{2}{c}{\textbf{JavaScript}} & \multicolumn{2}{c}{\textbf{TypeScript}} \\
     & & \% Insts. with & Avg. GT edits  & \% Insts. with & Avg. GT edits & \% Insts. with & Avg. GT edits & \% Insts. with & Avg. GT edits \\
     & & GT edits & per inst. & GT edits & per inst. & GT edits & per inst. & GT edits & per inst. \\
    \midrule
    \multirow{3}{*}{Function} & \swepoly{} & 85.93 & 3.73 & 70.30 & 5.52 & 53.98 & 1.98  & 29.63 & 2.68 \\
    & \multiswe{} & N/A & N/A & 59.54 & 2.46 & 81.74 & 3.88 & 27.68 & 2.29 \\
    & \sweverified{} & 89.00 & 1.82 & N/A & N/A & N/A & N/A & N/A & N/A \\
    \midrule
    \multirow{3}{*}{Class} & \swepoly{} & 75.38 & 2.9 & 87.27 & 3.07 & 20.45 & 1.37 & 19.07 & 1.91 \\
    & \multiswe{} & N/A & N/A & 81.68 & 1.58 & 9.55 & 1.06 & 3.13 & 1.57 \\
    & \sweverified{} & 74.00 & 1.40 & N/A & N/A & N/A & N/A & N/A & N/A \\
    \midrule
    \multirow{3}{*}{File} & \swepoly{} & 99.50 & 1.91 & 100.00 & 3.2 & 58.31 & 1.68 & 55.97 & 2.00 \\
    & \multiswe{} & N/A & N/A & 97.71 & 1.56 & 98.60 & 2.37 & 70.54 & 5.08 \\
    & \sweverified{} & 100.00 & 1.24 & N/A & N/A & N/A & N/A & N/A & N/A \\
    \bottomrule
    \end{tabular}
}
\label{tab:data_stats}
\end{table*}
\subsection{Building Code Graphs}
For a given codebase, we construct a code graph using only files written in the primary programming language (the language comprising the majority of the repository). Files in other languages are ignored to simplify construction; e.g., in a Python repository, we consider only Python source files.

We follow the graph schema proposed by \citet{chen2025locagent}, where each node represents a code entity—function, class, file, or directory—and edges encode structural relations of types \textit{imports}, \textit{invokes}, \textit{contains}, and \textit{inherits}. For Python, we use the built-in \texttt{ast} module\footnote{\url{https://docs.python.org/3/library/ast.html}}
 to parse source files and extract syntax trees. For Java, JavaScript, and TypeScript, we rely on Tree-sitter\footnote{\url{https://github.com/tree-sitter}} for syntax parsing.

Graph construction proceeds by traversing the repository hierarchy. Directory nodes are added to enable navigation across the project structure. File nodes are included only if they contain at least one function or class definition; files containing solely metadata or configuration are ignored. Function and class nodes, along with the edges connecting them, are extracted from the corresponding syntax trees. While directory nodes do not store code content, all other node types include their associated source code.

Constructing accurate graphs for Java, JavaScript and TypeScript presents additional challenges due to the flexible ways in which functions and classes can be declared/implemented. For example, class definitions and method implementations may be split across multiple locations or files. We explicitly handle such language-specific cases using Tree-sitter to ensure faithful graph construction. Refer App.~\ref{sec:graph_construction_details_app}.

Although the graph contains multiple node types, we focus on retrieving \emph{functions}, the most granular and most challenging unit. In \swepoly{}, $67\%$ of instances involve function-level edits (avg.\ $2.78$ edits/instance) versus $1.42\%$ class-only \citep{rashid2025swepolybenchmultilanguagebenchmarkrepository}, and accurate function retrieval naturally helps file-level localization as well.

Table~\ref{tab:data_stats} reports dataset statistics, including the proportion of retained instances per language. For function-level retrieval, we exclude instances involving only file- or class-level edits. Similarly, for class- and file-level retrieval, we ignore instances with edits exclusively at other granularities.

\subsection{Leveraging Code Graph via \ours{}}\label{sec:other_notations}

\cref{fig:spider_framework} provides a high-level overview of \ours{}, with algorithmic details in \cref{alg:spider}~(\cref{sec:algorithm_app}). Given an issue description $Q$ and a codebase represented as graph $\mathcal{G}$, \ours{} retrieves the top-$K$ functions most likely to require modification. The framework extends to class- and file-level retrieval by aggregating code chunks when full contexts exceed the LLM budget (see \cref{sec:class_level_performance_app}).

\ours{}'s design is motivated by the observation that, in instances requiring multiple edits, relevant functions tend to be structurally proximate in the graph due to call dependencies and containment relationships. Dense semantic retrieval provides broad coverage but often misses ``near-miss'' cases where the correct function lies close \emph{structurally} but not \emph{semantically} to top-ranked candidates.

\textbf{Semantic Retrieval} \quad We first compute the baseline dense retrieval
$\mathcal{S}_K(Q)$ as defined in \cref{sec:notations}.
Additionally, we compute $\mathcal{S}_N(Q)$ containing
the top-$N$ nodes ($N > K$) to constrain neighborhood exploration.

\textbf{Seed Selection} \quad From $\mathcal{S}_K(Q)$, we select the top-$C$ ranked nodes as seed centers for graph exploration:
$$\mathcal{C}_Q := {v \in \mathcal{S}_C(Q)}$$ where $C \leq K$. A seed anchors a local neighborhood search.

\textbf{Neighborhood Exploration} \quad For each seed, we perform breadth-first search up to depth $d$ along `contains' edges, collecting structurally proximate function nodes.
Formally, the $d$-hop neighborhood is:
$$\Gamma_d(\mathcal{C}_Q) := \{u \in \mathcal{V} \mid \mathrm{dist}_{\mathcal{G}}(u, \mathcal{C}_Q) \leq d\}$$
where $\mathrm{dist}_{\mathcal{G}}(u, \mathcal{C}_Q)$ denotes the shortest-path distance from $u$ to any node in $\mathcal{C}_Q$.
Two functions within the same class are 2 hops apart; functions in different classes across separate files are 4 hops apart.
While we use `contains' edges to capture hierarchical structure, the framework can incorporate other edge types like `invokes', `imports', and `inherits'.

\textbf{Two-Stage Neighbor Filtering} \quad
To control computational cost while maintaining quality, candidates are filtered in two stages.
\emph{Stage 1 (Semantic Filtering):} We restrict neighbors to those also appearing in the top-$N$ candidate pool: $$\hat{\mathcal{C}}_Q := \Gamma_d(\mathcal{C}_Q) \cap \mathcal{S}_N(Q)$$ This ensures exploration remains within semantically relevant regions of the graph.
\emph{Stage 2 (LLM-based Selection):} An LLM selector $\mathcal{L}$ evaluates each candidate $u \in \hat{\mathcal{C}}_Q$, receiving the source code content and issue description as context; the full prompt is in \cref{sec:prompt_app}.
The LLM returns a binary relevance decision $\mathcal{L}(u) \in \{0, 1\}$. Note that seed centers $\mathcal{C}_Q$ are always retained in $\mathcal{U}_Q$, where: $$\mathcal{U}_Q := {u \in \hat{\mathcal{C}}_Q \mid \mathcal{L}(u) = 1} \cup \mathcal{C}_Q$$

\textbf{Output Construction} \quad To maintain a fixed retrieval budget of $K$,
newly selected neighbors replace the lowest-ranked non-center nodes. Only
candidates the baseline missed consume budget, so let
$\gP := (\Gamma_d(\gC_Q) \cap \gS_N(Q)) \setminus \gS_K$ be the new proposals
and $\gR \subseteq \gS_K$ the protected items, namely the seed centers and any
accepted candidate already in $\gS_K$. Writing $\gA \subseteq \gP$ for the set
admitted after Stage-2 selection, deduplication and capacity truncation,
$J = |\gA| \le K - |\gR|$, and $\gD_J$ for the $J$ lowest-ranked members of
$\gS_K \setminus \gR$, the final retrieval set is
\begin{equation}
\label{eq:exchange_output}
\gS^{\mathrm{LLM}}_K(Q) := (\gS_K \setminus \gD_J) \cup \gA .
\end{equation}
Seed centers already lie in $\gS_K$ and displace nothing.

\textbf{Hyperparameters} \quad \ours{} introduces four hyperparameters:
retrieval budget $K$, number of seed centers $C$, exploration depth $d$,
and semantic filtering threshold $N$. We analyze their effects in \cref{sec:ablations}.

Let $\eta := \E|\gA \cap \gV^*| / \E|\gA|$ denote the \emph{admitted
precision} and $\tau := \E|\gD_J \cap \gV^*| / \E|\gD_J|$ the \emph{frontier
relevance}, both defined whenever $\E|\gA| > 0$.

\begin{proposition}[Exact exchange]
\label{prop:recall_condition}\label{prop:exchange}
Write $\Delta\Reck := \Reck(\gS^{\mathrm{LLM}}_K) - \Reck(\gS_K)$. For every
realization of $Q$, the labels and the policy,
\begin{equation}
\label{eq:exact_recall}
\Delta\Reck = \frac{|\gA \cap \gV^*| - |\gD_J \cap \gV^*|}{m},
\end{equation}
and hence, at fixed $m$ and with no further assumptions,
\begin{equation}
\label{eq:condition}
\E[\Delta\Reck] > 0 \iff \eta > \tau .
\end{equation}
\end{proposition}

\ours{} is therefore a strict exchange. A promotion helps only if the admitted
function is more likely to be edited than the item it displaces, so the
comparison is against the \emph{eviction frontier} $\tau$, the relevance of a
rank-$K$ dense hit, not the repository base rate. Graph proximity can clear
that base rate and still fall below $\tau$. Since \cref{eq:condition} is an
equivalence, neither locality nor a
selector with $\alpha > \beta$ implies improvement on its own. The selector
still governs the outcome, because $\eta$ is determined by $\alpha$ and
$\beta$ composed with the top-$N$ gate (\cref{sec:theory_rates}), but it must
move $\eta$ past $\tau$. \cref{sec:theory} proves these statements, treats
\acck{}, and models how candidate quality decays with distance, yielding conditions for a finite useful radius and an LLM-pivotal regime;
\cref{sec:edge_significance_app} reports edge types where it fails.

\section{Experiments and Results}
We organize our experiments around three research questions. \textbf{RQ1 (Exploiting Graph Structure):} Can incorporating structural relationships between code modules improve retrieval beyond what semantic similarity alone achieves? \textbf{RQ2 (Cross-Language Robustness):} Do graph-aware retrieval strategies generalize across programming languages and codebases of varying complexity? \textbf{RQ3 (Efficiency and Practicality):} Can they achieve improved coverage without excessive computational cost or LLM invocations?

\subsection{Experimental Setup}

\begin{figure*}
    \centering
    \includegraphics[width=0.9\linewidth]{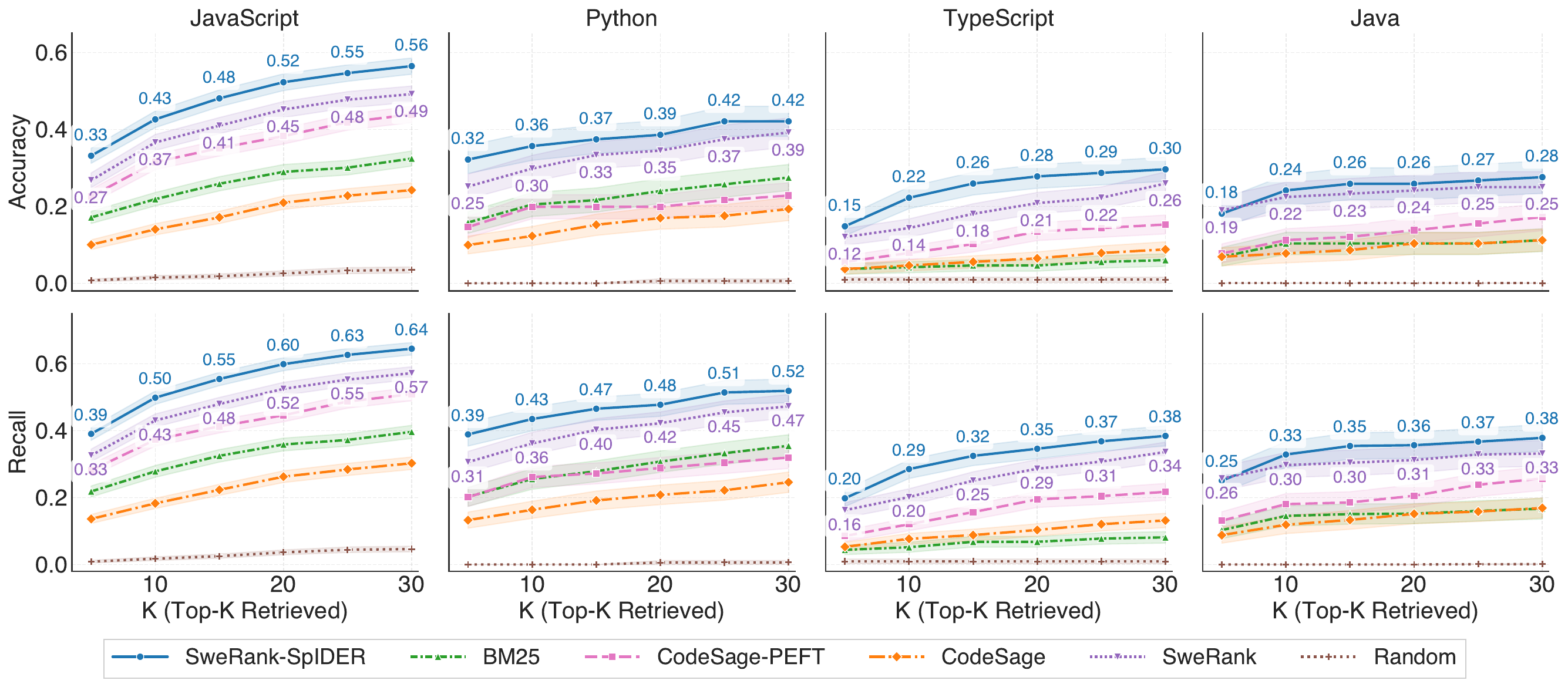}
    \caption{\textbf{Function-level retrieval performance on \swepoly{}.} SWERankEmbed-Small+\ours{} consistently outperforms dense and sparse baselines in Accuracy and Recall across all languages and retrieval budgets $K$. Shaded envelopes indicate standard deviations estimated via bootstrapping with \num{1000} draws.}
    \label{fig:der_performance}
\end{figure*}

\textbf{Datasets} \quad We focus on challenging multi-function edits and evaluate retrieval across multiple languages. We use \sweverified{}, \swepoly{}, and \multiswe{}, with patched functions as ground truth, and following \citet{suresh2025cornstack} exclude instances without function-level modifications. The evaluation spans Python, Java, TypeScript, and JavaScript repositories, assessing both the robustness of our approach and the generalization of baselines across languages and codebase complexities.

\textbf{Evaluation Metrics} \quad  Given a query $Q$ and a retrieved set $\mathcal{S}(Q)$ of size $K$, we evaluate retrieval quality using Recall@$K$ and Acc@$K$:
\begin{align*}
\text{Recall@}K &= \frac{|\mathcal{V}^* \cap \mathcal{S}(Q)|}{|\mathcal{V}^*|}, \\
\text{Acc@}K &= \mathbb{I}[\mathcal{V}^* \subseteq \mathcal{S}(Q)],
\end{align*}
where $\mathcal{V}^*$ denotes the ground-truth patched functions. Acc@$K$ is the stricter metric, equal to 1 only when \emph{all} of $\mathcal{V}^*$ is in the top-$K$ set. Both directly capture coverage under a fixed retrieval budget; we additionally report MRR in App.~\ref{sec:additional_exps_app}.

\textbf{Hyperparameters} \quad We analyze hyperparameters $C$, $d$, $N$, and $K$ in \cref{sec:ablations}. Following prior work \citep{reddy2025sweranksoftwareissuelocalization, fehr-etal-2025-coret}, we fix $K=20$ across datasets to ensure fair comparison with baselines sensitive to this choice (e.g., embedding-based retrieval with LLM reranking). The remaining \ours{} hyperparameters are selected via grid search on the Python split of \swepoly{}, yielding $C=5$, $d=4$, $N=500$; we hold these constant across all experiments to avoid overfitting, and App.~\ref{sec:sensitivity_app} verifies from the per-language sweeps that this configuration is at or near each language's optimum. Our ablations further examine how these parameters affect performance and practical considerations such as LLM context length and invocation budget. We use \texttt{Claude Sonnet~4} with temperature $0.1$ for consistent and reproducible evaluation; results with open-source Qwen-Coder variants as the LLM filter are reported in \cref{tab:spider_results}.

\textbf{Baselines} \quad
We compare \ours{} against representative state-of-the-art retrieval approaches spanning both dense and sparse methods. For dense retrieval, we consider \swerankszs{} and \codesageszs{}, which are widely used embedding-based models for software issue localization and code retrieval \citep{reddy2025sweranksoftwareissuelocalization, zhang2024codesage}. As a sparse retrieval baseline, we include BM25 \citep{Robertson1994OkapiATBM25}.

Since \codesageszs{} is not trained for GitHub-style problem-solving tasks, we additionally evaluate a parameter-efficiently fine-tuned variant using LoRA \citep{hu2022lora} on \num{2000} instances from three repositories in the \swebench{} training set, denoted as \codesagespeft{}. This adaptation budget is small, so \codesagespeft{} remains a comparatively weak dense baseline; the load-bearing dense comparison throughout is \swerankszs{}, which is trained on roughly \num{3300} repositories. \citet{fehr-etal-2025-coret} did not release their model or code, preventing direct comparison; \locagent{} \citep{chen2025locagent} is evaluated only on \sweverified{} and the Python split of \swepoly{} (tables~\ref{tab:locagent_performance} and~\ref{tab:locagent_vs_spider}) due to its iterative LLM-driven exploration. Beyond these, App.~\ref{sec:lexical_agentic_app} compares \ours{} against a lexical \texttt{grep} baseline and the \texttt{OpenHands} agent \citep{wang2025openhands} at matched $K$, and App.~\ref{sec:faq_app} explains why other graph-based localizers are excluded.

\subsection{Results}
\begin{table*}[!tbp]
\centering
\caption{\textbf{Main result.} Across all embedding models and languages, \ours{} consistently improves Recall and Accuracy over dense retrieval alone; gains are preserved and often amplified after LLM reranking. $K=20$, $N=500$, $C=5$, $d=4$; best per embedding model in light blue. \spisr{} is the spatially informed sparse retriever analogous to \ours{}. KDE plots in Figs.~\ref{fig:kde_plots_recall_swepolybench},~\ref{fig:kde_plots_acc_swepolybench}; confidence intervals in \cref{tab:swepolybench_ci_compact}.}
\resizebox{0.7\textwidth}{!}{
\begin{tabular}{l c c|cc|cc|cc}
\toprule
\multirow{3}{*}{\textbf{Language}} & \multirow{3}{*}{\textbf{Model}} & \multirow{3}{*}{\textbf{Retrieval}}
& \multicolumn{4}{c|}{\textbf{\swepoly{}}} & \multicolumn{2}{c}{\textbf{\sweverified{}}} \\
\cmidrule{4-9}
& &
& \multicolumn{2}{c|}{\textbf{Retrieval Only}}
& \multicolumn{2}{c|}{\textbf{Retrieval + LLM reranker}}
& \multicolumn{2}{c}{\textbf{Retrieval Only}}\\
 & &
 & \textbf{Recall@20} & \textbf{Accuracy@20}
 & \textbf{Recall@3} & \textbf{Accuracy@3}
 & \textbf{Recall@20} & \textbf{Accuracy@20} \\
\midrule
\multirow{8}{*}{Python}
 & \multirow{2}{*}{\texttt{\swerankszs}}
 & \texttt{\ours{}} & \cellcolor{winnerblue}0.49 & \cellcolor{winnerblue}0.40
 & \cellcolor{winnerblue}0.44 & \cellcolor{winnerblue}0.36 & \cellcolor{winnerblue}0.61 & \cellcolor{winnerblue}0.56 \\
 &  & \texttt{DER} & 0.42 & 0.35
 & 0.38 & 0.31 & 0.54 & 0.49 \\
 & \multirow{2}{*}{\texttt{\codesageszs{}}}
 & \texttt{\ours{}} & \cellcolor{winnerblue}0.27 & \cellcolor{winnerblue}0.21
 & \cellcolor{winnerblue}0.23 & \cellcolor{winnerblue}0.18 & \cellcolor{winnerblue}0.30 & \cellcolor{winnerblue}0.27 \\
 &  & \texttt{DER} & 0.21 & 0.17
 & 0.19 & 0.16 & 0.21 & 0.19 \\
 & \multirow{2}{*}{\texttt{\codesagespeft{}}}
 & \texttt{\ours{}} & \cellcolor{winnerblue}0.37 & \cellcolor{winnerblue}0.27
 & \cellcolor{winnerblue}0.32 & \cellcolor{winnerblue}0.25 & \cellcolor{winnerblue}0.55 & \cellcolor{winnerblue}0.49 \\
 &  & \texttt{DER} & 0.29 & 0.20
 & 0.26 & 0.20 & 0.43 & 0.38 \\
 & \multirow{2}{*}{\texttt{\bm}}
 & \texttt{\spisr{}} & \cellcolor{winnerblue}0.33 & \cellcolor{winnerblue}0.26
 & \cellcolor{winnerblue}0.30 & \cellcolor{winnerblue}0.23 & \cellcolor{winnerblue}0.45 & \cellcolor{winnerblue}0.41 \\
 &  & \texttt{\sr{}} & 0.31 & 0.24
 & 0.27 & 0.22 & 0.38 & 0.34 \\
\midrule
 \multicolumn{7}{c|}{} & \multicolumn{2}{c}{\textbf{\multiswe{}}}\\
\midrule
\multirow{8}{*}{Java}
 & \multirow{2}{*}{\texttt{\swerankszs}}
 & \texttt{\ours{}} & \cellcolor{winnerblue}0.36 & \cellcolor{winnerblue}0.27
 & \cellcolor{winnerblue}0.25 & \cellcolor{winnerblue}0.17 & \cellcolor{winnerblue}0.37 & \cellcolor{winnerblue}0.28 \\
 &  & \texttt{DER} & 0.31 & 0.24
 & 0.24 & 0.16 & 0.32 & 0.24 \\
 & \multirow{2}{*}{\texttt{\codesageszs{}}}
 & \texttt{\ours{}} & \cellcolor{winnerblue}0.23 & \cellcolor{winnerblue}0.17
 & \cellcolor{winnerblue}0.16 & \cellcolor{winnerblue}0.09 & \cellcolor{winnerblue}0.21 & \cellcolor{winnerblue}0.19 \\
 &  & \texttt{DER} & 0.15 & 0.10
 & 0.11 & 0.07 & 0.15 & 0.13 \\
 & \multirow{2}{*}{\texttt{\codesagespeft{}}}
 & \texttt{\ours{}} & \cellcolor{winnerblue}0.28 & \cellcolor{winnerblue}0.18
 & \cellcolor{winnerblue}0.20 & \cellcolor{winnerblue}0.14 & \cellcolor{winnerblue}0.31 & \cellcolor{winnerblue}0.25 \\
 &  & \texttt{DER} & 0.21 & 0.14
 & 0.17 & 0.11 & 0.23 & 0.17 \\
 & \multirow{2}{*}{\texttt{\bm}}
 & \texttt{\spisr{}} & \cellcolor{winnerblue}0.16 & \cellcolor{winnerblue}0.10
 & \cellcolor{winnerblue}0.12 & 0.08 & \cellcolor{winnerblue}0.23 & \cellcolor{winnerblue}0.17 \\
 &  & \texttt{\sr{}} & 0.15 & 0.10
 & \cellcolor{winnerblue}0.12 & \cellcolor{winnerblue}0.09 & 0.22 & \cellcolor{winnerblue}0.17 \\
\midrule
\multirow{8}{*}{JavaScript}
 & \multirow{2}{*}{\texttt{\swerankszs}}
 & \texttt{\ours{}} & \cellcolor{winnerblue}0.60 & \cellcolor{winnerblue}0.52
 & \cellcolor{winnerblue}0.43 & \cellcolor{winnerblue}0.37 & \cellcolor{winnerblue}0.39 & \cellcolor{winnerblue}0.30 \\
 &  & \texttt{DER} & 0.52 & 0.45
 & 0.41 & 0.34 & 0.31 & 0.23 \\
 & \multirow{2}{*}{\texttt{\codesageszs{}}}
 & \texttt{\ours{}} & \cellcolor{winnerblue}0.38 & \cellcolor{winnerblue}0.32
 & \cellcolor{winnerblue}0.30 & \cellcolor{winnerblue}0.25 & \cellcolor{winnerblue}0.17 & \cellcolor{winnerblue}0.12 \\
 &  & \texttt{DER} & 0.26 & 0.21
 & 0.22 & 0.17 & 0.09 & 0.06 \\
 & \multirow{2}{*}{\texttt{\codesagespeft{}}}
 & \texttt{\ours{}} & \cellcolor{winnerblue}0.55 & \cellcolor{winnerblue}0.48
 & \cellcolor{winnerblue}0.41 & \cellcolor{winnerblue}0.35 & \cellcolor{winnerblue}0.32 & \cellcolor{winnerblue}0.24 \\
 &  & \texttt{DER} & 0.45 & 0.38
 & 0.36 & 0.30 & 0.25 & 0.18 \\
 & \multirow{2}{*}{\texttt{\bm}}
 & \texttt{\spisr{}} & \cellcolor{winnerblue}0.46 & \cellcolor{winnerblue}0.39
 & \cellcolor{winnerblue}0.36 & \cellcolor{winnerblue}0.30 & \cellcolor{winnerblue}0.21 & \cellcolor{winnerblue}0.14 \\
 &  & \texttt{\sr{}} & 0.36 & 0.29
 & 0.31 & 0.25 & 0.14 & 0.10 \\
\midrule
\multirow{8}{*}{TypeScript}
 & \multirow{2}{*}{\texttt{\swerankszs}}
 & \texttt{\ours{}} & \cellcolor{winnerblue}0.35 & \cellcolor{winnerblue}0.28
 & \cellcolor{winnerblue}0.27 & \cellcolor{winnerblue}0.21 & \cellcolor{winnerblue}0.52 & \cellcolor{winnerblue}0.42 \\
 &  & \texttt{DER} & 0.29 & 0.21
 & 0.25 & 0.19 & 0.40 & 0.32 \\
 & \multirow{2}{*}{\texttt{\codesageszs{}}}
 & \texttt{\ours{}} & \cellcolor{winnerblue}0.16 & \cellcolor{winnerblue}0.10
 & \cellcolor{winnerblue}0.14 & \cellcolor{winnerblue}0.09 & \cellcolor{winnerblue}0.33 & \cellcolor{winnerblue}0.26 \\
 &  & \texttt{DER} & 0.10 & 0.06
 & 0.09 & 0.06 & 0.26 & 0.21 \\
 & \multirow{2}{*}{\texttt{\codesagespeft{}}}
 & \texttt{\ours{}} & \cellcolor{winnerblue}0.25 & \cellcolor{winnerblue}0.18
 & \cellcolor{winnerblue}0.22 & \cellcolor{winnerblue}0.16 & \cellcolor{winnerblue}0.30 & \cellcolor{winnerblue}0.26 \\
 &  & \texttt{DER} & 0.19 & 0.13
 & 0.18 & 0.12 & 0.19 & 0.15 \\
 & \multirow{2}{*}{\texttt{\bm}}
 & \texttt{\spisr{}} & \cellcolor{winnerblue}0.11 & \cellcolor{winnerblue}0.09
 & \cellcolor{winnerblue}0.10 & \cellcolor{winnerblue}0.07 & \cellcolor{winnerblue}0.32 & \cellcolor{winnerblue}0.27 \\
 &  & \texttt{\sr{}} & 0.07 & 0.05
 & 0.07 & 0.05 & 0.24 & 0.21 \\
\bottomrule
\end{tabular}
}
\label{tab:swepoly_combined}
\end{table*}

\begin{table*}[!tb]
\centering
\caption{\textbf{Ablation on the edge type used for neighborhood exploration}, on \swepoly{} with \texttt{SweRankEmbed-Small} at $K=20$, $N=500$, $C=5$ (single fixed-seed run). ``Tok'' is the average total (input $+$ output) LLM tokens per instance; \der{} issues no LLM call, hence $0$. Best Recall@20 and Acc@20 per language in light blue. For the denser \textit{invokes} graph we cap exploration at the $100$ highest-ranked neighbors per query before Stage-2 filtering, to bound the token budget. \textit{inherits} expansion is degenerate at function granularity: no \textit{inherits} edge in our schema is incident to a function node in any of the four languages, so it can surface no function beyond \der{}. We report the Python run as empirical confirmation, since it returns exactly the \der{} set, and mark the remaining languages N/A, since the configuration is inapplicable rather than unmeasured.}
\resizebox{\textwidth}{!}{
\begin{tabular}{l|ccc|ccc|ccc|ccc}
\toprule
\multirow{2}{*}{\textbf{Exploration edges / depth}}
& \multicolumn{3}{c|}{\textbf{Python}}
& \multicolumn{3}{c|}{\textbf{Java}}
& \multicolumn{3}{c|}{\textbf{JavaScript}}
& \multicolumn{3}{c}{\textbf{TypeScript}} \\
& Recall@20 & Acc@20 & Tok
& Recall@20 & Acc@20 & Tok
& Recall@20 & Acc@20 & Tok
& Recall@20 & Acc@20 & Tok \\
\midrule
\der{} (no exploration)
 & 0.42 & 0.35 & 0
 & 0.31 & 0.24 & 0
 & 0.52 & 0.45 & 0
 & 0.29 & 0.21 & 0 \\
\midrule
\textit{contains} / $d{=}4$
 & 0.49 & 0.40 & 12.9K
 & 0.36 & 0.27 & 7.8K
 & 0.60 & 0.52 & 14.9K
 & 0.35 & 0.28 & 8.3K \\
\textit{invokes} / $d{=}2$
 & 0.56 & 0.50 & 109.9K
 & \cellcolor{winnerblue}0.42 & \cellcolor{winnerblue}0.30 & 87.4K
 & 0.65 & \cellcolor{winnerblue}0.58 & 55.5K
 & 0.47 & 0.41 & 92.7K \\
\textit{imports} / $d{=}2$
 & 0.44 & 0.38 & 7.9K
 & 0.31 & 0.24 & 4.6K
 & 0.52 & 0.45 & 5.2K
 & 0.29 & 0.21 & 4.9K \\
\textit{inherits} / $d{=}2$
 & 0.42 & 0.35 & 6.8K
 & N/A & N/A & N/A
 & N/A & N/A & N/A
 & N/A & N/A & N/A \\
\midrule
\textit{contains} / $d{=}4$ + \textit{invokes} / $d{=}2$
 & \cellcolor{winnerblue}0.60 & \cellcolor{winnerblue}0.52 & 110.1K
 & \cellcolor{winnerblue}0.42 & 0.29 & 90.1K
 & \cellcolor{winnerblue}0.66 & \cellcolor{winnerblue}0.58 & 60.8K
 & \cellcolor{winnerblue}0.48 & \cellcolor{winnerblue}0.42 & 98.6K \\
\bottomrule
\end{tabular}
}
\label{tab:edge_ablation}
\end{table*}

\textbf{Leveraging code graphs improves retrieval} \quad \cref{tab:swepoly_combined} demonstrates the benefit of incorporating code graph structure across Python, TypeScript, JavaScript, and Java repositories from \swepoly{}, \multiswe{}, and \sweverified{} at $K=20$. Fig.~\ref{fig:der_performance} confirms the trend on \swepoly{} as $K$ varies: \texttt{SweRankEmbed-Small + SpIDER} consistently outperforms all baselines across languages and $K$ values. The zero-shot \texttt{SweRankEmbed-Small}, trained on roughly 3,300 Python repositories, also remains competitive on non-Python codebases, outperforming \bm{} and \codesage{} in both zero-shot and PEFT settings. Across datasets, \ours{} improves Recall@20 and Acc@20 of \texttt{SweRankEmbed-Small} by at least $13\%$ and $14\%$ \emph{relative} (minimum over language/dataset cells in \cref{tab:swepoly_combined}; corresponding absolute gains range from $+0.05$ to $+0.12$ points), indicating that graph-aware retrieval yields substantial gains even when paired with already strong semantic embeddings. MRR@20 is comparable to \der{} but drops in a few cells where the first ground-truth function is a non-center node; we discuss this and the underlying mechanism in App.~\ref{sec:additional_exps_app}.

\textbf{Which edge types carry the structural signal} \quad \cref{tab:edge_ablation} ablates the edge type used for neighborhood exploration on \swepoly{} at $K=20$, $N=500$, $C=5$. Expansion along `invokes' edges outperforms `contains'-only expansion in every language, and combining the two is best overall, raising the minimum relative Recall@20 gain over \der{} across the four languages from $13\%$ to $27\%$ (Python $0.42 \to 0.60$, Java $0.31 \to 0.42$, JavaScript $0.52 \to 0.66$, TypeScript $0.29 \to 0.48$). This ordering follows the locality analysis in \cref{tab:locality_app}: Java combines the highest invoke-connectivity ($0.63$) with the lowest dense top-$100$ coverage ($29\%$), and is correspondingly the split where `contains'-only exploration cannot structurally reach cross-file ground truth. `imports' and `inherits', by contrast, add nothing over \der{} outside Python, since no \textit{inherits} edge in our schema is incident to a function node and \textit{imports} edges touch function nodes only in Python. The gains are not free: routing through `invokes' raises per-instance token consumption by roughly an order of magnitude. We therefore retain `contains' as the default configuration reported in \cref{tab:swepoly_combined}, and treat `invokes' as the setting of choice when the token budget allows. Gains from structural edges are significant under paired per-instance tests. Paired bootstrap ($B{=}10{,}000$) and two-sided paired permutation agree in every cell, giving $p<0.01$ on Recall@20 in all four languages, with Java Acc@20 under `contains' alone the single exception, resolved once `invokes' edges are added. Meanwhile `imports' and `inherits' are uniformly non-significant (App.~\cref{tab:paired_significance_app}). App.~\ref{sec:case_study_app} works through cases where each edge type recovers ground truth that dense retrieval misses, and where the Stage-2 filter errs.

\textbf{Retrieval strategy is agnostic to downstream rerankers} \quad
To validate that \ours{} is independent of downstream localization components, we pair every retrieval method with an LLM reranker following \citet{reddy2025sweranksoftwareissuelocalization}, which selects the top-3 functions from $K=20$ candidates. As shown in \cref{tab:swepoly_combined}, the improved top-20 coverage from \ours{} consistently translates into higher Recall@3 and Acc@3 after reranking, indicating that retrieval-stage gains are preserved and often amplified downstream.

\begin{table}[!tb]
\centering
\caption{\textbf{Hands-tied downstream code generation.} \texttt{mini-swe-agent} is restricted to the top-$K$ retrieved functions with no open grep, and run once with \der{} and once with \ours{} at the same budget on a \sweverified{} subset of \oursbench{}. Resolve rate is the fraction of instances whose generated patch passes; best per $K$ in light blue. Tokens are the total per instance for \der{} and for the better \ours{} variant at that $K$.}
\resizebox{\columnwidth}{!}{
\begin{tabular}{c|ccc|cc}
\toprule
\multirow{2}{*}{$\boldsymbol{K}$}
& \multicolumn{3}{c|}{\textbf{Resolve rate} $\uparrow$}
& \multicolumn{2}{c}{\textbf{Tokens/inst.} $\downarrow$} \\
& \der{} & \textit{contains}/$4$ & \textit{invokes}/$2$ & \der{} & best \ours{} \\
\midrule
5  & 0.584 & \cellcolor{winnerblue}0.623 & 0.611 & 102.6K & 62.2K \\
10 & 0.611 & \cellcolor{winnerblue}0.681 & 0.658 & 96.4K & 81.5K \\
15 & 0.626 & 0.638 & \cellcolor{winnerblue}0.646 & 122.3K & 97.6K \\
20 & 0.650 & 0.673 & \cellcolor{winnerblue}0.677 & 185.3K & 87.6K \\
\bottomrule
\end{tabular}
}
\label{tab:downstream_handstied}
\end{table}

\textbf{Improved retrieval translates to better code generation} \quad
We test whether retrieval gains survive to the end task under a \emph{hands-tied} protocol: \texttt{mini-swe-agent}~\citep{yang2024swe} is restricted to the top-$K$ retrieved functions with no open grep, so the retrieved set is the only thing that varies. Running it once with \der{} and once with \ours{} at matched budget (\cref{tab:downstream_handstied}), the better \ours{} variant resolves more instances than \der{} at every $K$, with the largest margin at $K{=}10$ ($+7.0$ pp, $18$ additional resolved instances). The margin compresses at larger $K$, where a wider budget lets \der{} include more of the ground truth on its own. The effect is strongest on efficiency: \ours{} with \textit{contains} expansion at $K{=}10$ resolves more than \der{} at $K{=}20$ ($0.681$ vs.\ $0.650$) while using $44\%$ of the tokens ($81.5$K vs.\ $185.3$K), because it reaches comparable coverage at a smaller budget. App.~\ref{sec:additional_exps_app} reports the complementary setting in which the agent keeps full tool access.

\subsection{Ablation Experiments}
\label{sec:ablations}

We ablate four hyperparameters of \ours{} to guide downstream users: the number of seed centers $C$, the retrieval budget $K$, the BFS exploration depth $d$, and the primary neighborhood filtering threshold $N$.

\textbf{Number of Top $C$ Centers} \quad Each center triggers one LLM call, so larger $C$ trades LLM cost for broader exploration. App.~\cref{tab:bfs_C_ablation} shows performance saturates at $C=5$ for $N=500$, $d=4$, $K=20$.

\textbf{Retrieval Budget $K$} \quad Performance improves monotonically with $K$ (\cref{fig:der_performance,fig:k_ablation_grid}), since a larger pool raises the chance of including relevant nodes; $K$ can be selected to match downstream task constraints.

\textbf{Exploration Depth $d$} \quad Larger $d$ widens the candidate neighborhood and improves Recall@20 and Acc@20 (App.~\cref{tab:bfs_d_ablation}), at the cost of more LLM tokens during filtering. Input tokens grow $8.2\times$ from $d=4$ to $d=8$ on Python (App.~\cref{tab:bfs_d_ablation_tokens_app}), so the deployed $d=4$ is a budget choice rather than a measured optimum, which \cref{sec:theory_llm} makes precise: pricing inference per candidate moves the useful radius inward.

\textbf{Primary Neighborhood Filtering Threshold $N$} \quad Higher $N$ admits more neighbors before LLM refinement, with saturation around $N=300$ for $C=3$, $d=4$, $K=20$ (\cref{tab:bfs_N_ablation}); like $d$, $N$ can be tuned to the LLM token budget.

\textbf{Stage-2 Filtering} \quad Disabling the selector promotes every Stage-1 neighbor and abandons the fixed budget, so its recall bounds the \emph{reach} of graph expansion rather than anything achievable at $K$. \ours{} captures $79$--$89\%$ of that reach while returning only $K$ functions (App.~\cref{tab:stage2_ceiling_app}), so the selector's admission rate $\alpha$, not the graph, is what limits $\eta$. Removing the filter would not help, since at fixed $K$ the sign is set by $\eta > \tau$.

\textbf{Takeaways} \quad Settings $C=5$, $d=4$, $N=500$, $K=20$ give strong performance across \swepoly{} at a per-query cost of $C{=}5$ LLM calls at $\sim 3$--$6$K input tokens each (\cref{tab:bfs_d_ablation_tokens_app,tab:bfs_N_ablation}); \ours{} sits roughly two orders of magnitude below iterative LLM-agent localization (\cref{tab:locagent_performance}) and one LLM round-trip slower than embedding retrieval (App.~\ref{sec:faq_app}).

\section{Conclusion}
We introduced \ours{}, a graph-aware dense retrieval framework that augments embedding similarity with structural neighborhood exploration and LLM-based filtering over per-repository code graphs.
Through rigorous empirical analysis, we demonstrated its efficacy and efficiency over existing methods in the challenging setting of function-level retrieval.
\ours{}'s exploration depth can be efficiently controlled by adjusting the subgraph size during retrieval, maintaining robust performance even with compact subgraph configurations for token-constrained applications.
While we pioneer retrieval evaluation for non-Python languages such as Java, JavaScript and TypeScript, \ours{} can be readily adapted to others as it builds on the \texttt{Tree-sitter} library.

\section*{Limitations}
\ours{} extracts a graph from the primary language of each repository; files in other languages are ignored, leaving cross-language edits out of reach. Neighborhood exploration is configurable over edge types, but the configuration reported throughout \cref{tab:swepoly_combined} uses `contains' edges alone; the edge-type comparison in \cref{tab:edge_ablation}, which finds `invokes' expansion stronger in every language, was run on \swepoly{} only, so the \sweverified{} and \multiswe{} cells do not reflect it. Inheritance-based proximity is not exploited at all, since no `inherits' edge is incident to a function node under our schema. Stage~2 filtering depends on an LLM selector, so cost and latency are bounded by per-instance LLM invocations ($C$ calls per query), and the selector rather than the graph is what now bounds achievable recall within the candidate pool: \ours{} reaches $79$--$89\%$ of the unfiltered neighborhood's recall (\cref{tab:stage2_ceiling_app}), so the remainder is lost to filtering rather than to unreachable structure; App.~\ref{sec:case_study_app} characterizes the two failure modes, an over-restrictive selector that discards ground truth reached from a tangential seed and an over-permissive one that lets neighbor flooding evict it. For very large $K$ or $d$, token consumption grows accordingly (see \cref{tab:bfs_d_ablation,tab:bfs_N_ablation}). A more fundamental ceiling comes from the initial dense ranker: $38\%$ of \sweverified{} instances and $51\%/62\%/22\%$ of \swepoly{} Python/Java/JavaScript instances contain at least one ground-truth function ranked beyond the top-$500$ (\cref{tab:locality_app}), and these targets fall outside \ours{}'s candidate pool $\mathcal{S}_N(Q)$ and cannot be recovered regardless of graph structure. Finally, evaluation focuses on Python, Java, JavaScript and TypeScript from \swepoly{}, \sweverified{} and \multiswe{}; performance on other languages or proprietary monorepos remains to be characterized.

\bibliography{custom}

@misc{reddy2025sweranksoftwareissuelocalization,
      title={SweRank: Software Issue Localization with Code Ranking}, 
      author={Revanth Gangi Reddy and Tarun Suresh and JaeHyeok Doo and Ye Liu and Xuan Phi Nguyen and Yingbo Zhou and Semih Yavuz and Caiming Xiong and Heng Ji and Shafiq Joty},
      year={2025},
      eprint={2505.07849},
      archivePrefix={arXiv},
      primaryClass={cs.SE},
      url={https://arxiv.org/abs/2505.07849}, 
}

@inproceedings{chen2025locagent,
    title = "{L}oc{A}gent: Graph-Guided {LLM} Agents for Code Localization",
    author = "Chen, Zhaoling  and
      Tang, Robert  and
      Deng, Gangda  and
      Wu, Fang  and
      Wu, Jialong  and
      Jiang, Zhiwei  and
      Prasanna, Viktor  and
      Cohan, Arman  and
      Wang, Xingyao",
    editor = "Che, Wanxiang  and
      Nabende, Joyce  and
      Shutova, Ekaterina  and
      Pilehvar, Mohammad Taher",
    booktitle = "Proceedings of the 63rd Annual Meeting of the Association for Computational Linguistics (Volume 1: Long Papers)",
    month = jul,
    year = "2025",
    address = "Vienna, Austria",
    publisher = "Association for Computational Linguistics",
    url = "https://aclanthology.org/2025.acl-long.426/",
    doi = "10.18653/v1/2025.acl-long.426",
    pages = "8697--8727",
    ISBN = "979-8-89176-251-0",
}

@inproceedings{fehr-etal-2025-coret,
    title = "{C}o{R}et: Improved Retriever for Code Editing",
    author = "Fehr, Fabio James  and
      Teja S, Prabhu  and
      Franceschi, Luca  and
      Zappella, Giovanni",
    editor = "Che, Wanxiang  and
      Nabende, Joyce  and
      Shutova, Ekaterina  and
      Pilehvar, Mohammad Taher",
    booktitle = "Proceedings of the 63rd Annual Meeting of the Association for Computational Linguistics (Volume 2: Short Papers)",
    month = jul,
    year = "2025",
    address = "Vienna, Austria",
    publisher = "Association for Computational Linguistics",
    url = "https://aclanthology.org/2025.acl-short.62/",
    doi = "10.18653/v1/2025.acl-short.62",
    pages = "775--789",
    ISBN = "979-8-89176-252-7",
}

@inproceedings{
suresh2025cornstack,
title={Co{RNS}tack: High-Quality Contrastive Data for Better Code Retrieval and Reranking},
author={Tarun Suresh and Revanth Gangi Reddy and Yifei Xu and Zach Nussbaum and Andriy Mulyar and Brandon Duderstadt and Heng Ji},
booktitle={The Thirteenth International Conference on Learning Representations},
year={2025},
url={https://openreview.net/forum?id=iyJOUELYir}
}

@inproceedings{
ouyang2025repograph,
title={RepoGraph: Enhancing {AI} Software Engineering with Repository-level Code Graph},
author={Siru Ouyang and Wenhao Yu and Kaixin Ma and Zilin Xiao and Zhihan Zhang and Mengzhao Jia and Jiawei Han and Hongming Zhang and Dong Yu},
booktitle={The Thirteenth International Conference on Learning Representations},
year={2025},
url={https://openreview.net/forum?id=dw9VUsSHGB}
}

@inproceedings{
jimenez2024swebench,
title={{SWE}-bench: Can Language Models Resolve Real-world Github Issues?},
author={Carlos E Jimenez and John Yang and Alexander Wettig and Shunyu Yao and Kexin Pei and Ofir Press and Karthik R Narasimhan},
booktitle={The Twelfth International Conference on Learning Representations},
year={2024},
url={https://openreview.net/forum?id=VTF8yNQM66}
}

@inproceedings{luo2024repoagent,
    title = "{R}epo{A}gent: An {LLM}-Powered Open-Source Framework for Repository-level Code Documentation Generation",
    author = "Luo, Qinyu  and
      Ye, Yining  and
      Liang, Shihao  and
      Zhang, Zhong  and
      Qin, Yujia  and
      Lu, Yaxi  and
      Wu, Yesai  and
      Cong, Xin  and
      Lin, Yankai  and
      Zhang, Yingli  and
      Che, Xiaoyin  and
      Liu, Zhiyuan  and
      Sun, Maosong",
    editor = "Hernandez Farias, Delia Irazu  and
      Hope, Tom  and
      Li, Manling",
    booktitle = "Proceedings of the 2024 Conference on Empirical Methods in Natural Language Processing: System Demonstrations",
    month = nov,
    year = "2024",
    address = "Miami, Florida, USA",
    publisher = "Association for Computational Linguistics",
    url = "https://aclanthology.org/2024.emnlp-demo.46/",
    doi = "10.18653/v1/2024.emnlp-demo.46",
    pages = "436--464"
}

@misc{xia2024agentlessdemystifyingllmbasedsoftware,
      title={Agentless: Demystifying LLM-based Software Engineering Agents}, 
      author={Chunqiu Steven Xia and Yinlin Deng and Soren Dunn and Lingming Zhang},
      year={2024},
      eprint={2407.01489},
      archivePrefix={arXiv},
      primaryClass={cs.SE},
      url={https://arxiv.org/abs/2407.01489}, 
}

@misc{aider,
  author       = {Paul Gauthier},
  howpublished = {\url{https://github.com/Aider-AI/aider}},
  title        = {Aider is AI pair programming in your terminal},
  year         = {2024}
}

@misc{yang2024swe,
      title={SWE-agent: Agent-Computer Interfaces Enable Automated Software Engineering}, 
      author={John Yang and Carlos E. Jimenez and Alexander Wettig and Kilian Lieret and Shunyu Yao and Karthik Narasimhan and Ofir Press},
      year={2024},
      eprint={2405.15793},
      archivePrefix={arXiv},
      primaryClass={cs.SE},
      url={https://arxiv.org/abs/2405.15793}, 
}

@misc{rashid2025swepolybenchmultilanguagebenchmarkrepository,
      title={SWE-PolyBench: A multi-language benchmark for repository level evaluation of coding agents}, 
      author={Muhammad Shihab Rashid and Christian Bock and Yuan Zhuang and Alexander Buchholz and Tim Esler and Simon Valentin and Luca Franceschi and Martin Wistuba and Prabhu Teja Sivaprasad and Woo Jung Kim and Anoop Deoras and Giovanni Zappella and Laurent Callot},
      year={2025},
      eprint={2504.08703},
      archivePrefix={arXiv},
      primaryClass={cs.SE},
      url={https://arxiv.org/abs/2504.08703}, 
}

@misc{zan2025multiswebenchmultilingualbenchmarkissue,
      title={Multi-SWE-bench: A Multilingual Benchmark for Issue Resolving}, 
      author={Daoguang Zan and Zhirong Huang and Wei Liu and Hanwu Chen and Linhao Zhang and Shulin Xin and Lu Chen and Qi Liu and Xiaojian Zhong and Aoyan Li and Siyao Liu and Yongsheng Xiao and Liangqiang Chen and Yuyu Zhang and Jing Su and Tianyu Liu and Rui Long and Kai Shen and Liang Xiang},
      year={2025},
      eprint={2504.02605},
      archivePrefix={arXiv},
      primaryClass={cs.SE},
      url={https://arxiv.org/abs/2504.02605}, 
}

@misc{jiang2025cosilsoftwareissuelocalization,
      title={Issue Localization via LLM-Driven Iterative Code Graph Searching}, 
      author={Zhonghao Jiang and Xiaoxue Ren and Meng Yan and Wei Jiang and Yong Li and Zhongxin Liu},
      year={2025},
      eprint={2503.22424},
      archivePrefix={arXiv},
      primaryClass={cs.SE},
      url={https://arxiv.org/abs/2503.22424}, 
}

@inproceedings{
zhang2024codesage,
title={{CODE} {REPRESENTATION} {LEARNING} {AT} {SCALE}},
author={Dejiao Zhang and Wasi Uddin Ahmad and Ming Tan and Hantian Ding and Ramesh Nallapati and Dan Roth and Xiaofei Ma and Bing Xiang},
booktitle={The Twelfth International Conference on Learning Representations},
year={2024},
url={https://openreview.net/forum?id=vfzRRjumpX}
}

@inproceedings{feng-etal-2020-codebert,
    title = "{C}ode{BERT}: A Pre-Trained Model for Programming and Natural Languages",
    author = "Feng, Zhangyin  and
      Guo, Daya  and
      Tang, Duyu  and
      Duan, Nan  and
      Feng, Xiaocheng  and
      Gong, Ming  and
      Shou, Linjun  and
      Qin, Bing  and
      Liu, Ting  and
      Jiang, Daxin  and
      Zhou, Ming",
    editor = "Cohn, Trevor  and
      He, Yulan  and
      Liu, Yang",
    booktitle = "Findings of the Association for Computational Linguistics: EMNLP 2020",
    month = nov,
    year = "2020",
    address = "Online",
    publisher = "Association for Computational Linguistics",
    url = "https://aclanthology.org/2020.findings-emnlp.139/",
    doi = "10.18653/v1/2020.findings-emnlp.139",
    pages = "1536--1547"
}

@inproceedings{
guo2021graphcodebertpretrainingcoderepresentations,
title={GraphCode{\{}BERT{\}}: Pre-training Code Representations with Data Flow},
author={Daya Guo and Shuo Ren and Shuai Lu and Zhangyin Feng and Duyu Tang and Shujie LIU and Long Zhou and Nan Duan and Alexey Svyatkovskiy and Shengyu Fu and Michele Tufano and Shao Kun Deng and Colin Clement and Dawn Drain and Neel Sundaresan and Jian Yin and Daxin Jiang and Ming Zhou},
booktitle={International Conference on Learning Representations},
year={2021},
url={https://openreview.net/forum?id=jLoC4ez43PZ}
}

@inproceedings{Robertson1994OkapiATBM25,
  title={Okapi at TREC-3},
  author={Stephen E. Robertson and Steve Walker and Susan Jones and Micheline Hancock-Beaulieu and Mike Gatford},
  booktitle={Text Retrieval Conference},
  year={1994},
  url={https://api.semanticscholar.org/CorpusID:41563977}
}

@inproceedings{
hu2022lora,
title={Lo{RA}: Low-Rank Adaptation of Large Language Models},
author={Edward J Hu and yelong shen and Phillip Wallis and Zeyuan Allen-Zhu and Yuanzhi Li and Shean Wang and Lu Wang and Weizhu Chen},
booktitle={International Conference on Learning Representations},
year={2022},
url={https://openreview.net/forum?id=nZeVKeeFYf9}
}

@book{shaked2007stochastic,
  title={Stochastic Orders},
  author={Shaked, M. and Shanthikumar, J.G.},
  isbn={9780387346755},
  lccn={2006927724},
  series={Springer Series in Statistics},
  url={https://books.google.com/books?id=rPiToBK2rwwC},
  year={2007},
  publisher={Springer New York}
}

@article{Kang_2024_llm-based_faultlocalization,
author = {Kang, Sungmin and An, Gabin and Yoo, Shin},
title = {A Quantitative and Qualitative Evaluation of LLM-Based Explainable Fault Localization},
year = {2024},
issue_date = {July 2024},
publisher = {Association for Computing Machinery},
address = {New York, NY, USA},
volume = {1},
number = {FSE},
url = {https://doi.org/10.1145/3660771},
doi = {10.1145/3660771},
journal = {Proc. ACM Softw. Eng.},
month = jul,
articleno = {64},
numpages = {23}
}

@inproceedings{
wang2025openhands,
title={OpenHands: An Open Platform for {AI} Software Developers as Generalist Agents},
author={Xingyao Wang and Boxuan Li and Yufan Song and Frank F. Xu and Xiangru Tang and Mingchen Zhuge and Jiayi Pan and Yueqi Song and Bowen Li and Jaskirat Singh and Hoang H. Tran and Fuqiang Li and Ren Ma and Mingzhang Zheng and Bill Qian and Daniel Shao and Niklas Muennighoff and Yizhe Zhang and Binyuan Hui and Junyang Lin and Robert Brennan and Hao Peng and Heng Ji and Graham Neubig},
booktitle={The Thirteenth International Conference on Learning Representations},
year={2025},
url={https://openreview.net/forum?id=OJd3ayDDoF}
}

@misc{
orwall2024moatless,
title={Moatless tools},
author={Albert Örwall},
year={2024},
url={https://github.com/aorwall/moatless-tools}
}

@misc{ma2025repodeepsearch,
      title={Tool-integrated Reinforcement Learning for Repo Deep Search}, 
      author={Zexiong Ma and Chao Peng and Qunhong Zeng and Pengfei Gao and Yanzhen Zou and Bing Xie},
      year={2025},
      eprint={2508.03012},
      archivePrefix={arXiv},
      primaryClass={cs.SE},
      url={https://arxiv.org/abs/2508.03012}, 
}

@inproceedings{
zhang2026reponavigator,
title={One Tool Is Enough: Reinforcement Learning of {LLM} Agents for Repository-Level Code Navigation},
author={Zhaoxi Zhang and Yitong Duan and Yanzhi Zhang and Yiming Xu and Zhixiang Wang and Kun Liang and Weikang Li and Jiahui Liang and Deguo Xia and Jizhou Huang and Jiyan He and Shuxin Zheng and Yunfang Wu},
booktitle={Forty-third International Conference on Machine Learning},
year={2026},
url={https://openreview.net/forum?id=WTziQZdpTV}
}

@article{evans2000broadcasting,
  author  = {Evans, William and Kenyon, Claire and Peres, Yuval and Schulman, Leonard J.},
  title   = {Broadcasting on Trees and the {I}sing Model},
  journal = {The Annals of Applied Probability},
  volume  = {10}, number = {2}, year = {2000},
  doi     = {10.1214/aoap/1019487349}
}

@article{lyons1989ising,
  author  = {Lyons, Russell},
  title   = {The {I}sing Model and Percolation on Trees and Tree-Like Graphs},
  journal = {Communications in Mathematical Physics},
  volume  = {125}, pages = {337--353}, year = {1989},
  doi     = {10.1007/BF01217911}
}

@article{robertson1977prp,
  author  = {Robertson, Stephen E.},
  title   = {The Probability Ranking Principle in {IR}},
  journal = {Journal of Documentation},
  volume  = {33}, number = {4}, pages = {294--304}, year = {1977},
  doi     = {10.1108/eb026647}
}

@article{thompson1990adaptive,
  author  = {Thompson, Steven K.},
  title   = {Adaptive Cluster Sampling},
  journal = {Journal of the American Statistical Association},
  volume  = {85}, number = {412}, pages = {1050--1059}, year = {1990},
  doi     = {10.1080/01621459.1990.10474975}
}

@inproceedings{macavaney2022adaptive,
  author    = {MacAvaney, Sean and Tonellotto, Nicola and Macdonald, Craig},
  title     = {Adaptive Re-Ranking with a Corpus Graph},
  booktitle = {CIKM}, pages = {1491--1500}, year = {2022},
  doi       = {10.1145/3511808.3557231}
}

\appendix
\section{Appendix}
\label{sec:appendix}

\etocsetnexttocdepth{subsection}
\localtableofcontents

\subsection{Notation Summary}\label{sec:notation_app}
\cref{tab:notation_app} summarizes the symbols used throughout the main paper. Symbols introduced only inside proofs (\cref{sec:theory}) are defined locally there.

\begin{table*}[!ht]
\centering
\caption{Symbols and notation used in the main paper.}
\label{tab:notation_app}
\renewcommand{\arraystretch}{1.15}
\begin{tabular}{ll}
\toprule
\textbf{Symbol} & \textbf{Description} \\
\midrule
\multicolumn{2}{l}{\textit{Graph and codebase}} \\
$\mathcal{G} = (\mathcal{V}, \mathcal{E})$ & Code graph with nodes and edges \\
$\mathcal{V}$ & Code entity nodes (functions, classes, files, directories) \\
$\mathcal{E}$ & Edges of type \textit{contains}, \textit{invokes}, \textit{imports}, \textit{inherits} \\
$n$ & Number of nodes in $\mathcal{V}$ \\
$\mathrm{dist}_{\mathcal{G}}(\cdot,\cdot)$ & Shortest-path distance on $\mathcal{G}$ \\
\midrule
\multicolumn{2}{l}{\textit{Query and ground truth}} \\
$Q$ & Issue description (query) \\
$\mathcal{V}^* \subseteq \mathcal{V}$ & Ground-truth function nodes for $Q$ \\
$m$ & Number of ground-truth nodes, $m = |\mathcal{V}^*|$ \\
\midrule
\multicolumn{2}{l}{\textit{Scoring and dense retrieval}} \\
$\mathcal{F}(\cdot)$ & Bi-modal encoder mapping code/text to a shared embedding space \\
$s_Q(v) = \cos(\mathcal{F}(v), \mathcal{F}(Q))$ & Relevance score of node $v$ for $Q$ \\
$K$ & Retrieval budget \\
$\mathcal{S}_K(Q)$ & Dense-retrieval (\der{}) top-$K$ set \\
\midrule
\multicolumn{2}{l}{\textit{\ours{} hyperparameters and intermediate sets}} \\
$N$ & Primary neighborhood filtering threshold, $N > K$ \\
$\mathcal{S}_N(Q)$ & Top-$N$ candidate pool \\
$C$ & Number of seed centers, $C \le K$ \\
$\mathcal{C}_Q$ & Top-$C$ seed centers drawn from $\mathcal{S}_K(Q)$ \\
$d$ & BFS exploration depth \\
$\Gamma_d(\mathcal{C}_Q)$ & $d$-hop neighborhood of $\mathcal{C}_Q$ along \textit{contains} edges \\
$\hat{\mathcal{C}}_Q$ & Stage-1 filtered candidates: $\Gamma_d(\mathcal{C}_Q) \cap \mathcal{S}_N(Q)$ \\
$\mathcal{L}(u)$ & LLM selector, returns $\{0,1\}$ \\
$\mathcal{U}_Q$ & LLM-selected candidates union seed centers \\
$\mathcal{S}^{\mathrm{LLM}}_K(Q)$ & Final \ours{} retrieval set of size $K$ \\
\midrule
\multicolumn{2}{l}{\textit{Evaluation and theory}} \\
Recall@$K$ & $|\mathcal{V}^* \cap \mathcal{S}(Q)| / |\mathcal{V}^*|$ \\
Acc@$K$ & $\mathbb{I}[\mathcal{V}^* \subseteq \mathcal{S}(Q)]$ \\
MRR@$K$ & Mean reciprocal rank of the first ground-truth node in $\mathcal{S}(Q)$ \\
$\gP$ & New graph proposals, $(\Gamma_d(\gC_Q) \cap \gS_N(Q)) \setminus \gS_K$ \\
$\gR$ & Protected baseline items: seeds, plus accepted candidates already in $\gS_K$ \\
$\gA$ & Admitted set, the candidates that consume retrieval budget \\
$\gD_J$ & The $J = |\gA|$ displaced baseline items \\
$\eta$ & Admitted precision, $\E|\gA \cap \gV^*| / \E|\gA|$ (Proposition~\ref{prop:recall_condition}) \\
$\tau$ & Frontier relevance of a displaced item, $\E|\gD_J \cap \gV^*| / \E|\gD_J|$ (Proposition~\ref{prop:recall_condition}) \\
$h(v),\ n_h$ & Distance from $v$ to its nearest seed, and the number of functions at distance $h$ \\
$p_h$ & Target prevalence at distance $h$, $\pi + \Delta\lambda^h$ (Assumption~\ref{as:decay}) \\
$\pi,\ \Delta,\ \lambda$ & Background rate, seeding gain, and per-hop persistence \\
$a,\ c$ & Admission rates for edited and non-edited candidates (Assumption~\ref{as:rates}) \\
$b$ & Branching factor of the shell counts, $n_h = k b^{h-1}$ \\
$\theta,\ d^\star$ & Profitability threshold and useful exploration radius (\cref{eq:dstar}) \\
\bottomrule
\end{tabular}
\end{table*}


\subsection{Theoretical Analysis}
\label{sec:theory}

\subsubsection{Setup and exact accounting}
\label{sec:theory_setup}

Fix an issue $Q$ with edited set $\gV^*$, $m=|\gV^*|\ge1$. Recall
$\gP=(\Gamma_d(\gC_Q)\cap\gS_N(Q))\setminus\gS_K$, the protected baseline
set $\gR\subseteq\gS_K$ (seed centers and accepted candidates already in
$\gS_K$), the admitted set $\gA\subseteq\gP$ with $J=|\gA|\le K-|\gR|$, and
the displaced set $\gD_J$ consisting of the $J$ lowest-ranked members of
$\gS_K\setminus\gR$, so that
$\gS^{\mathrm{LLM}}_K=(\gS_K\setminus\gD_J)\cup\gA$ has size $K$.

Three modelling points deserve emphasis, because the earlier version of this
analysis did not enforce them. First, the seed centers already belong to
$\gS_K$; treating $\gU_Q$ as new insertions double-counts them, and the
budget-consuming set is $\gU_Q\setminus\gS_K$. Second, $J$ is a
\emph{random} quantity determined by the query, and cannot be replaced by
$K$. Third, $\gU_Q$ is already post-selection, so its counts must not be
multiplied by $\alpha$ and $\beta$ a second time. Restricting attention to
$\gA$ and $\gD_J$ removes all three issues.

\begin{proof}[Proof of Proposition~\ref{prop:exchange}]
For the first claim, note that by construction $\gA\cap\gS_K=\varnothing$ and
$\gD_J\subseteq\gS_K$, so
$|\gS^{\mathrm{LLM}}_K\cap\gV^*|
=|\gS_K\cap\gV^*|-|\gD_J\cap\gV^*|+|\gA\cap\gV^*|$.
Dividing by $m$ gives \cref{eq:exact_recall}. For the second, take
expectations in \cref{eq:exact_recall} at fixed $m$ and use
$|\gD_J|=|\gA|=J$, which gives
$\E[\Delta\Reck]=\bigl(\E|\gA\cap\gV^*|-\E|\gD_J\cap\gV^*|\bigr)/m
=\bigl(\eta-\tau\bigr)\E|\gA|/m$. This is positive if and only if
$\eta>\tau$.
\end{proof}

The proof requires neither label independence across functions nor any
distributional assumption, since only linearity of expectation is used.

The same accounting gives the exact condition for full recovery. Writing
$\gM:=\gV^*\setminus\gS_K$ for the targets missed by dense retrieval,
\begin{equation}
\label{eq:acc_indicator}
\Acck(\gS^{\mathrm{LLM}}_K)
=\mathbbm{1}\{\gM\subseteq\gA\ \text{and}\ \gD_J\cap\gV^*=\varnothing\},
\end{equation}
since the output contains all of $\gV^*$ exactly when every baseline miss is
admitted and no baseline target is displaced. \acck{} is therefore not an
average-precision quantity, and \cref{eq:acc_indicator} remains the exact
requirement.

Nothing above assumes anything about the score distribution. The depth
analysis in the rest of this section does, so we restate the paper's score
assumption here; it is used in Remark~\ref{rem:seed_condition} and
Proposition~\ref{prop:prefix_robust}, not in
Proposition~\ref{prop:exchange}.

\begin{assumption}[Score distributions]
\label{as:score_dominance}
There exist distributions $F_1,F_0$ on $\mathbb{R}$ with $s_Q(v)\sim F_1$ for
$v\in\gV^*$ and $s_Q(v)\sim F_0$ otherwise, such that $F_1$ first-order
stochastically dominates $F_0$, i.e.\ $F_1(t)\le F_0(t)$ for all $t$
\citep{shaked2007stochastic}.
\end{assumption}

\paragraph{Roadmap for the rest of this section.}
Proposition~\ref{prop:exchange} settles \emph{whether} an exchange helps, but
it is silent on the two questions that decide how \ours{} should be
configured: when $\eta>\tau$ actually holds, and how far from a seed it is
still worth exploring. Both need a model of how candidate quality varies with
graph distance, and the remainder of this section builds one and then draws
its consequences.

\cref{sec:theory_model} derives the model from a latent issue-bearing context
that propagates along graph edges. It yields a shell-prevalence curve
$p_h=\pi+\Delta\lambda^{h}$, and relates the seeding condition it requires to
the paper's existing score assumption, while being explicit that the two are
not equivalent without a further condition on the ranking.
\cref{sec:theory_rates} decomposes the admission rates into the top-$N$ gate,
the Stage-2 selector, and capacity truncation, which is what connects
$\alpha$ and $\beta$ to $\eta$.

\cref{sec:theory_proof} proves Theorem~\ref{thm:depth}, the destination of the
analysis. It states that a shell is profitable exactly when its prevalence
exceeds a threshold $\theta$ determined by $\tau$ and the admission rates,
that deeper shells can expose strictly more edited functions even as
per-candidate precision falls, and that the profitable shells form a prefix,
so that whenever the threshold sits strictly above the background
prevalence the useful exploration radius $d^\star$ is finite, and is
language-dependent rather than universal. When it does not, every shell
remains profitable and exploration is bounded by token budget instead.

The three remaining parts qualify that result.
\cref{sec:theory_hetero} shows the prefix property survives when the rates
vary with depth, so only the closed form for $d^\star$ depends on holding them
constant. \cref{sec:theory_llm} identifies the regime in which the Stage-2
filter is pivotal rather than redundant, and prices its token cost.
\cref{sec:theory_tests} states what would have to be measured to test the
model on held-out repositories and what would falsify it, and
\cref{sec:theory_scope} bounds where the branching approximation applies. We
stress throughout that this half of the analysis is a falsifiable hypothesis
about our data, not something our experiments establish.
Across issues with different $m$, per-issue gains must be weighted by $1/m$
before aggregation, which is exactly macro-averaged \recallk{}.

\begin{remark}[Why the base rate is the wrong comparison]
\label{rem:baserate}
The locality analysis of \cref{sec:empirical_locality_app} establishes the
\emph{necessary} precondition for \ours{}: edited functions are mutually
reachable at small radius, so a seed landing on one of them can in principle
reach the others. Proposition~\ref{prop:exchange} shows what must be added for
sufficiency. Because the exchange is against a rank-$K$ dense hit rather than
a random function, the relevant comparison is $\eta>\tau$, and $\tau$ is far
above the repository base rate; establishing
$\pi_\Gamma\ge\Pr(u\in\gV^*)$ therefore does not by itself imply improvement.
Concretely, two edited functions among $100$ siblings of one file are $2$
hops apart, satisfying every locality statistic, yet their neighborhood has
prevalence $2/100$, which may lie below $\tau$: reachability between targets
does not count the non-edited functions in the same ball. The two analyses
are complementary. \cref{sec:empirical_locality_app} shows the targets are
reachable, and Proposition~\ref{prop:exchange} states the precision they must clear once
reached.
\end{remark}

\subsubsection{A latent-context model for \texorpdfstring{Assumption~\ref{as:decay}}{Assumption 1}}
\label{sec:theory_model}

Equating ``the seed is edited'' with ``the seed is useful'' is too
restrictive: it cannot explain recovering a single edited function from a
semantically related but \emph{unedited} seed, which is the regime \ours{}
targets. We instead posit a latent regional state.

Fix the implementation's distance convention and a deterministic BFS parent
for every reached node, yielding a rooted analysis tree whose geometry is
independent of the labels. Let $Z_v\in\{0,1\}$ indicate that $v$ lies in an
\emph{issue-bearing structural context} (the right class, file, or local
dependency region). Assume the tree-indexed conditional Markov property
\begin{align}
\label{eq:transition}
\Pr(Z_{\text{child}}{=}1\mid Z_{\text{parent}}{=}1)&=\rho_1,\\
\Pr(Z_{\text{child}}{=}1\mid Z_{\text{parent}}{=}0)&=\rho_0,
\end{align}
with $0\le\rho_0<\rho_1\le1$ and $(\rho_0,\rho_1)\neq(0,1)$. We further assume
that the seed-selection event $S$ is conditionally independent of a node's
context given its parent's,
\begin{equation}
\label{eq:seed_ci}
S \perp Z_{\text{child}} \mid Z_{\text{parent}},
\end{equation}
so that conditioning on $S$ leaves the transition kernel in
\cref{eq:transition} unchanged. Write
\begin{equation}
\label{eq:lambda_zeta}
\lambda=\rho_1-\rho_0\in(0,1),
\qquad
\zeta=\frac{\rho_0}{1-\rho_1+\rho_0},
\end{equation}
the per-hop persistence of regional information and the stationary
probability of the issue-bearing state. Edit labels are emissions,
\begin{align}
\label{eq:emission}
\Pr(v\in\gV^*\mid Z_v{=}1)&=r_1,\\
\Pr(v\in\gV^*\mid Z_v{=}0)&=r_0,\qquad r_1>r_0,
\end{align}
and are conditionally independent of ancestry and of the seed-selection
event $S$ given $Z_v$. Let $q:=\Pr(Z_{\text{seed}}{=}1\mid S)$ be the
context posterior induced by dense retrieval selecting the seed.

\begin{lemma}[Seeded spatial mixing]
\label{lem:mixing}
Under \cref{eq:transition,eq:emission} with $q>\zeta$, every nonempty shell
$h\ge0$ satisfies
\begin{equation}
\label{eq:mixing}
\begin{split}
p_h:=\Pr\bigl(w\in\gV^*\mid w\ \text{uniform at distance}\ h,\ S\bigr)\\
=\pi+\Delta\lambda^{h},
\end{split}
\end{equation}
where $\pi=r_0+(r_1-r_0)\zeta$ and $\Delta=(r_1-r_0)(q-\zeta)>0$. Hence
enrichment above the local background decays geometrically with graph
distance.
\end{lemma}

\begin{proof}
Let $z_h=\Pr(Z{=}1\mid h,S)$. By \cref{eq:seed_ci} the transition
probabilities are unchanged by conditioning on $S$, so averaging
\cref{eq:transition} over the parent state gives the affine recurrence
$z_{h+1}=\rho_0+(\rho_1-\rho_0)z_h$ with $z_0=q$. Its fixed point is $\zeta$
from \cref{eq:lambda_zeta}, so $z_h-\zeta=\lambda(z_{h-1}-\zeta)$ and
$z_h=\zeta+(q-\zeta)\lambda^{h}$. Applying \cref{eq:emission},
$p_h=r_0+(r_1-r_0)z_h=\pi+\Delta\lambda^{h}$.
\end{proof}

\cref{eq:seed_ci} is the transition-side counterpart of the conditional
independence already imposed on the emissions, and it is what lets a posterior
formed at the seed be propagated with the unconditioned kernel. Substantively
it says that dense retrieval's decision to select this seed carries
information about the surrounding region only through the seed's own context
state. It would fail if the ranker scored a node using signal that also
predicts the context of nodes deeper in the tree without passing through the
parent, for instance a repository-level or file-level feature shared by a
whole subtree. As with the other conditions in this section, we regard it as a
restriction on the analyzed regime rather than something the setup supplies.

Lemma~\ref{lem:mixing} yields exactly Assumption~\ref{as:decay}. The observable content is
the curve $p_h=\pi+\Delta\lambda^h$; the latent state is one mechanism that
produces it, and the theory does not require $Z_v$ to be a literal causal
variable.

\begin{remark}[Interpreting the seeding condition]
\label{rem:seed_condition}
The requirement $q>\zeta$ has a direct operational reading. Since
$\Pr(\text{seed}\in\gV^*\mid S)=r_0+(r_1-r_0)q$ and the unconditional rate
is $\pi=r_0+(r_1-r_0)\zeta$, we have
\begin{equation}
q>\zeta
\iff
\Pr(\text{seed}\in\gV^*\mid S)>\Pr(v\in\gV^*),
\end{equation}
i.e.\ $q>\zeta$ holds exactly when a dense-selected seed is more likely to be
edited than a uniformly random function. This is the property
Assumption~\ref{as:score_dominance} is intended to capture, but it does
\emph{not} follow from that assumption alone. First-order stochastic dominance
constrains the two marginal score distributions, whereas a seed is the
$\arg\max$ of scores across nodes, so whether it is enriched depends on their
joint behaviour. Under ties, or under strong dependence between target and
non-target scores, a dominating marginal can still yield a top-ranked item that
is no more likely to be edited than a random function, and passing from the
marginals to the selection event requires an additional exchangeability or
conditional-independence condition. We therefore treat $q>\zeta$ the same way
we treat the per-rank profile in \cref{sec:theory_hetero}, as an empirical
property rather than a theorem. It is also the cheapest of our quantities to
check, since $q$ is identified by seed precision, which is directly measurable
from the runs we already report.
\end{remark}

\begin{remark}[Unedited seeds]
\label{rem:unedited_seed}
If one additionally observes that the seed is \emph{not} edited, its context
posterior drops to
$q_0=q(1-r_1)\big/\bigl(q(1-r_1)+(1-q)(1-r_0)\bigr)$, and such a seed
remains informative only when $q_0>\zeta$: the dense evidence must outweigh
the negative update from the seed's own label. \ours{} does not observe that
label at retrieval time, so its operative ex-ante parameter is $q$.
\end{remark}

\begin{proposition}[Mixing over multiple seeds]
\label{prop:multiseed}
Let $\gC_Q$ contain $C$ seeds with context posteriors $q_1,\dots,q_C$, assign
each reached function to its nearest seed, and let $w_{c,h}$ be the share of
shell-$h$ functions assigned to seed $c$, determined by tree geometry alone.
Then
\begin{equation}
\label{eq:multiseed}
\begin{split}
p_h&=\pi+\bar\Delta_h\lambda^{h},
\qquad
\bar\Delta_h=(r_1-r_0)(\bar q_h-\zeta),\\
\bar q_h&=\textstyle\sum_c w_{c,h}\,q_c .
\end{split}
\end{equation}
If in addition the shares are depth-invariant over the analyzed range,
$w_{c,h}=w_c$ for all $h$, then $\bar\Delta_h\equiv\bar\Delta$ and the curve
takes the single-seed form assumed in Assumption~\ref{as:decay}.
\end{proposition}

\begin{proof}
By Lemma~\ref{lem:mixing}, $p_h(q)=\pi+(r_1-r_0)(q-\zeta)\lambda^{h}$ is affine in
$q$. Averaging over the shell-$h$ seed assignment therefore replaces $q$ by
$\bar q_h$, giving \cref{eq:multiseed}. Depth-invariant shares make $\bar q_h$
and hence $\bar\Delta_h$ constant in $h$.
\end{proof}

\begin{remark}[Depth-invariance is an extra condition]
\label{rem:multiseed_gap}
Depth-invariance of the shares does \emph{not} follow from disjointness of the
seed neighborhoods. Disjointness removes double counting, but the shells around
different seeds can still grow at different rates, so a seed sitting in a large
class may supply most of the functions at $h{=}2$ while a seed in a small one
supplies more at $h{=}4$. In that case $\bar q_h$ drifts with depth and there is
no single $\bar\Delta$. Two consequences follow. First,
Assumption~\ref{as:decay} should be read, under $C>1$, as holding exactly only
when the shares are depth-invariant, and otherwise as an approximation whose
adequacy is part of what the protocol in \cref{sec:theory_tests} tests, since
$w_{c,h}$ is directly countable from the graph. Second, the prefix property in
Theorem~\ref{thm:depth} relies on $p_h$ decreasing in $h$, which
\cref{eq:multiseed} no longer gives automatically. A sufficient condition is
that $\bar\Delta_h$ be nonincreasing, since the product of a nonincreasing
positive sequence with the strictly decreasing $\lambda^{h}$ is strictly
decreasing. This is the multi-seed analogue of the monotonicity condition on
$\theta_h$ in \cref{sec:theory_hetero}, and we treat it the same way, as a
checkable condition rather than a consequence.
\end{remark}

Subject to that condition, Proposition~\ref{prop:multiseed} is why the
single-seed analysis carries over to the deployed $C{=}5$ configuration, with
an effective seeding gain $\bar\Delta$. Assigning each node to its nearest seed
also discards evidence when neighborhoods overlap, and the direction of that
bias is known:
a node reachable from a second, farther seed $c'$ receives additional positive
evidence whenever $q_{c'}>\zeta$, and discarding it understates $\bar q$, hence
$\bar\Delta$, hence $p_h$. The approximation therefore biases $d^\star$
\emph{downward} and under-promises reach, so a cutoff derived under it is
conservative. Adding seeds raises $\bar q$ only if the new seeds
are themselves well-placed, which is one mechanism for the saturation in $C$
observed in \cref{sec:ablations}.

\subsubsection{Graph volume and admission rates}
\label{sec:theory_rates}

Let $H_h$ be the distinct function nodes at distance $h$ and
$n_h=|H_h|$. Theorem~\ref{thm:depth} uses these observed counts and does \emph{not}
require geometric growth; the regular-growth form $n_h=kb^{h-1}$ is invoked
only to interpret target volume. In pure \texttt{contains} graphs,
function-to-function distances are typically even (siblings are $2$ hops
apart through their container), so odd shells may be empty; the theorem is
stated over \emph{nonempty} shells and the last beneficial one is
$\bar d^\star=\max\{h\le d^\star: n_h>0\}$.

Decompose the admission pipeline by ground-truth label, at depth $h$:
\begin{equation}
\label{eq:rates}
\begin{array}{l}
t=\Pr(v\in\gS_N\setminus\gS_K\mid v\in\gV^*),\\[2pt]
f=\Pr(v\in\gS_N\setminus\gS_K\mid v\notin\gV^*),\\[2pt]
\alpha=\Pr(\gL(v){=}1\mid v\in\gV^*,\cdot),\\[2pt]
\beta=\Pr(\gL(v){=}1\mid v\notin\gV^*,\cdot),\\[2pt]
\gamma_1=\Pr(v\in\gA\mid \gL(v){=}1,v\in\gV^*),\\[2pt]
\gamma_0=\Pr(v\in\gA\mid \gL(v){=}1,v\notin\gV^*),
\end{array}
\end{equation}
so that the effective rates of Assumption~\ref{as:rates} are
$a=t\alpha\gamma_1$ and $c=f\beta\gamma_0$, with $\gamma_1,\gamma_0$
absorbing priority, deduplication and capacity truncation. Depth comparisons
use a nested, shell-first policy: increasing $d$ from $h-1$ to $h$ leaves
earlier admissions unchanged and causes only additional baseline-prefix
evictions. Once capacity binds the effective rates are zero, a saturated
shell has zero incremental gain, and the depth statement no longer applies.

\ifdepthappendix
Let $h(v)$ be the distance from $v$ to its nearest seed and $n_h$ the number
of distinct functions in shell $h$.

\begin{assumption}[Seeded spatial decay]
\label{as:decay}
The shell-$h$ target prevalence is
$p_h:=\Pr(v\in\gV^*\mid h(v)=h)=\pi+\Delta\lambda^{h}$, for a background rate
$\pi$, a seeding gain $\Delta>0$ and a per-hop persistence
$\lambda\in(0,1)$. Lemma~\ref{lem:mixing} derives this form.
\end{assumption}

\begin{assumption}[Admission and exchange rates]
\label{as:rates}
Over the analyzed depth range, a shell-$h$ candidate is admitted with
probability $a>0$ if edited and $c>0$ if not (composing top-$N$ retention,
the LLM decision, deduplication and capacity). Moreover the frontier
relevance is constant \emph{conditional on the number of evictions},
\begin{equation}
\label{eq:tau_conditional}
\E\bigl[|\gD_J\cap\gV^*| \bigm| J\bigr]=\tau J,
\qquad \tau\in(0,1).
\end{equation}
\end{assumption}

\begin{remark}[Why \cref{eq:tau_conditional} is stronger than the definition of $\tau$]
\label{rem:tau_conditional}
In Proposition~\ref{prop:exchange}, $\tau$ is a ratio of marginal
expectations, $\E|\gD_J\cap\gV^*|/\E|\gD_J|$, and that identity holds with no
assumption because the proposition aggregates over the whole exchange.
Theorem~\ref{thm:depth} instead charges an eviction cost to each shell
separately, which requires the per-eviction relevance to be unaffected by how
many candidates a query happens to admit. Without
\cref{eq:tau_conditional} the two can differ: if $J$ and the relevance of the
displaced frontier are correlated, for instance because queries that admit
many neighbors are also queries whose rank-$K$ items are unusually relevant,
then $\E[\tau_J J]\neq\tau\,\E[J]$ and the shell decomposition below fails.
We therefore state the conditional form explicitly rather than deriving it
from the definition. Like the constancy of $a$ and $c$, it is an assumption
about our operating range and not a consequence of the setup.
\end{remark}

\begin{theorem}[Seeded spatial-completion law]
\label{thm:depth}
Under Assumptions~\ref{as:decay} and~\ref{as:rates}, shell $h$ contributes $n_h a p_h$ targets and
$n_h c(1-p_h)$ distractors at precision
$\eta_h=a p_h/(a p_h+c(1-p_h))$, and its expected contribution to the recall
numerator is $G_h=n_h[((1-\tau)a+\tau c)p_h-\tau c]$. Then:
(i) with $\theta:=\tau c/((1-\tau)a+\tau c)$,
\begin{equation}
\label{eq:threshold}
G_h>0 \iff \eta_h>\tau \iff p_h>\theta;
\end{equation}
(ii) the shell's excess target mass is $n_h\Delta\lambda^{h}$, which under
regular growth $n_h=kb^{h-1}$ equals $k\Delta\lambda(b\lambda)^{h-1}$, so if
$b\lambda>1$ deeper shells expose strictly \emph{more} edited functions even
though $\eta_h$ strictly \emph{decreases}; and
(iii) beneficial shells form a prefix, so when $\pi<\theta<p_1$ the expected
recall numerator is maximized at the useful radius
\begin{equation}
\label{eq:dstar}
d^\star=\bigl\lceil
\log((\theta-\pi)/\Delta)\big/\log\lambda\bigr\rceil-1 .
\end{equation}
\end{theorem}
\fi

\subsubsection{Proof of \texorpdfstring{Theorem~\ref{thm:depth}}{Theorem 1}}
\label{sec:theory_proof}

\begin{proof}
\emph{(Counts and precision.)} By linearity of expectation, and again without
requiring independence within a shell, the expected admitted target and
distractor counts from shell $h$ are $T_h=n_h a p_h$ and
$F_h=n_h c(1-p_h)$, and dividing target admissions by all admissions gives
$\eta_h=a p_h/(a p_h+c(1-p_h))$.

\emph{(Threshold.)} By \cref{eq:tau_conditional}, the expected relevant loss
accompanying these admissions is $\tau(T_h+F_h)$, so
\begin{align}
G_h &= T_h-\tau(T_h+F_h) \nonumber\\
    &= n_h\bigl[(1-\tau)a p_h-\tau c(1-p_h)\bigr] \nonumber\\
    &= n_h\bigl[\bigl((1-\tau)a+\tau c\bigr)p_h-\tau c\bigr].
\end{align}
Since $(1-\tau)a+\tau c>0$, $G_h>0\iff p_h>\theta$ with $\theta$ as in
\cref{eq:threshold}. Dividing $\eta_h>\tau$ through by $c(1-p_h)>0$ yields
$(a/c)\,p_h/(1-p_h)>\tau/(1-\tau)$, which rearranges to the same inequality,
giving the equivalence $G_h>0\iff\eta_h>\tau\iff p_h>\theta$.

\emph{(Reach vs.\ precision.)} Substituting $p_h-\pi=\Delta\lambda^{h}$ gives
excess mass $n_h\Delta\lambda^h$, and $n_h=kb^{h-1}$ gives
$k\Delta\lambda(b\lambda)^{h-1}$, which increases in $h$ iff $b\lambda>1$.
Meanwhile $\eta_h$ is strictly increasing in $p_h$ and $p_h$ is strictly
decreasing, so $\eta_h$ strictly decreases regardless of $b$.

\emph{(Useful radius.)} As $0<\lambda<1$ and $\Delta>0$, $p_h$ decreases
strictly to $\pi$, so $\{h:p_h>\theta\}$ is a prefix. In the interior case
$\pi<\theta<p_1$,
\begin{equation}
\begin{split}
p_h>\theta
&\iff \lambda^{h}>\frac{\theta-\pi}{\Delta}\\
&\iff h<\frac{\log((\theta-\pi)/\Delta)}{\log\lambda},
\end{split}
\end{equation}
the inequality reversing because $\log\lambda<0$. The largest integer
strictly below the right-hand side is $d^\star$ of \cref{eq:dstar}, and
$\theta<p_1$ forces $d^\star\ge1$. Shell increments are positive up to
$d^\star$ and nonpositive after, so the cumulative expectation is maximized
there; if the first excluded shell has $G_h=0$ then $d^\star+1$ attains the
same value but no larger.
\end{proof}

\ifcompacttheory
\begin{corollary}[Accuracy is harder than recall]
\label{cor:accuracy}
If the baseline is already complete ($\gV^*\subseteq\gS_K$), then
$\E[\Acck(\gS^{\mathrm{LLM}}_K)]-\E[\Acck(\gS_K)]
=-\Pr(\gD_J\cap\gV^*\neq\varnothing)\le 0$, with strict inequality whenever
$\tau>0$ and $J\ge1$.
\end{corollary}
\fi

\begin{proof}[Proof of Corollary~\ref{cor:accuracy}]
If $\gV^*\subseteq\gS_K$ then $\gM=\varnothing$, so by \cref{eq:acc_indicator}
$\Acck(\gS^{\mathrm{LLM}}_K)=\mathbbm{1}\{\gD_J\cap\gV^*=\varnothing\}$
while $\Acck(\gS_K)=1$. Taking expectations gives the claim; the inequality
is strict when some eviction can be a target.
\end{proof}

Corollary~\ref{cor:accuracy} is a \emph{stratified} prediction and we are careful not to
overstate it. It says that on issues whose baseline already contained every
edited function, any insertion can only lose \acck{}, through a single bad
eviction. It does \emph{not} predict that aggregate \acck{} falls, and
aggregate \acck{} is not a valid test of it: gains on incomplete-baseline
issues can dominate, and in our main results \acck{} rises alongside
\recallk{} in almost every cell. The corresponding test is to restrict to
instances where the dense baseline already attains $\acck{}=1$ and measure the
eviction rate there; we have not run it, and report Corollary~\ref{cor:accuracy} as a
prediction rather than a confirmed finding.

\subsubsection{Relaxing the constant-rate assumption}
\label{sec:theory_hetero}

The closed form \cref{eq:dstar} comes from holding $a,c,\tau$ fixed. If they
vary with depth, define
$\theta_h=\tau_h c_h/\bigl((1-\tau_h)a_h+\tau_h c_h\bigr)$.

\begin{proposition}[Robustness of the prefix property]
\label{prop:prefix_robust}
Write $\theta_h=\tau_h/\bigl((1-\tau_h)x_h+\tau_h\bigr)$ with
$x_h=a_h/c_h$. Then $\theta_h$ is increasing in $\tau_h$ and decreasing in
$x_h$. Consequently, if $(\tau_h)$ is nondecreasing and $(x_h)$ is
nonincreasing in $h$, then $\theta_h$ is nondecreasing and $p_h-\theta_h$ is
strictly decreasing, so the beneficial shells still form a prefix. That prefix
is finite if and only if $\theta_h>\pi$ for some $h$, equivalently
$\liminf_h\theta_h>\pi$. If instead $\theta_h\le\pi$ at every depth, then
$p_h>\pi\ge\theta_h$ for all $h$ and every shell remains beneficial, so no
finite radius exists.
\end{proposition}

\begin{proof}
Write $\theta_h=\bigl(1+(1-\tau_h)x_h/\tau_h\bigr)^{-1}$. The map
$\tau\mapsto(1-\tau)/\tau$ is decreasing on $(0,1)$, so $\theta_h$ increases
in $\tau_h$; it plainly decreases in $x_h$. Under the stated monotonicity
$\theta_h$ is nondecreasing while $p_h$ strictly decreases, so
$p_h-\theta_h$ is strictly decreasing and changes sign at most once, from
positive to negative, which gives the prefix property. For finiteness, recall
$p_h\downarrow\pi$ strictly. If $\theta_H>\pi$ for some $H$ then
$\theta_h\ge\theta_H>\pi$ for all $h\ge H$ by monotonicity, while $p_h\to\pi$,
so $p_h<\theta_H$ for all large $h$ and the prefix terminates. Conversely if
$\theta_h\le\pi$ for every $h$ then $p_h>\pi\ge\theta_h$ at every depth and the
prefix is the whole sequence.
\end{proof}

Both hypotheses are the \emph{natural} directions of violation, though we
assume rather than derive them. Under the shell-first policy, admitting more
candidates evicts progressively further up the dense ranking, so $\tau_h$ is
nondecreasing whenever the baseline's per-rank target probability is
nonincreasing in rank. We note that this does \emph{not} follow from
Assumption~\ref{as:score_dominance} alone: first-order stochastic dominance of the score
distributions constrains the marginals, not the per-rank posterior, and an
additional exchangeability condition is required to pass from one to the other.
The per-rank profile is directly estimable from the existing dense rankings at
no additional inference cost, and we regard it as an empirical check rather
than a theorem. Likewise deeper candidates are less semantically similar to
$Q$, so both the top-$N$ gate and the LLM plausibly discriminate less well and
$x_h=a_h/c_h$ degrades. Proposition~\ref{prop:prefix_robust} therefore says
the qualitative conclusion survives the simplification, and only the closed
form \cref{eq:dstar} depends on it. The finiteness clause is the varying-rate
analogue of the trichotomy in Theorem~\ref{thm:depth}: just as a constant
$\theta\le\pi$ makes every shell profitable there, a sequence with
$\theta_h\le\pi$ throughout leaves the useful radius unbounded here, and
exploration is then limited by token budget rather than by precision.
Without monotone $\theta_h$, a later shell can be useful after a harmful one
and no single-cutoff theorem holds.

\subsubsection{LLM pivotality and token cost}
\label{sec:theory_llm}

\ifcompacttheory
\begin{corollary}[When is the LLM pivotal?]
\label{cor:llm}
Compare a graph-only policy with rates $(t,f)$ against the LLM-filtered
policy with rates $(t\alpha,f\beta)$ at a common, count-matched frontier
$\tau$. If $\alpha>\beta>0$ then $\theta_L<\theta_G$, and shell $h$ is
\emph{graph-sufficient} if $p_h>\theta_G$, \emph{LLM-pivotal} if
$\theta_L<p_h\le\theta_G$, and unprofitable for both if $p_h\le\theta_L$.
\end{corollary}
\fi

\begin{proof}[Proof of Corollary~\ref{cor:llm}]
With nonbinding capacity the graph-only policy has effective rates
$(a_G,c_G)=(t,f)$ and the filtered policy $(a_L,c_L)=(t\alpha,f\beta)$.
Writing $\theta=\tau/\bigl((1-\tau)x+\tau\bigr)$ with $x=a/c$, the map
$x\mapsto\theta$ is strictly decreasing for $x>0$. The graph-only ratio is
$t/f$ and the filtered ratio is $(t/f)(\alpha/\beta)>t/f$ since
$\alpha>\beta$, hence $\theta_L<\theta_G$. The three regimes follow by
applying \cref{eq:threshold} with each threshold.
\end{proof}

Two caveats bound this result. It does not establish that the evaluated LLM
\emph{is} pivotal: that requires measuring $\alpha,\beta$ and running a
count-matched graph-only ablation, since an unfiltered policy admits more
candidates and would otherwise face a different frontier, which is why
Corollary~\ref{cor:llm} fixes a common $\tau$. Nor does it make an LLM uniquely
necessary: any calibrated query--candidate interaction model with the same
likelihood ratio $\alpha/\beta$ plays the same role. Finally, if LLM
inference is priced at $\kappa$ utility per raw candidate, the numerator of
$\theta$ becomes $\tau c+\kappa$, so a better likelihood ratio need not imply
better total utility.

\subsubsection{What to measure, and what would falsify the model}
\label{sec:theory_tests}

\begin{remark}[What the existing ablations can and cannot show]
\label{rem:evidence}
Theorem~\ref{thm:depth} predicts a finite useful radius in its interior case. Our present ablations cannot
confirm or refute one, and we state this explicitly rather than claim support.
The depth and top-$N$ sweeps (\cref{tab:bfs_d_ablation,tab:bfs_N_ablation_app})
are single-seed runs reported to two decimals, with no confidence intervals;
the differences that would bear on \cref{eq:threshold}, namely the flat Java
depth column, the TypeScript plateau from $d{=}6$ to $d{=}8$, and the Python
top-$N$ maximum near $N{=}500$, are all of size $0.01$. That is roughly one sixth of
the $95\%$ interval half-width we report for the main results and below the
seed-to-seed drift we document for the Stage-2 selector, which runs at nonzero
temperature. Pairing across settings would tighten the interval on a
\emph{difference}, but it does not remove that selector noise, since each run
draws fresh LLM decisions. We therefore treat these sweeps as compatible with
either a cutoff beyond the measured range or saturation within it, and as
distinguishing neither.

One measurement in this family does clear the noise floor, because it is
deterministic given the candidate set: inference \emph{cost}. Input tokens per
query in \cref{tab:bfs_d_ablation_tokens_app} grow $3.6\mathrm{K}\to
29.7\mathrm{K}$ (Python) from $d{=}4$ to $d{=}8$, an $8.2\times$ increase.
Pricing inference at $\kappa$ per raw candidate replaces the numerator of
$\theta$ by $\tau c+\kappa$ (\cref{sec:theory_llm}), which moves the useful
radius inward; this is what makes the deployed $d{=}4$ a budget choice rather
than a measured performance optimum. We emphasise that these token counts do
\emph{not} estimate the shell sizes $n_h$. They are recorded after Stage-1
gating, and the depth ablation runs at $N{=}100$, so the evaluated set
$\Gamma_d(\gC_Q)\cap\gS_N$ is capped at $100$ candidates irrespective of $d$;
the curve therefore reflects a bounded pool filling up, not raw shell volume.
Consistently, the per-two-hop growth ratios are not constant and Java's
decreases from $d{=}2$ to $d{=}4$. Estimating $n_h$ requires the
before-gating counts specified in \cref{sec:theory_tests}.
\end{remark}

The theory is useful only as a held-out curve prediction. Its observables are
$n_h$ (distinct shell functions before dense-rank gating), $p_h$
(count-weighted target prevalence before top-$N$), $t_h,f_h$
($\gS_N\setminus\gS_K$ retention by label), $\alpha_h,\beta_h$ (raw LLM
decisions by label), $\gamma_{1h},\gamma_{0h}$ (post-LLM survival), and
$\tau_j$ (the ground-truth rate of the actual $j$-th eviction). Across
variable-size shells, count weighting is essential:
$n_h=\E N_h$ and $p_h=\E R_h/\E N_h$ for raw and target shell counts
$N_h,R_h$; an unweighted mean of per-query shell precisions does not satisfy
$\E R_h=n_h p_h$. For macro \recallk{} across varying $m$, use
$n_h^{\mathrm{macro}}=\E[N_h/m]$ and
$p_h^{\mathrm{macro}}=\E[R_h/m]/\E[N_h/m]$.

The protocol is to fit $\pi,\Delta,\lambda$ on shallow shells and $k,b$ on
distinct non-backtracking shell counts, freeze them, and \emph{predict}
deeper-shell prevalence, the sign of $G_h$, and $d^\star$ on held-out
repositories, rather than estimating a free prevalence at every depth.
Repository-held-out estimation is required because functions within a
repository, and candidates within one LLM batch, are not independent samples.
The model is falsified, or must be replaced by a relation-specific one, if:
(i) shell prevalence is not approximately geometric after controlling for
dense rank, node type and degree; (ii) candidate quality has multiple peaks
across depth, so beneficial shells are not a prefix; (iii) degree-preserving
graph rewiring does not reduce seeded enrichment; (iv) a count-matched
graph-only policy matches the LLM policy in the purported pivotal region;
(v) actual relevant evictions do not track the frontier model; or (vi) the
predicted cutoff does not transfer from shallow shells to held-out deeper
shells and repositories.

These quantities are complementary to, not substitutes for,
\cref{tab:locality_app}: target-to-target distances establish reachability,
whereas Theorem~\ref{thm:depth} additionally requires all-candidate shell prevalence
and the realized exchange frontier (Remark~\ref{rem:baserate}). Likewise the depth
and top-$N$ ablations in \cref{sec:ablations} are outcomes of the full
pipeline, so while they are consistent with the model
(Remark~\ref{rem:evidence}) they do not identify which term of $G_h$ changed.

\subsubsection{Scope}
\label{sec:theory_scope}

The regular-branching model is deliberately small enough to make a
falsifiable prediction. Real containment graphs mix directories, files,
classes and functions, admit multiple paths under the chosen distance
convention, and have overlapping seed neighborhoods
(Proposition~\ref{prop:multiseed}). A multitype extension replaces $b\lambda$ by the
spectral radius of a type-specific branching--persistence matrix; it is
warranted only if the single-decay model fails and enough data exist to
estimate the extra parameters. The branching approximation must also stop
before graph collisions, finite-target depletion, or insertion capacity
invalidate geometric shell growth.

We note that $b\lambda$ governs growth of the \emph{expected number} of
excess targets and is not the Kesten--Stigum root-reconstruction threshold
$b\lambda^2$ for broadcasting on trees. The theorem concerns expected hit
count, hence \recallk{} at fixed $m$; it does not analyze MRR, because
\cref{eq:exchange_output} does not fully determine the final ordering.

The hidden-context model is related to broadcasting on trees, where a noisy
state propagates through a branching structure
\citep{evans2000broadcasting}, and the first-moment $b\lambda$ behaviour has
precedent in Ising and percolation analysis on trees
\citep{lyons1989ising}. The use here is narrower: the relevant question is
not reconstruction of the root state, but whether a decaying spatial signal
can support a profitable fixed-budget exchange. The filtered-precision
comparison instantiates posterior ranking under additive retrieval utility
\citep{robertson1977prp}, and triggering local exploration from an initially
promising unit resembles adaptive cluster sampling
\citep{thompson1990adaptive}; adaptive corpus-graph re-ranking
\citep{macavaney2022adaptive} is the closest system analogue. The individual
mathematical ingredients are standard; the contribution is their coupling to
\ours{}'s spatial mixing, filtering, and fixed-budget eviction.

\subsection{Empirical Support for Graph Locality}
\label{sec:empirical_locality_app}

Proposition~\ref{prop:exchange} needs no locality assumption, since it is an identity about the exchange. Graph locality enters one step later, through Assumption~\ref{as:decay}, which posits that target prevalence decays with distance from a seed as $p_h = \pi + \Delta\lambda^h$ with $\Delta > 0$. That form implies $p_h > \pi$ at every finite $h$, which is exactly the requirement that relevance be non-negatively correlated with graph proximity. Locality is therefore necessary for the depth analysis to have any content, and, as Remark~\ref{rem:baserate} notes, necessary but not sufficient for a profitable exchange. To check that it is not merely a convenient modeling choice, we analyze the spatial distribution of ground-truth functions in \sweverified{}.

\paragraph{Per-instance distances.} We restrict attention to instances with more than one ground-truth function edit, since locality is only meaningful when multiple targets are involved. Of the $469$ retained \sweverified{} instances (out of $500$), $190$ contain multi-function edits. Among these, $169$ ($88.9\%$) have all ground-truth functions within a maximum graph distance of $d \le 4$ hops along \texttt{contains} edges, with $119$ ($62.6\%$) within $\le 2$ hops. The median per-instance \emph{mean} distance is $1.92$ hops and the median per-instance \emph{maximum} distance is $2$ hops; no instance has a disconnected ground-truth pair.

\paragraph{Pairwise distances.} At the pairwise level, we examine all $3{,}869$ ground-truth function pairs across these $193$ instances (each pair drawn from the same instance). Of these, $30\%$ ($1{,}171$) are sibling functions in the same file or class at distance $2$ hops, $45$ pairs have distance $1$ (nested containment), and $733$ pairs are cross-class within the same file at distance $3$--$4$. The remaining $48\%$ ($1{,}851$) are cross-file or cross-package pairs with distance $\ge 6$.

\paragraph{Implication for Proposition~\ref{prop:recall_condition}.} A clear majority of multi-edit instances admit a small $d$ that captures all ground-truth targets, and the bulk of within-instance ground-truth pairs concentrate at small hop counts. Both observations are direct empirical instantiations of the decay form assumed in Assumption~\ref{as:decay}: conditioning on membership in a small-radius neighborhood $\Gamma_d(\{v\})$ of a ground-truth node $v$ raises the probability that a candidate $u$ is itself ground truth, relative to the unconditional base rate over the codebase. The precondition that the depth analysis needs therefore holds in practice. It does not by itself imply that expansion is profitable, which by Proposition~\ref{prop:exchange} additionally requires $\eta>\tau$.

\paragraph{Cross-dataset extension.}
To confirm that the locality precondition is not specific to Python or to \sweverified{}, we run the same analysis on the Python, Java, and JavaScript splits of \swepoly{} and consolidate it with the \sweverified{} numbers above. \cref{tab:locality_app} reports per-dataset statistics. The median maximum within-instance ground-truth distance is exactly $2$ hops in all four splits, the mean shortest-path distance ranges from $1.71$ to $2.14$ hops, and no split contains a disconnected ground-truth pair. The locality that Assumption~\ref{as:decay} presumes therefore holds uniformly across the multilingual benchmark.

\begin{table*}[!ht]
\centering
\caption{Locality and dense-retrieval coverage statistics on \sweverified{} and the Python, Java, and JavaScript splits of \swepoly{}. Multi-func rows are computed over instances with $>1$ ground-truth function. ``Tightly clustered'' denotes maximum within-instance ground-truth-pair distance $\le 2$ hops. Dense retrieval refers to \texttt{SweRankEmbed-Small}.}
\label{tab:locality_app}
\renewcommand{\arraystretch}{1.1}
\resizebox{\textwidth}{!}{%
\begin{tabular}{lcccc}
\toprule
& \textbf{\sweverified{}} & \textbf{\swepoly{} Python} & \textbf{\swepoly{} Java} & \textbf{\swepoly{} JavaScript} \\
\midrule
\multicolumn{5}{l}{\textit{Instance / ground-truth statistics}} \\
\# instances                       & 469      & 182      & 158      & 573 \\
\# multi-func instances            & 190      & 115      & 123      & 367 \\
Mean / median GT funcs/inst        & 2.25 / 1 & 5.02 / 2 & 7.93 / 3 & 3.39 / 2 \\
Mean / median GT files/inst        & 1.23 / 1 & 1.79 / 1 & 2.79 / 2 & 1.58 / 1 \\
\midrule
\multicolumn{5}{l}{\textit{Graph locality (multi-func instances)}} \\
Mean shortest-path (hops)               & 1.92 & 2.14 & 2.00 & 1.71 \\
Median max shortest-path (hops)         & 2.0  & 2.0  & 2.0  & 2.0 \\
Disconnected GT pairs (\%)              & 0    & 0    & 0    & 0 \\
Mean invoke-connected ratio             & 0.57 & 0.31 & 0.63 & 0.59 \\
Tightly / mod.\ / spread (\%)           & 84.9 / 9.4 / 5.8 & 64.8 / 14.8 / 20.3 & 46.8 / 20.3 / 32.9 & 71.4 / 15.4 / 13.3 \\
\midrule
\multicolumn{5}{l}{\textit{Dense retrieval coverage (top-$K$)}} \\
All GT in top-100 (\%)                  & 44.8 & 36.3 & 29.1 & 53.6 \\
All GT in top-500 (\%)                  & 61.8 & 48.9 & 38.0 & 78.2 \\
Any GT beyond top-500 (\%)              & 38.2 & 51.1 & 62.0 & 21.8 \\
All in top-100, multi-func (\%)         & ---  & 18.3 & 15.4 & 41.4 \\
All in top-100, tightly clustered (\%)  & ---  & 34.0 & 30.2 & 53.1 \\
\bottomrule
\end{tabular}}
\end{table*}

\paragraph{Where dense retrieval falls short.}
Two complementary observations from \cref{tab:locality_app} explain why the graph expansion step matters even though this locality holds. First, single-function instances are easily handled by dense retrieval alone (the single ground truth is in the top-$100$ for $67\%$, $77\%$, and $75\%$ of \swepoly{} Python, Java, and JavaScript instances respectively), but multi-function instances collapse to $18\%$, $15\%$, and $41\%$. The dense ranker reliably surfaces at least one seed, which is the seeding condition $q>\zeta$ of Remark~\ref{rem:seed_condition}, but it does not co-rank distant siblings. Second, even among tightly-clustered multi-function instances (max GT-pair distance $\le 2$ hops), only $34\%$ (Python), $30\%$ (Java), $53\%$ (JavaScript) recover all ground truth in the top-$100$. Structural proximity exists in the graph but is not captured by embedding similarity alone; this is precisely the gap closed by the BFS-and-LLM-filter expansion in \ours{}.

\paragraph{Dataset-level reading of \cref{tab:locality_app}.}
The four splits stratify cleanly along the locality--coverage axes the method exploits. (a) \sweverified{} is the tightest dataset of the four (smallest fan-out at $1.23$ files and $2.25$ funcs/inst, $84.9\%$ tightly-clustered), which explains why \texttt{contains}-edge expansion is so effective on \sweverified{} in \cref{tab:swepoly_combined}. (b) \swepoly{} Java is the spread-out case ($32.9\%$ spread-out, $2.79$ files/inst) yet has the highest \texttt{invokes}-connectivity ($0.63$); BFS along \texttt{contains} alone structurally cannot reach cross-file ground truth in this regime, and routing through \texttt{invokes} is the most promising direction noted in our Limitations. (c) \swepoly{} JavaScript is dense-ranker-dominated: $53.6\%$ of instances are already fully covered in the top-$100$ and only $21.8\%$ have any ground truth beyond top-$500$, leaving little headroom for any graph mechanism. (d) \swepoly{} Python sits in between, with \sweverified{}-like dense-rank quality but messier structure ($5.02$ funcs/inst across $1.79$ files), which is where \ours{} adds the most absolute headroom.

\paragraph{Empirical ceiling under the current initial ranker.}
\cref{tab:locality_app} also shows that $38\%$ (\sweverified{}), $51\%$ (Python), $62\%$ (Java), and $22\%$ (JavaScript) of instances contain at least one ground-truth function ranked \emph{beyond} the top-$500$ by the dense retriever. Since \ours{}'s neighborhood exploration is restricted to the candidate pool $\mathcal{S}_N(Q)$ with $N{=}500$, these out-of-pool ground-truth functions cannot be recovered regardless of graph structure. This places a quantifiable upper bound on \ours{}'s reachable Recall@$K$ under the current initial ranker, and motivates the limitation noted in the main text: closing this gap requires either a stronger initial ranker or a larger candidate pool $N$ (with a corresponding LLM-token cost from \cref{tab:bfs_N_ablation}). The same ranker artifact also helps explain the relative gain ordering across languages, since the per-language coverage rates correlate with where \ours{}'s absolute Recall@$20$ gains sit in \cref{tab:swepoly_combined}.

\subsection{Algorithm}\label{sec:algorithm_app}

\begin{algorithm}[!h]
\caption{SpIDER}
\label{alg:spider}
\begin{algorithmic}[1]
\REQUIRE Code graph $\mathcal{G} = (\mathcal{V}, \mathcal{E})$, issue description $Q$,
         bi-modal encoder $\mathcal{F}(\cdot)$
\REQUIRE Parameters: retrieval budget $K$, filtering threshold $N$,
         number of centers $C$, search depth $d$

\STATE \textit{// Semantic Retrieval}
\STATE Compute node embeddings: $\{\mathcal{F}(v)\}_{v \in \mathcal{V}}$
\STATE Compute query embedding: $\mathcal{F}(Q)$
\STATE Compute relevance scores: $s_Q(v) = \cos(\mathcal{F}(v), \mathcal{F}(Q))$ for all $v \in \mathcal{V}$
\STATE $\mathcal{S}_K(Q) \gets \mathrm{arg\,top}_K s_Q(v)$ top-$K$ nodes ranked by $s_Q(v)$ \COMMENT{Baseline retrieval}
\STATE $\mathcal{S}_N(Q) \gets$ top-$N$ nodes ranked by $s_Q(v)$ \COMMENT{Candidate pool}

\STATE \textit{// Seed Selection}
\STATE $\mathcal{C}_Q \gets$ top-$C$ nodes from $\mathcal{S}_K(Q)$ ranked by $s_Q(v)$ \COMMENT{Centers for BFS}

\STATE \textit{// Neighborhood Exploration}
\STATE $\Gamma_d(\mathcal{C}_Q) \gets \{u \in \mathcal{V} : \mathrm{dist}_{\mathcal{G}}(u, \mathcal{C}_Q) \leq d\}$ via BFS along `\textit{contains}' edges

\STATE \textit{// Two-Stage Neighbor Filtering}
\STATE $\hat{\mathcal{C}}_Q \gets \Gamma_d(\mathcal{C}_Q) \cap \mathcal{S}_N(Q)$ \COMMENT{Stage 1: Semantic filtering}
\STATE $\mathcal{U}_Q \gets \{u \in \hat{\mathcal{C}}_Q : \mathcal{L}(u) = 1\} \cup \mathcal{C}_Q$ \COMMENT{Stage 2: LLM selection}

\STATE \textit{// Output Construction}
\STATE $\mathcal{S}^{\mathrm{LLM}}_K(Q) \gets \mathcal{U}_Q \cup \mathcal{S}_{K-|\mathcal{U}_Q|}(Q)$ \COMMENT{where $ \mathcal{S}_{K-|\mathcal{U}_Q|}(Q) = \underset{v \in \mathcal{\mathcal{S}_K(Q)}}{\mathrm{arg\,top}_{K - |\mathcal{U}_Q|}}\; s_Q(v)$}

\STATE \textbf{return} $\mathcal{S}(Q)$
\end{algorithmic}
\end{algorithm}

\textbf{Note:} The LLM selector $\mathcal{L}$ receives the source code content of all candidates $u \in \hat{\mathcal{C}}_Q$ for a given center, along with the issue description $Q$.
It returns $\mathcal{L}(u) = 1$ for candidates deemed relevant to resolving the issue, and $0$ otherwise.

\subsection{Stage-2 LLM Filter Prompt}\label{sec:prompt_app}

The Stage-2 selector is driven by the system and user prompts reproduced below. The prompt is deliberately simple, to limit prompt-sensitivity, and is applied independently to each of the $C$ seed centers.

\begin{lstlisting}[style=promptstyle,title=\textbf{System prompt.}]
You are a code localization agent. You are examining the graph neighborhood of a {node_type} to find related {node_type}s that MUST be modified to fix a GitHub issue.
You will see:
- The center {node_type}
- Its graph neighbors (connected via {edge_phrase})
BE VERY RESTRICTIVE. Only select neighbors where you have HIGH CONFIDENCE that the code MUST be modified to fix the issue. When in doubt, do NOT select. It is better to return an empty list than to include a neighbor that probably does not need changes.
Do NOT include the center itself. Do NOT include neighbors that are merely related, called by, or similar to code in the issue - only include neighbors whose own code must change.
Respond with ONLY a JSON object:
{{"selected": ["node_id_1", "node_id_2", ...]}}
If no neighbor clearly needs changes, respond with:
{{"selected": []}}
\end{lstlisting}

\begin{lstlisting}[style=promptstyle,title=\textbf{User prompt.}]
## GitHub Issue
{PROBLEM_STATEMENT}

## Center function (1/5)

### [center_i] {CENTER_NODE_ID}
{CENTER_CODE}

## Neighbors

### [nbr_i] {NEIGHBOR_i_ID}
{NEIGHBOR_i_CODE}

Select ONLY the functions that you are highly confident MUST be modified to fix the issue.
\end{lstlisting}

\paragraph{Placeholders.} \texttt{node\_type} is the retrieval granularity (\texttt{function} for all function-level results in this paper) and \texttt{edge\_phrase} names the edge types traversed for that run. \texttt{PROBLEM\_STATEMENT} is the raw issue title and body. The center header is numbered $(i{+}1)/C$ over the $C$ seed centers, shown above for the default $C{=}5$. \texttt{CENTER\_CODE} and each \texttt{NEIGHBOR\_i\_CODE} carry the node's source, truncated to $1500$ characters with a trailing \texttt{... [truncated]} marker when longer. Neighbors are enumerated with the graph node ids that the model echoes back in its JSON response. Doubled braces are Python format-string escapes; the model receives single braces. The template contains no in-context examples, so the selector is used zero-shot.

\subsection{Graph construction details}\label{sec:graph_construction_details_app}
JavaScript did not have a unique class data type until 2015 as it was only introduced ECMAScript 2015 also known as ES6. Hence most of the functionality of classes were implemented through functions, causing increased variations in the definition styles of a function (e.g. a nested function with the definition of container function at some other location in the  same/different file).

\paragraph{Artifacts and license.}
\oursbench{} is derived from publicly released benchmarks (\sweverified{}, \swepoly{}, and \multiswe{}). Ground-truth labels are inherited from those upstream sources, and construction is deterministic given the repository state, using no manual annotation. We release the dataset, the graph-construction code, and the Stage-2 LLM-filter prompt through the links on the first page. Because \oursbench{} embeds source code drawn from many upstream repositories, it carries no single license. Individual code segments remain under the licenses of the projects they were adapted from, which include MIT, Apache~2.0, BSD and GPL, and the release ships a notice file identifying the third-party sources. Users must comply with the license attached to each segment they use.

\subsection{Extension to Class-level edits}\label{sec:class_level_performance_app}
\begin{table*}[htbp]
\centering
\caption{Class-level performance on \swepoly{} for $K=20$, $N=500$, $C=5$, $d=2$. Winners per metric are in \textbf{bold}.}
\resizebox{0.7\textwidth}{!}{
\begin{tabular}{lllccc}
\toprule
\textbf{Language} & \textbf{Model} & \textbf{Retrieval} & \textbf{Recall@20} & \textbf{Accuracy@20} & \textbf{MRR@20} \\
\midrule
\multirow{8}{*}{Python}
 & \texttt{\swerankszs} & \ours{} & \textbf{0.65} & \textbf{0.54} & 0.36 \\
 &  & DER & 0.59 & 0.48 & \textbf{0.45} \\
 & \texttt{\codesageszs{}} & \ours{} & \textbf{0.39}  & \textbf{0.33}  & 0.16  \\
 &  & DER & 0.33  & 0.25  & \textbf{0.19}  \\
 & \texttt{\codesagespeft{}} & \ours{} & \textbf{0.31}  & \textbf{0.26}  & 0.16  \\
 &  & DER & 0.30  & 0.22  & \textbf{0.19}  \\
 & \texttt{\bm} & \spisr{} & \textbf{0.60}  & \textbf{0.52}  & 0.28  \\
 &  & \sr{} & 0.58  & 0.48  & \textbf{0.35}  \\
\midrule
\multirow{8}{*}{Java}
 & \texttt{\swerankszs} & \ours{} & \textbf{0.48}  & \textbf{0.37}  & \textbf{0.34}  \\
 &  & DER & 0.46  & \textbf{0.37}  & \textbf{0.34}  \\
 & \texttt{\codesageszs{}} & \ours{} & \textbf{0.25}  & \textbf{0.17}  & \textbf{0.11}  \\
 &  & DER & 0.23  & \textbf{0.17}  & 0.10  \\
 & \texttt{\codesagespeft{}} & \ours{} & 0.26  & 0.18  & 0.12  \\
 &  & DER & \textbf{0.28}  & \textbf{0.21}  & \textbf{0.14}  \\
 & \texttt{\bm} & \spisr{} & \textbf{0.33}  & \textbf{0.26}  & 0.22  \\
 &  & \sr{} & \textbf{0.33}  & \textbf{0.26}  & \textbf{0.25}  \\
\bottomrule
\end{tabular}}
\label{tab:swepolybench_class}
\end{table*}

In \cref{tab:swepolybench_class}, we demonstrate the efficacy of the proposed approach at a class level. Here, we include only Python and Java here, excluding JavaScript, as most JavaScript repositories in the \swepoly{} dataset lack significant use of class syntax (introduced in ECMAScript 2015), resulting in very few instances with class ground truth nodes. A ground truth edit is classified as a class-level edit (and the corresponding class as a ground truth class node), if the patch introduces a modification within any part of the class declaration or definition (e.g., constructors, destructors, or methods) to address the issue description.

\subsection{Additional Experiments}\label{sec:additional_exps_app}

\begin{figure}[ht]
    \centering
    \includegraphics[width=\linewidth]{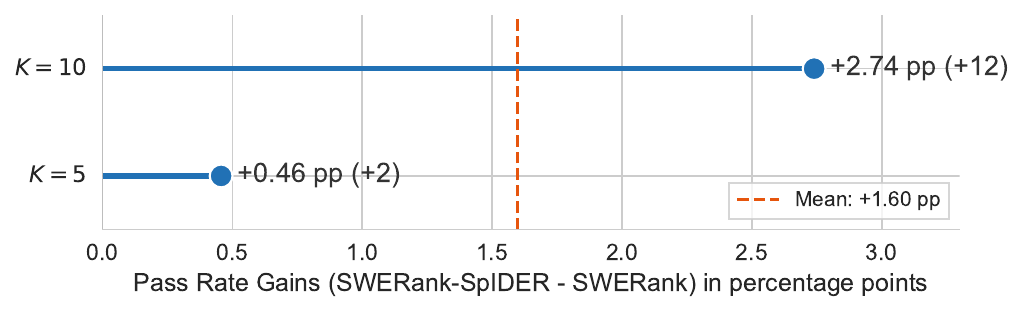}
    \caption{Pass rate improvement of \texttt{SWERankEmbed-Small-SpIDER} over \texttt{SWERank} on \texttt{SWE-bench Verified} with $C=5, d=4, N=500$.}
    \label{fig:sweverif-gains}
\end{figure}

\paragraph{Downstream code generation with a tool-equipped agent.}
\cref{fig:sweverif-gains} reports the complementary setting to \cref{tab:downstream_handstied}, in which the protocol is \emph{not} hands-tied: \texttt{mini-swe-agent}~\citep{yang2024swe} with \texttt{Claude Sonnet~4.5} receives the retrieved function paths but retains full grep and file-navigation tools, so it can partially compensate for a poor retrieval list on its own. In this setting \ours{} improves the pass rate on \sweverified{} by $+0.46$ percentage points (pp) at $K{=}5$ and $+2.74$ pp at $K{=}10$, corresponding to $2$ and $12$ additional resolved instances. Two caveats apply. First, these are single runs, and the $K{=}5$ margin of $2$ instances is within run-to-run variation, so we do not read a positive effect into it. Second, and more fundamentally, an agent that can already locate code independently compresses the gap between any two retrieval lists, so this comparison measures only what \ours{} contributes \emph{on top of} a competent search capability rather than the contribution of retrieval quality itself. \cref{tab:downstream_handstied} isolates the latter by removing the agent's independent search, which is why we report it as the primary downstream result.
\begin{figure*}
    \centering
    \includegraphics[width=\linewidth]{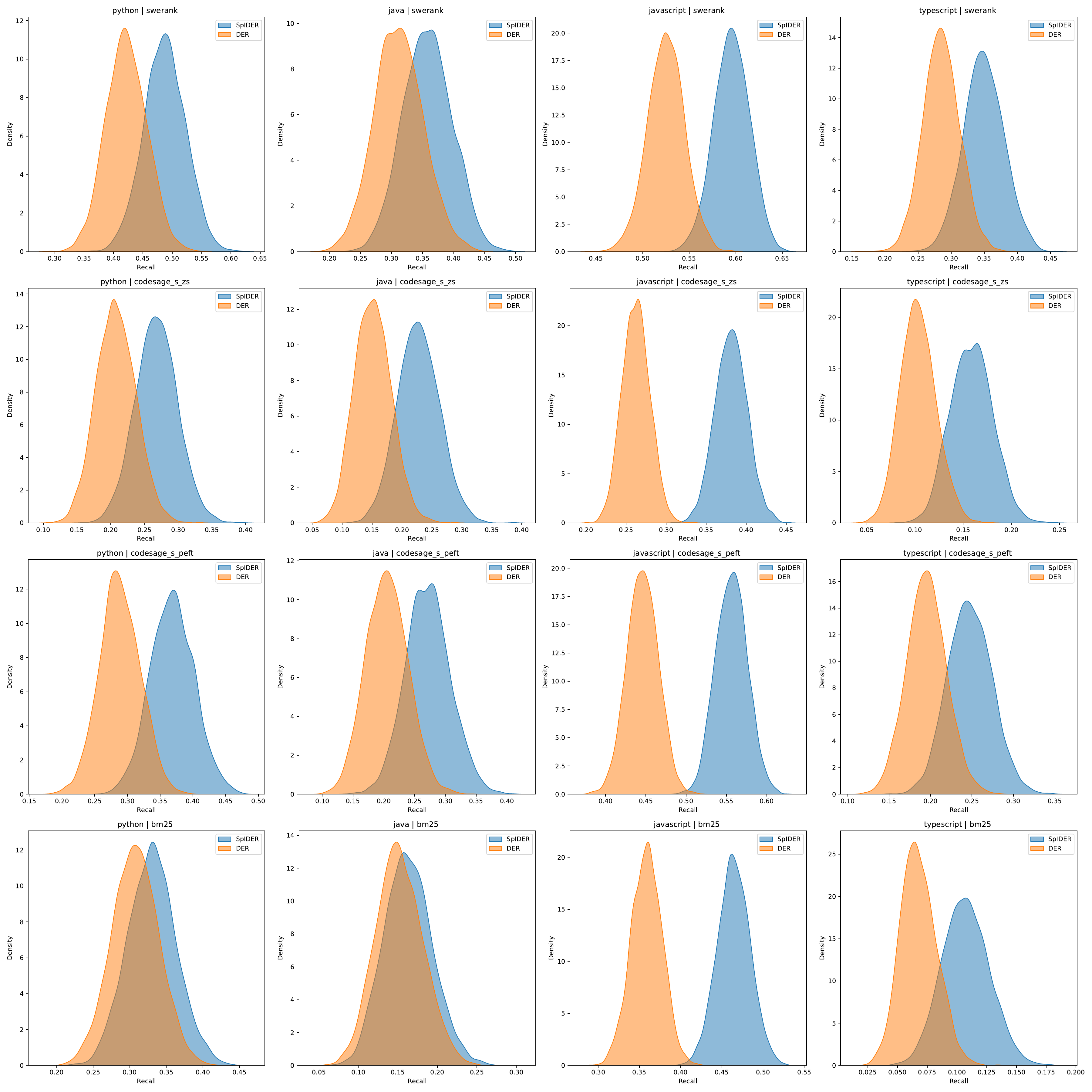}
    \caption{KDE (Kernel Density Estimate) plots with bootstrapped results of \swepoly{} benchmark for Recall@20 performance across various retrieval methods.}
    \label{fig:kde_plots_recall_swepolybench}
\end{figure*}

\begin{figure*}
    \centering
    \includegraphics[width=\linewidth]{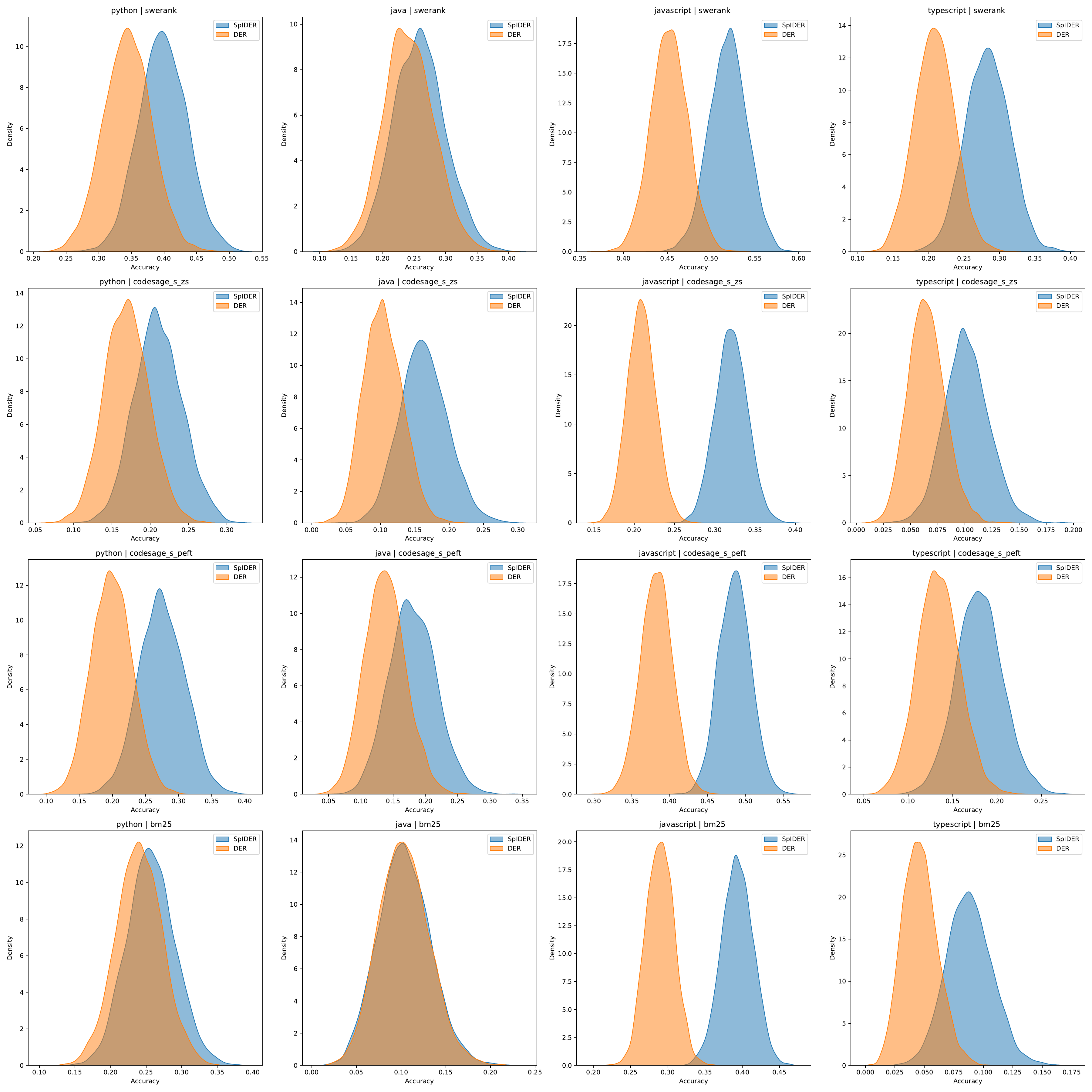}
    \caption{KDE (Kernel Density Estimate) plots with bootstrapped results of \swepoly{} benchmark for Acc@20 performance across various retrieval methods.}
    \label{fig:kde_plots_acc_swepolybench}
\end{figure*}

\begin{table*}[htbp]
\centering
\caption{Token usage for the BFS depth ($d$) ablation from \cref{tab:bfs_d_ablation} ($K=20$, $N=100$, $C=3$). Average per-call LLM input/output tokens grow with $d$ across languages.}
\resizebox{\textwidth}{!}{%
\begin{tabular}{c|cc|cc|cc|cc}
\toprule
\multirow{2}{*}{$\boldsymbol{d}$}
& \multicolumn{2}{c|}{\textbf{Python}}
& \multicolumn{2}{c|}{\textbf{JavaScript}}
& \multicolumn{2}{c|}{\textbf{Java}}
& \multicolumn{2}{c}{\textbf{TypeScript}} \\
& Input & Output & Input & Output & Input & Output & Input & Output \\
\midrule
2 & 2.6K & 32 & 2.0K & 50 & 2.2K & 22 & 1.5K & 21 \\
4 & 3.6K & 35 & 3.1K & 25 & 2.1K & 54 & 2.0K & 24 \\
6 & 11.9K & 48 & 10.1K & 42 & 3.6K & 67 & 4.3K & 33 \\
8 & 29.7K & 69 & 15.0K & 48 & 5.6K & 82 & 5.6K & 37 \\
\bottomrule
\end{tabular}
}
\label{tab:bfs_d_ablation_tokens_app}
\end{table*}

\begin{table*}[htbp]
\centering
\caption{Performance of \ours{} + \swerankszs{} encoder with MRR for different BFS depths $d$ with $K=20$, $N=100$, $C=3$}
\resizebox{\textwidth}{!}{%
\begin{tabular}{lccc|ccc|ccc|ccc}
\hline
\multirow{2}{*}{$\boldsymbol{d}$} & \multicolumn{3}{c|}{\textbf{Python}} & \multicolumn{3}{c|}{\textbf{JavaScript}} & \multicolumn{3}{c|}{\textbf{Java}} & \multicolumn{3}{c}{\textbf{TypeScript}} \\
 & Recall@20 & Accuracy@20 & MRR@20 & Recall@20 & Accuracy@20 & MRR@20 & Recall@20 & Accuracy@20 & MRR@20 & Recall@20 & Accuracy@20 & MRR@20 \\
\hline
2 & 0.45 & 0.36 & 0.28 & 0.54 & 0.46 & 0.17 & 0.33 & 0.25 & 0.18 & 0.31 & 0.23 & 0.06 \\
4 & 0.46 & 0.37 & 0.30 & 0.57 & 0.49 & 0.18 & 0.33 & 0.25 & 0.18 & 0.30 & 0.24 & 0.06 \\
6 & 0.50 & 0.41 & 0.33 & 0.62 & 0.53 & 0.20 & 0.33 & 0.25 & 0.18 & 0.36 & 0.28 & 0.07 \\
8 & 0.55 & 0.44 & 0.38 & 0.67 & 0.58 & 0.22 & 0.34 & 0.27 & 0.20 & 0.36 & 0.27 & 0.07 \\
\hline
\end{tabular}
}
\label{tab:bfs_d_ablation_app}\label{tab:bfs_d_ablation}
\end{table*}

\begin{table*}[htbp]
\centering
\caption{\ours{} + \swerankszs{} encoder performance with varying neighbor filtering threshold $N$ for $K=20$, $C=3$, $d=4$}
\resizebox{\textwidth}{!}{
\begin{tabular}{lccc|ccc|ccc|ccc}
\hline
\multirow{2}{*}{$\boldsymbol{N}$} & \multicolumn{3}{c|}{\textbf{Python}} & \multicolumn{3}{c|}{\textbf{JavaScript}} & \multicolumn{3}{c|}{\textbf{Java}} & \multicolumn{3}{c}{\textbf{TypeScript}} \\
 & Recall@20 & Accuracy@20 & MRR@20 & Recall@20 & Accuracy@20 & MRR@20 & Recall@20 & Accuracy@20 & MRR@20 & Recall@20 & Accuracy@20 & MRR@20 \\
\hline
50 & 0.45 & 0.37 & 0.29 & 0.56 & 0.48 & 0.18 & 0.32 & 0.24 & 0.19 & 0.31 & 0.24 & 0.06 \\
100 & 0.46 & 0.37 & 0.30 & 0.57 & 0.49 & 0.18 & 0.33 & 0.25 & 0.18 & 0.30 & 0.24 & 0.06 \\
300 & 0.48 & 0.39 & 0.30 & 0.58 & 0.50 & 0.18 & 0.35 & 0.26 & 0.19 & 0.32 & 0.25 & 0.06 \\
500 & 0.49 & 0.39 & 0.30 & 0.59 & 0.51 & 0.18 & 0.35 & 0.26 & 0.19 & 0.31 & 0.24 & 0.06 \\
700 & 0.48 & 0.39 & 0.31 & 0.59 & 0.51 & 0.18 & 0.35 & 0.25 & 0.18 & 0.31 & 0.23 & 0.06 \\
900 & 0.48 & 0.39 & 0.31 & 0.58 & 0.50 & 0.18 & 0.34 & 0.25 & 0.19 & 0.31 & 0.24 & 0.06 \\
1000 & 0.48 & 0.39 & 0.31 & 0.59 & 0.51 & 0.18 & 0.36 & 0.27 & 0.19 & 0.32 & 0.24 & 0.06 \\
\hline
\end{tabular}
}
\label{tab:bfs_N_ablation_app}
\end{table*}

\begin{table*}[htbp]
\centering
\caption{LocAgent performance on \sweverified{} and a subset of Python instances in \swepoly{}. Running LocAgent took over 48 hours on \sweverified{} and approximately 24 hours on \swepoly{}, highlighting a key limitation of this approach. In addition to the long runtime, LocAgent also requires significantly more LLM invocations per instance compared to dense embedding retrieval and \bm{}, resulting in higher computational costs.}
\resizebox{\textwidth}{!}{
    \begin{tabular}{l|l|cccccccccccc}
    \hline
    \multirow{2}{*}{\textbf{Dataset}} & \multirow{2}{*}{\textbf{Metrics}} & \multicolumn{12}{c}{\textbf{K}} \\
    \cline{3-14}
     &  & \textbf{3} &
    \textbf{5} & \textbf{10} & \textbf{20} & \textbf{30} & \textbf{40} & \textbf{50} & \textbf{60} & \textbf{70} & \textbf{80} & \textbf{90} & \textbf{100} \\
    \hline
    \multirow{3}{*}{\sweverified{}}
     & \textbf{Recall@K} & 0.18 & 0.26 & 0.41 & 0.54 & 0.62 & 0.68 & 0.69 & 0.71 & 0.72 & 0.73 & 0.74 & 0.74 \\
     & \textbf{Acc@K} & 0.17 & 0.23 & 0.37 & 0.48 & 0.57 & 0.61 & 0.63 & 0.64 & 0.66 & 0.66 & 0.67 & 0.67 \\
     & \textbf{MRR@K} & 0.12 & 0.14 & 0.16 & 0.17 & 0.17 & 0.17 & 0.17 & 0.17 & 0.17 & 0.17 & 0.18 & 0.18 \\

    \cline{1-14}
    \multirow{2}{*}{\swepoly{}}
     & \textbf{Recall@K} & 0.23 & 0.30 & 0.41 & 0.49 & 0.53 & 0.58 & 0.58 & 0.59 & 0.59 & 0.60 & 0.61 & 0.61 \\
     & \textbf{Acc@K} & 0.17 & 0.23 & 0.31 & 0.37 & 0.41 & 0.46 & 0.46 & 0.46 & 0.46 & 0.47 & 0.48 & 0.49 \\
     (Python subset) & \textbf{MRR@K} & 0.21 & 0.23 & 0.25 & 0.25 & 0.25 & 0.26 & 0.26 & 0.26 & 0.26 & 0.26 & 0.26 & 0.26 \\
    \hline
    \end{tabular}
}
\label{tab:locagent_performance}
\end{table*}

\begin{table*}[htbp]
\centering
\caption{Comparison of LocAgent with dense retrieval methods + LLM reranker for Python repositories in \swepoly{}}
\resizebox{\textwidth}{!}{
    \begin{tabular}{ll|ccc}
    \hline
    \textbf{Dataset} & \textbf{Model} & \textbf{Recall@3} & \textbf{Accuracy@3} & \textbf{MRR@3} \\
    \hline
    \multirow{5}{*}{\swepoly{}} & \locagent{} & 0.18  & 0.17  & 0.12  \\
    \cline{2-5}
     & \swerankszs + \ours{} & 0.44  & 0.36  & 0.54  \\
     & + LLM reranker & & & \\
     & \codesageszs{} + \ours{} & 0.23  & 0.18  & 0.31  \\
     & + LLM reranker & & & \\
     & \codesagespeft{} + \ours{} & 0.32  & 0.25  & 0.42  \\
     & + LLM reranker & & & \\
     & \bm + \spisr{} & 0.30 & 0.23  & 0.38  \\
     \hline
    \end{tabular}
}
\label{tab:locagent_vs_spider}
\end{table*}

\begin{table*}[htbp]
\centering
\caption{\swepoly{} retrieval performance with 95\% confidence intervals ($\mu \pm$ CI) for $K=20$, $N=500$, $C=5$, $d=4$}
\resizebox{\textwidth}{!}{
\begin{tabular}{llllcc}
\toprule
\textbf{Language} & \textbf{Model} & \textbf{Retrieval} &
\textbf{Recall@20} & \textbf{Accuracy@20} \\
\midrule
\multirow{8}{*}{Python}
 & \multirow{2}{*}{\texttt{\swerankszs}} & \texttt{\ours{}} & 0.49 $\pm$ 0.06 & 0.40 $\pm$ 0.07 \\
 &  & \texttt{DER} & 0.42 $\pm$ 0.06 & 0.35 $\pm$ 0.07 \\
 & \multirow{2}{*}{\texttt{\codesageszs{}}} & \texttt{\ours{}} & 0.27 $\pm$ 0.06 & 0.21 $\pm$ 0.06 \\
 &  & \texttt{DER} & 0.21 $\pm$ 0.06 & 0.17 $\pm$ 0.06 \\
 & \multirow{2}{*}{\texttt{\codesagespeft{}}} & \texttt{\ours{}} & 0.37 $\pm$ 0.06 & 0.27 $\pm$ 0.07 \\
 &  & \texttt{DER} & 0.29 $\pm$ 0.06 & 0.20 $\pm$ 0.06 \\
 & \multirow{2}{*}{\texttt{\bm}} & \texttt{\spisr{}} & 0.33 $\pm$ 0.07 & 0.26 $\pm$ 0.07 \\
 &  & \texttt{\sr{}} & 0.31 $\pm$ 0.06 & 0.24 $\pm$ 0.07 \\
\midrule
\multirow{8}{*}{Java}
 & \multirow{2}{*}{\texttt{\swerankszs}} & \texttt{\ours{}} & 0.36 $\pm$ 0.08 & 0.27 $\pm$ 0.08 \\
 &  & \texttt{DER} & 0.31 $\pm$ 0.07 & 0.24 $\pm$ 0.08 \\
 & \multirow{2}{*}{\texttt{\codesageszs{}}} & \texttt{\ours{}} & 0.23 $\pm$ 0.07 & 0.17 $\pm$ 0.07 \\
 &  & \texttt{DER} & 0.15 $\pm$ 0.06 & 0.10 $\pm$ 0.05 \\
 & \multirow{2}{*}{\texttt{\codesagespeft{}}} & \texttt{\ours{}} & 0.28 $\pm$ 0.08 & 0.18 $\pm$ 0.07 \\
 &  & \texttt{DER} & 0.21 $\pm$ 0.07 & 0.14 $\pm$ 0.06 \\
 & \multirow{2}{*}{\texttt{\bm}} & \texttt{\spisr{}} & 0.16 $\pm$ 0.06 & 0.10 $\pm$ 0.05 \\
 &  & \texttt{\sr{}} & 0.15 $\pm$ 0.06 & 0.10 $\pm$ 0.05 \\
\midrule
\multirow{8}{*}{JavaScript}
 & \multirow{2}{*}{\texttt{\swerankszs}} & \texttt{\ours{}} & 0.60 $\pm$ 0.04 & 0.52 $\pm$ 0.04 \\
 &  & \texttt{DER} & 0.52 $\pm$ 0.04 & 0.45 $\pm$ 0.04 \\
 & \multirow{2}{*}{\texttt{\codesageszs{}}} & \texttt{\ours{}} & 0.38 $\pm$ 0.04 & 0.32 $\pm$ 0.04 \\
 &  & \texttt{DER} & 0.26 $\pm$ 0.04 & 0.21 $\pm$ 0.04 \\
 & \multirow{2}{*}{\texttt{\codesagespeft{}}} & \texttt{\ours{}} & 0.55 $\pm$ 0.04 & 0.48 $\pm$ 0.05 \\
 &  & \texttt{DER} & 0.45 $\pm$ 0.04 & 0.38 $\pm$ 0.04 \\
 & \multirow{2}{*}{\texttt{\bm}} & \texttt{\spisr{}} & 0.46 $\pm$ 0.04 & 0.39 $\pm$ 0.04 \\
 &  & \texttt{\sr{}} & 0.36 $\pm$ 0.04 & 0.29 $\pm$ 0.04 \\
\midrule
\multirow{8}{*}{TypeScript}
 & \multirow{2}{*}{\texttt{\swerankszs}} & \texttt{\ours{}} & 0.35 $\pm$ 0.06 & 0.28 $\pm$ 0.05 \\
 &  & \texttt{DER} & 0.29 $\pm$ 0.05 & 0.21 $\pm$ 0.05 \\
 & \multirow{2}{*}{\texttt{\codesageszs{}}} & \texttt{\ours{}} & 0.16 $\pm$ 0.04 & 0.10 $\pm$ 0.04 \\
 &  & \texttt{DER} & 0.10 $\pm$ 0.04 & 0.06 $\pm$ 0.04 \\
 & \multirow{2}{*}{\texttt{\codesagespeft{}}} & \texttt{\ours{}} & 0.25 $\pm$ 0.05 & 0.18 $\pm$ 0.05 \\
 &  & \texttt{DER} & 0.19 $\pm$ 0.05 & 0.13 $\pm$ 0.05 \\
 & \multirow{2}{*}{\texttt{\bm}} & \texttt{\spisr{}} & 0.11 $\pm$ 0.04 & 0.09 $\pm$ 0.04 \\
 &  & \texttt{\sr{}} & 0.07 $\pm$ 0.03 & 0.05 $\pm$ 0.03 \\
\bottomrule
\end{tabular}
}
\label{tab:swepolybench_ci_compact}
\end{table*}

\begin{table*}[htbp]
\centering
\caption{\textbf{Ablation on primary neighborhood filtering threshold $N$} using \ours{} + \texttt{SweRankEmbed-Small} for $K=20$, $C=3$, $d=4$. Increasing $N$ improves coverage and Recall@20 with modest increases in LLM token usage. Avg. tokens indicate average consumption per LLM call; number of calls equals $C$. Best Recall and Accuracy per language are highlighted.}
\resizebox{\textwidth}{!}{%
\begin{tabular}{c|ccc|ccc|ccc|ccc}
\toprule
\multirow{2}{*}{$\boldsymbol{N}$}
& \multicolumn{3}{c|}{\textbf{Python}}
& \multicolumn{3}{c|}{\textbf{JavaScript}}
& \multicolumn{3}{c|}{\textbf{Java}}
& \multicolumn{3}{c}{\textbf{TypeScript}} \\
\cmidrule(lr){2-4} \cmidrule(lr){5-7} \cmidrule(lr){8-10} \cmidrule(lr){11-13}
& Recall & Accuracy & Tokens
& Recall & Accuracy & Tokens
& Recall & Accuracy & Tokens
& Recall & Accuracy & Tokens \\
& & & (Input / Output)
& & & (Input / Output)
& & & (Input / Output)
& & & (Input / Output) \\
\midrule
50   & 0.45 & 0.37 & 3108 / 33  & 0.56 & 0.48 & 2383 / 23  & 0.32 & 0.24 & 1845 / 49  & 0.31 & 0.24 & 1763 / 23 \\
100  & 0.46 & 0.37 & 3591 / 35  & 0.57 & 0.49 & 3115 / 25  & 0.33 & 0.25 & 2083 / 54  & 0.30 & 0.24 & 1994 / 24 \\
300  & 0.48 & 0.39 & 4848 / 37  & 0.58 & 0.50 & 5119 / 29  & 0.35 & 0.26 & 2644 / 56  & 0.32 & 0.25 & 2459 / 26 \\
500  & \cellcolor{winnerblue}0.49 & \cellcolor{winnerblue}0.39 & 5708 / 39  & 0.59 & 0.51 & 6775 / 30  & 0.35 & 0.26 & 2978 / 57  & 0.31 & 0.24 & 2687 / 26 \\
700  & 0.48 & 0.39 & 6361 / 39  & 0.59 & 0.51 & 7998 / 31  & 0.35 & 0.25 & 3203 / 59  & 0.31 & 0.23 & 2843 / 27 \\
900  & 0.48 & 0.39 & 6903 / 40  & 0.58 & 0.50 & 8780 / 32  & 0.35 & 0.26 & 3343 / 58  & 0.31 & 0.24 & 2943 / 27 \\
1000 & 0.48 & 0.39 & 7180 / 40  & \cellcolor{winnerblue}0.59 & \cellcolor{winnerblue}0.51 & 9098 / 32  & \cellcolor{winnerblue}0.36 & \cellcolor{winnerblue}0.27 & 3441 / 59  & \cellcolor{winnerblue}0.32 & \cellcolor{winnerblue}0.25 & 3001 / 27 \\
\bottomrule
\end{tabular}
}
\label{tab:bfs_N_ablation}
\end{table*}

\begin{table*}[htbp]
\centering
\caption{Effect of retrieval budget $K$ on \ours{} + \texttt{SweRankEmbed-Small} performance for $N=500$, $C=5$, $d=4$. Increasing $K$ consistently improves Recall@K and Accuracy@K, while MRR@K remains stable. Best values per metric are highlighted.}
\resizebox{\textwidth}{!}{%
\begin{tabular}{c|ccc|ccc|ccc|ccc}
\toprule
\multirow{2}{*}{\textbf{K}}
& \multicolumn{3}{c|}{\textbf{Python}}
& \multicolumn{3}{c|}{\textbf{Java}}
& \multicolumn{3}{c|}{\textbf{JavaScript}}
& \multicolumn{3}{c}{\textbf{TypeScript}} \\
\cmidrule(lr){2-4} \cmidrule(lr){5-7} \cmidrule(lr){8-10} \cmidrule(lr){11-13}
& Recall & Accuracy & MRR
& Recall & Accuracy & MRR
& Recall & Accuracy & MRR
& Recall & Accuracy & MRR \\
\midrule
3   & 0.31 & 0.25 & 0.27 & 0.17 & 0.09 & 0.17 & 0.32 & 0.26 & 0.16 & 0.17 & 0.12 & 0.06 \\
5   & 0.39 & 0.32 & 0.30 & 0.24 & 0.16 & 0.18 & 0.39 & 0.34 & 0.17 & 0.19 & 0.14 & 0.06 \\
10  & 0.44 & 0.36 & 0.30 & 0.33 & 0.23 & 0.19 & 0.49 & 0.41 & 0.17 & 0.29 & 0.23 & 0.06 \\
20  & 0.49 & 0.39 & 0.29 & 0.35 & 0.26 & 0.19 & 0.59 & 0.51 & 0.18 & 0.35 & 0.28 & 0.06 \\
30  & 0.52 & 0.42 & 0.30 & 0.37 & 0.27 & 0.19 & 0.62 & 0.54 & 0.18 & 0.40 & 0.31 & 0.06 \\
40  & 0.52 & 0.43 & 0.30 & 0.38 & 0.28 & 0.19 & 0.65 & 0.57 & 0.18 & 0.42 & 0.33 & 0.07 \\
50  & 0.54 & 0.44 & 0.30 & 0.42 & 0.31 & 0.19 & 0.68 & 0.59 & 0.18 & 0.42 & 0.33 & 0.06 \\
60  & 0.58 & 0.47 & 0.31 & 0.43 & 0.32 & 0.19 & 0.70 & 0.61 & 0.18 & 0.45 & 0.36 & 0.06 \\
70  & 0.58 & 0.46 & 0.30 & 0.45 & 0.34 & 0.19 & 0.72 & 0.63 & 0.18 & 0.47 & 0.38 & 0.07 \\
80  & 0.60 & 0.48 & 0.30 & 0.46 & 0.34 & 0.19 & 0.73 & 0.64 & 0.18 & 0.48 & 0.38 & 0.07 \\
90  & 0.61 & 0.50 & 0.30 & 0.46 & 0.34 & 0.19 & 0.75 & 0.67 & 0.18 & 0.48 & 0.39 & 0.06 \\
100 & \cellcolor{winnerblue}0.62 & \cellcolor{winnerblue}0.50 & 0.30 & 0.46 & 0.34 & 0.19 & \cellcolor{winnerblue}0.77 & \cellcolor{winnerblue}0.68 & 0.18 & \cellcolor{winnerblue}0.52 & \cellcolor{winnerblue}0.43 & 0.07 \\
\bottomrule
\end{tabular}
}
\label{tab:bfs_K_ablation}
\end{table*}

\begin{table*}[htbp]
\centering
\caption{\texttt{SweRankEmbed-Small} zero-shot performance across varying retrieval budgets $K$ for $N=500$, $C=5$, $d=4$. Best Recall and Accuracy per language are highlighted.}
\resizebox{\textwidth}{!}{%
\begin{tabular}{c|ccc|ccc|ccc|ccc}
\toprule
\multirow{2}{*}{\textbf{K}}
& \multicolumn{3}{c|}{\textbf{Python}}
& \multicolumn{3}{c|}{\textbf{Java}}
& \multicolumn{3}{c|}{\textbf{JavaScript}}
& \multicolumn{3}{c}{\textbf{TypeScript}} \\
\cmidrule(lr){2-4} \cmidrule(lr){5-7} \cmidrule(lr){8-10} \cmidrule(lr){11-13}
& Recall & Accuracy & MRR
& Recall & Accuracy & MRR
& Recall & Accuracy & MRR
& Recall & Accuracy & MRR \\
\midrule
3   & 0.24 & 0.19 & 0.27 & 0.18 & 0.12 & 0.24 & 0.26 & 0.22 & 0.27 & 0.11 & 0.07 & 0.15 \\
5   & 0.31 & 0.25 & 0.29 & 0.26 & 0.19 & 0.26 & 0.33 & 0.27 & 0.28 & 0.16 & 0.12 & 0.16 \\
10  & 0.36 & 0.30 & 0.30 & 0.30 & 0.22 & 0.26 & 0.43 & 0.37 & 0.30 & 0.20 & 0.14 & 0.17 \\
20  & 0.42 & 0.35 & 0.30 & 0.31 & 0.24 & 0.26 & 0.52 & 0.45 & 0.30 & 0.29 & 0.21 & 0.18 \\
30  & 0.47 & 0.39 & 0.30 & 0.33 & 0.25 & 0.26 & 0.57 & 0.49 & 0.31 & 0.34 & 0.26 & 0.18 \\
40  & 0.49 & 0.40 & 0.30 & 0.35 & 0.26 & 0.27 & 0.61 & 0.53 & 0.31 & 0.37 & 0.29 & 0.18 \\
50  & 0.51 & 0.42 & 0.31 & 0.39 & 0.29 & 0.27 & 0.65 & 0.56 & 0.31 & 0.38 & 0.30 & 0.18 \\
60  & 0.55 & 0.45 & 0.31 & 0.41 & 0.30 & 0.27 & 0.67 & 0.58 & 0.31 & 0.41 & 0.32 & 0.18 \\
70  & 0.56 & 0.45 & 0.31 & 0.43 & 0.32 & 0.27 & 0.68 & 0.59 & 0.31 & 0.42 & 0.33 & 0.18 \\
80  & 0.57 & 0.46 & 0.31 & 0.43 & 0.32 & 0.27 & 0.70 & 0.61 & 0.31 & 0.43 & 0.34 & 0.18 \\
90  & 0.58 & 0.47 & 0.31 & 0.43 & 0.32 & 0.27 & 0.72 & 0.63 & 0.31 & 0.46 & 0.36 & 0.18 \\
100 & \cellcolor{winnerblue}0.59 & \cellcolor{winnerblue}0.48 & 0.31
    & \cellcolor{winnerblue}0.44 & \cellcolor{winnerblue}0.33 & 0.27
    & \cellcolor{winnerblue}0.73 & \cellcolor{winnerblue}0.64 & 0.31
    & \cellcolor{winnerblue}0.48 & \cellcolor{winnerblue}0.39 & 0.18 \\
\bottomrule
\end{tabular}
}
\label{tab:zs_K_ablation}
\end{table*}

\begin{figure*}
    \centering
    \includegraphics[width=\linewidth]{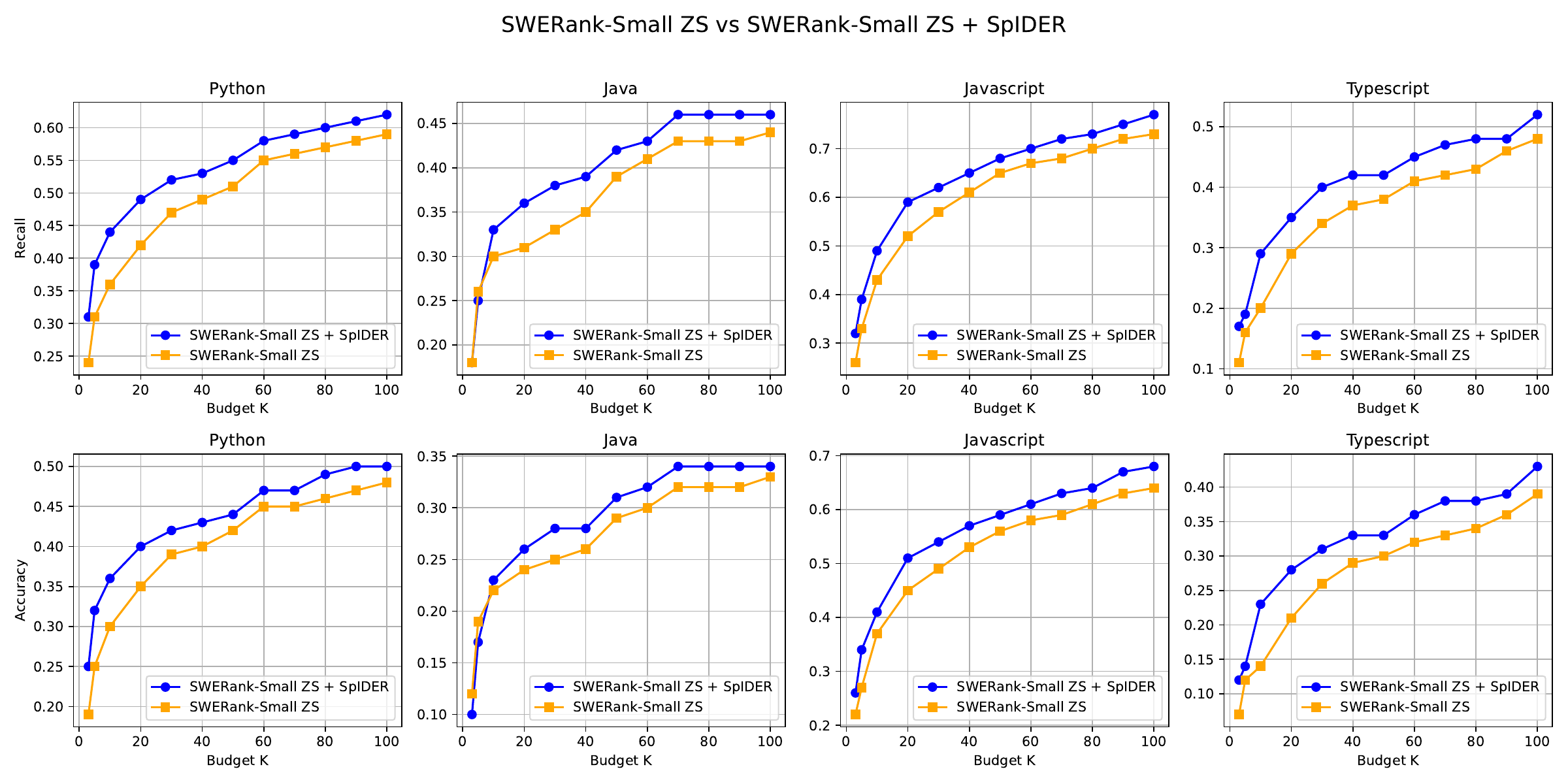}
    \caption{\ours{} vs \der{} performance at various values of $K$ on \swepoly{} benchmark for \swerankszs{} embedding model, $N=500$, $C=5$, $d=4$.}
    \label{fig:k_ablation_grid}
\end{figure*}

\begin{figure*}
    \centering
    \includegraphics[width=\linewidth]{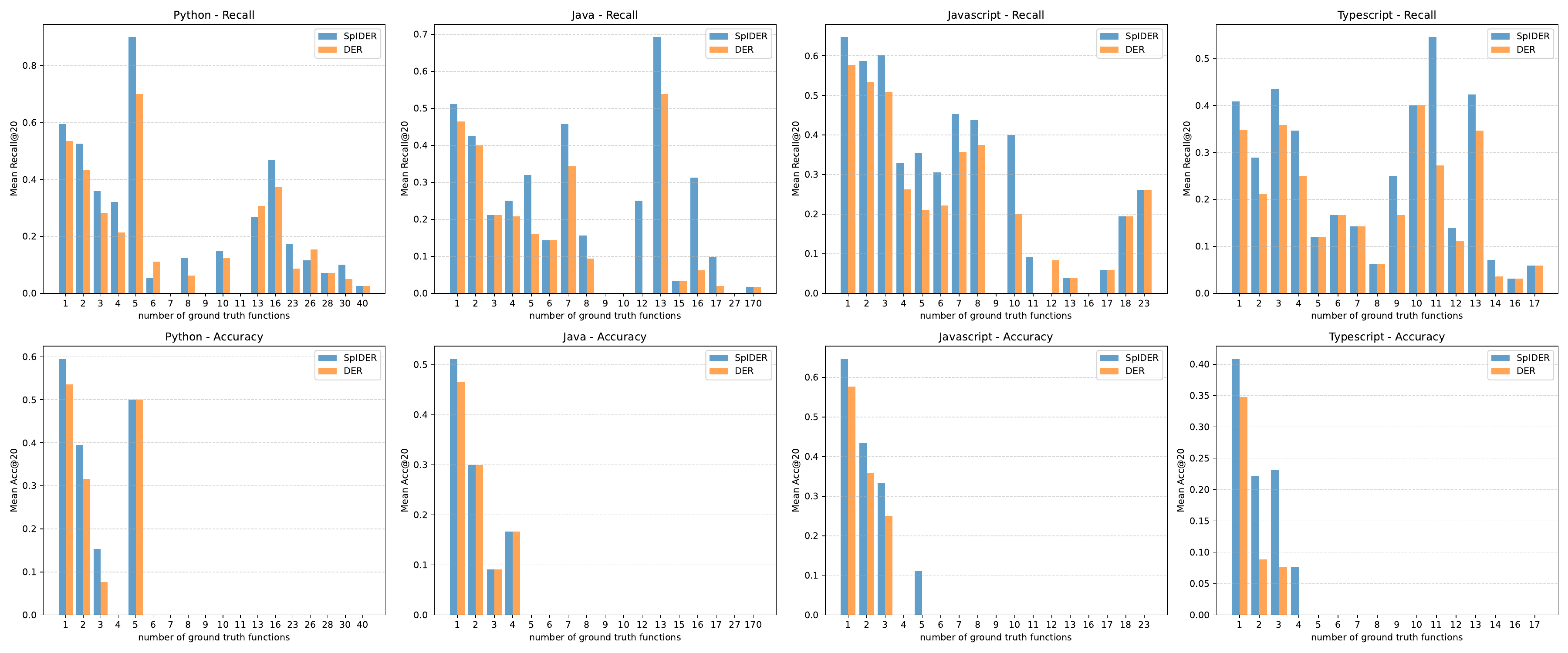}
    \caption{Comparing retrieval Recall@20 and Accuracy vs. number of edits for \der{} vs \ours{} on \swerankszs{} using \swepoly{} across various languages for $N=500$, $C=5$, $d=4$. Top row: Recall@20. Bottom row: Accuracy.}
    \label{fig:performance_vs_edits}
\end{figure*}

\begin{table*}[!tbp]
\centering
\caption{Comparison of dense retrieval and dense retrieval + LLM reranker for $K=20$ and $N=500$, $C=5$, $d=4$ on \swepoly{}. Winners per metric per embedding model are highlighted in light blue.}
\resizebox{\textwidth}{!}{
\begin{tabular}{l c c|ccc|ccc}
\toprule
\textbf{Language} & \textbf{Model} & \textbf{Retrieval}
& \multicolumn{3}{c|}{\textbf{Dense Retrieval Only}}
& \multicolumn{3}{c}{\textbf{Dense + LLM reranker}} \\
 & &
 & \textbf{Recall@20} & \textbf{Accuracy@20} & \textbf{MRR@20}
 & \textbf{Recall@3} & \textbf{Accuracy@3} & \textbf{MRR@3} \\
\midrule
\multirow{8}{*}{Python}
 & \multirow{2}{*}{\texttt{\swerankszs}}
 & \texttt{\ours{}} & \cellcolor{winnerblue}0.49 & \cellcolor{winnerblue}0.40 & \cellcolor{winnerblue}0.30
 & \cellcolor{winnerblue}0.44 & \cellcolor{winnerblue}0.36 & \cellcolor{winnerblue}0.54 \\
 &  & \texttt{DER} & 0.42 & 0.35 & 0.26
 & 0.38 & 0.31 & 0.48 \\
 & \multirow{2}{*}{\texttt{\codesageszs{}}}
 & \texttt{\ours{}} & \cellcolor{winnerblue}0.27 & \cellcolor{winnerblue}0.21 & \cellcolor{winnerblue}0.13
 & \cellcolor{winnerblue}0.23 & \cellcolor{winnerblue}0.18 & \cellcolor{winnerblue}0.31 \\
 &  & \texttt{DER} & 0.21 & 0.17 & 0.11
 & 0.19 & 0.16 & 0.25 \\
 & \multirow{2}{*}{\texttt{\codesagespeft{}}}
 & \texttt{\ours{}} & \cellcolor{winnerblue}0.37 & \cellcolor{winnerblue}0.27 & \cellcolor{winnerblue}0.24
 & \cellcolor{winnerblue}0.32 & \cellcolor{winnerblue}0.25 & \cellcolor{winnerblue}0.42 \\
 &  & \texttt{DER} & 0.29 & 0.20 & 0.21
 & 0.26 & 0.20 & 0.38 \\
 & \multirow{2}{*}{\texttt{\bm}}
 & \texttt{\spisr{}} & \cellcolor{winnerblue}0.33 & \cellcolor{winnerblue}0.26 & \cellcolor{winnerblue}0.19
 & \cellcolor{winnerblue}0.30 & \cellcolor{winnerblue}0.23 & \cellcolor{winnerblue}0.38 \\
 &  & \texttt{\sr{}} & 0.31 & 0.24 & 0.18
 & 0.27 & 0.22 & 0.35 \\
\midrule
\multirow{8}{*}{Java}
 & \multirow{2}{*}{\texttt{\swerankszs}}
 & \texttt{\ours{}} & \cellcolor{winnerblue}0.36 & \cellcolor{winnerblue}0.27 & \cellcolor{winnerblue}0.18
 & \cellcolor{winnerblue}0.25 & \cellcolor{winnerblue}0.17 & \cellcolor{winnerblue}0.36 \\
 &  & \texttt{DER} & 0.31 & 0.24 & 0.18
 & 0.24 & 0.16 & 0.33 \\

 & \multirow{2}{*}{\texttt{\codesageszs{}}}
 & \texttt{\ours{}} & \cellcolor{winnerblue}0.23 & \cellcolor{winnerblue}0.17 & \cellcolor{winnerblue}0.10
 & \cellcolor{winnerblue}0.16 & \cellcolor{winnerblue}0.09 & \cellcolor{winnerblue}0.26 \\
 &  & \texttt{DER} & 0.15 & 0.10 & 0.07
 & 0.11 & 0.07 & 0.19 \\

 & \multirow{2}{*}{\texttt{\codesagespeft{}}}
 & \texttt{\ours{}} & \cellcolor{winnerblue}0.28 & \cellcolor{winnerblue}0.18 & \cellcolor{winnerblue}0.15
 & \cellcolor{winnerblue}0.20 & \cellcolor{winnerblue}0.14 & \cellcolor{winnerblue}0.28 \\
 &  & \texttt{DER} & 0.21 & 0.14 & 0.13
 & 0.17 & 0.11 & 0.25 \\

 & \multirow{2}{*}{\texttt{\bm}}
 & \texttt{\spisr{}} & \cellcolor{winnerblue}0.16 & \cellcolor{winnerblue}0.10 & \cellcolor{winnerblue}0.10
 & \cellcolor{winnerblue}0.12 & 0.08 & \cellcolor{winnerblue}0.20 \\
 &  & \texttt{\sr{}} & 0.15 & 0.10 & 0.10
 & \cellcolor{winnerblue}0.12 & \cellcolor{winnerblue}0.09 & 0.19 \\
\midrule
\multirow{8}{*}{JavaScript}
 & \multirow{2}{*}{\texttt{\swerankszs}}
 & \texttt{\ours{}} & \cellcolor{winnerblue}0.60 & \cellcolor{winnerblue}0.52 & \cellcolor{winnerblue}0.18
 & \cellcolor{winnerblue}0.43 & \cellcolor{winnerblue}0.37 & \cellcolor{winnerblue}0.43 \\
 &  & \texttt{DER} & 0.52 & 0.45 & 0.16
 & 0.41 & 0.34 & 0.42 \\

 & \multirow{2}{*}{\texttt{\codesageszs{}}}
 & \texttt{\ours{}} & \cellcolor{winnerblue}0.38 & \cellcolor{winnerblue}0.32 & \cellcolor{winnerblue}0.09
 & \cellcolor{winnerblue}0.30 & \cellcolor{winnerblue}0.25 & \cellcolor{winnerblue}0.31 \\
 &  & \texttt{DER} & 0.26 & 0.21 & 0.07
 & 0.22 & 0.17 & 0.24 \\

 & \multirow{2}{*}{\texttt{\codesagespeft{}}}
 & \texttt{\ours{}} & \cellcolor{winnerblue}0.55 & \cellcolor{winnerblue}0.48 & \cellcolor{winnerblue}0.15
 & \cellcolor{winnerblue}0.41 & \cellcolor{winnerblue}0.35 & \cellcolor{winnerblue}0.41 \\
 &  & \texttt{DER} & 0.45 & 0.38 & 0.14
 & 0.36 & 0.30 & 0.39 \\

 & \multirow{2}{*}{\texttt{\bm}}
 & \texttt{\spisr{}} & \cellcolor{winnerblue}0.46 & \cellcolor{winnerblue}0.39 & \cellcolor{winnerblue}0.13
 & \cellcolor{winnerblue}0.36 & \cellcolor{winnerblue}0.30 & \cellcolor{winnerblue}0.38 \\
 &  & \texttt{\sr{}} & 0.36 & 0.29 & 0.11
 & 0.31 & 0.25 & 0.34 \\
\midrule
\multirow{8}{*}{TypeScript}
 & \multirow{2}{*}{\texttt{\swerankszs}}
 & \texttt{\ours{}} & \cellcolor{winnerblue}0.35 & \cellcolor{winnerblue}0.28 & 0.06
 & \cellcolor{winnerblue}0.27 & \cellcolor{winnerblue}0.21 & 0.29 \\
 &  & \texttt{DER} & 0.29 & 0.21 & \cellcolor{winnerblue}0.18
 & 0.25 & 0.19 & \cellcolor{winnerblue}0.30 \\

 & \multirow{2}{*}{\texttt{\codesageszs{}}}
 & \texttt{\ours{}} & \cellcolor{winnerblue}0.16 & \cellcolor{winnerblue}0.10 & 0.03
 & \cellcolor{winnerblue}0.14 & \cellcolor{winnerblue}0.09 & \cellcolor{winnerblue}0.19 \\
 &  & \texttt{DER} & 0.10 & 0.06 & \cellcolor{winnerblue}0.06
 & 0.09 & 0.06 & 0.14 \\

 & \multirow{2}{*}{\texttt{\codesagespeft{}}}
 & \texttt{\ours{}} & \cellcolor{winnerblue}0.25 & \cellcolor{winnerblue}0.18 & \cellcolor{winnerblue}0.04
 & \cellcolor{winnerblue}0.22 & \cellcolor{winnerblue}0.16 & \cellcolor{winnerblue}0.27 \\
 &  & \texttt{DER} & 0.19 & 0.13 & 0.11
 & 0.18 & 0.12 & 0.24 \\

 & \multirow{2}{*}{\texttt{\bm}}
 & \texttt{\spisr{}} & \cellcolor{winnerblue}0.11 & \cellcolor{winnerblue}0.09 & 0.02
 & \cellcolor{winnerblue}0.10 & \cellcolor{winnerblue}0.07 & \cellcolor{winnerblue}0.11 \\
 &  & \texttt{\sr{}} & 0.07 & 0.05 & \cellcolor{winnerblue}0.04
 & 0.07 & 0.05 & 0.09 \\
\bottomrule
\end{tabular}
}
\label{tab:swepoly_combined_app}
\end{table*}

\begin{table*}
\caption{Dense retrieval performance across benchmarks for $K=20$, $N=500$, $C=5$, $d=4$. Winners per metric per embedding model highlighted in light blue.}
\resizebox{\textwidth}{!}{
    \begin{tabular}{cccccc}
    \toprule
    \textbf{Language} & \textbf{Model} & \textbf{Retrieval} & \textbf{Recall@20} & \textbf{Accuracy@20} & \textbf{MRR@20} \\
    \midrule
    \multicolumn{6}{c}{\textbf{\multiswe{}}} \\
    \midrule
    \multirow{8}{*}{Java}
     & \multirow{2}{*}{\texttt{\swerankszs}} & \texttt{\ours{}} & \cellcolor{winnerblue}0.37 & \cellcolor{winnerblue}0.28 & \cellcolor{winnerblue}0.08 \\
     &  & \texttt{DER} & 0.32 & 0.24 & \cellcolor{winnerblue}0.08 \\
     & \multirow{2}{*}{\texttt{\codesageszs{}}} & \texttt{\ours{}} & \cellcolor{winnerblue}0.21 & \cellcolor{winnerblue}0.19 & \cellcolor{winnerblue}0.05 \\
     &  & \texttt{DER} & 0.15 & 0.13 & 0.03 \\
     & \multirow{2}{*}{\texttt{\codesagespeft{}}} & \texttt{\ours{}} & \cellcolor{winnerblue}0.31 & \cellcolor{winnerblue}0.25 & \cellcolor{winnerblue}0.10 \\
     &  & \texttt{DER} & 0.23 & 0.17 & 0.09 \\
     & \multirow{2}{*}{\texttt{\bm}} & \texttt{\spisr{}} & \cellcolor{winnerblue}0.23 & \cellcolor{winnerblue}0.17 & \cellcolor{winnerblue}0.07 \\
     &  & \texttt{\sr{}} & 0.22 & \cellcolor{winnerblue}0.17 & 0.05 \\
    \midrule
    \multirow{8}{*}{JavaScript}
     & \multirow{2}{*}{\texttt{\swerankszs}} & \texttt{\ours{}} & \cellcolor{winnerblue}0.39 & \cellcolor{winnerblue}0.30 & \cellcolor{winnerblue}0.18 \\
     &  & \texttt{DER} & 0.31 & 0.23 & 0.15 \\
     & \multirow{2}{*}{\texttt{\codesageszs{}}} & \texttt{\ours{}} & \cellcolor{winnerblue}0.17 & \cellcolor{winnerblue}0.12 & \cellcolor{winnerblue}0.08 \\
     &  & \texttt{DER} & 0.09 & 0.06 & 0.05 \\
     & \multirow{2}{*}{\texttt{\codesagespeft{}}} & \texttt{\ours{}} & \cellcolor{winnerblue}0.32 & \cellcolor{winnerblue}0.24 & \cellcolor{winnerblue}0.13 \\
     &  & \texttt{DER} & 0.25 & 0.18 & 0.11 \\
     & \multirow{2}{*}{\texttt{\bm}} & \texttt{\spisr{}} & \cellcolor{winnerblue}0.21 & \cellcolor{winnerblue}0.14 & \cellcolor{winnerblue}0.08 \\
     &  & \texttt{\sr{}} & 0.14 & 0.10 & 0.06 \\
    \midrule
    \multirow{8}{*}{TypeScript}
     & \multirow{2}{*}{\texttt{\swerankszs}} & \texttt{\ours{}} & \cellcolor{winnerblue}0.52 & \cellcolor{winnerblue}0.42 & 0.08 \\
     &  & \texttt{DER} & 0.40 & 0.32 & \cellcolor{winnerblue}0.23 \\
     & \multirow{2}{*}{\texttt{\codesageszs{}}} & \texttt{\ours{}} & \cellcolor{winnerblue}0.33 & \cellcolor{winnerblue}0.26 & 0.04 \\
     &  & \texttt{DER} & 0.26 & 0.21 & \cellcolor{winnerblue}0.13 \\
     & \multirow{2}{*}{\texttt{\codesagespeft{}}} & \texttt{\ours{}} & \cellcolor{winnerblue}0.30 & \cellcolor{winnerblue}0.26 & 0.05 \\
     &  & \texttt{DER} & 0.19 & 0.15 & \cellcolor{winnerblue}0.12 \\
     & \multirow{2}{*}{\texttt{\bm}} & \texttt{\spisr{}} & \cellcolor{winnerblue}0.32 & \cellcolor{winnerblue}0.27 & 0.04 \\
     &  & \texttt{\sr{}} & 0.24 & 0.21 & \cellcolor{winnerblue}0.11 \\
    \midrule
    \multicolumn{6}{c}{\textbf{\sweverified{}}} \\
    \midrule
    \multirow{8}{*}{Python}
     & \multirow{2}{*}{\texttt{\swerankszs}} & \texttt{\ours{}} & \cellcolor{winnerblue}0.61 & \cellcolor{winnerblue}0.56 & \cellcolor{winnerblue}0.32 \\
     &  & \texttt{DER} & 0.54 & 0.49 & 0.29 \\
     & \multirow{2}{*}{\texttt{\codesageszs{}}} & \texttt{\ours{}} & \cellcolor{winnerblue}0.30 & \cellcolor{winnerblue}0.27 & \cellcolor{winnerblue}0.13 \\
     &  & \texttt{DER} & 0.21 & 0.19 & 0.10 \\
     & \multirow{2}{*}{\texttt{\codesagespeft{}}} & \texttt{\ours{}} & \cellcolor{winnerblue}0.55 & \cellcolor{winnerblue}0.49 & \cellcolor{winnerblue}0.27 \\
     &  & \texttt{DER} & 0.43 & 0.38 & 0.22 \\
     & \multirow{2}{*}{\texttt{\bm}} & \texttt{\spisr{}} & \cellcolor{winnerblue}0.45 & \cellcolor{winnerblue}0.41 & \cellcolor{winnerblue}0.25 \\
     &  & \texttt{\sr{}} & 0.38 & 0.34 & 0.22 \\
    \bottomrule
    \end{tabular}
    }
\label{tab:multiswe_sweverified_combined_app}
\end{table*}


\begin{table*}[!htbp]
\centering
\caption{Impact of the number of center nodes ($C$) on retrieval performance using \ours{} + \texttt{SweRankEmbed-Small}. Increasing $C$ generally improves Recall@20 and Acc@20 across languages, while MRR@20 shows modest variation.}
\resizebox{\textwidth}{!}{
\begin{tabular}{c|ccc|ccc|ccc|ccc}
\toprule
\multirow{2}{*}{$\boldsymbol{C}$}
& \multicolumn{3}{c|}{\textbf{Python}}
& \multicolumn{3}{c|}{\textbf{JavaScript}}
& \multicolumn{3}{c|}{\textbf{Java}}
& \multicolumn{3}{c}{\textbf{TypeScript}} \\
& Recall@20 & Acc@20 & MRR@20
& Recall@20 & Acc@20 & MRR@20
& Recall@20 & Acc@20 & MRR@20
& Recall@20 & Acc@20 & MRR@20 \\
\midrule
1 & 0.47 & 0.38 & 0.29 & 0.56 & 0.48 & 0.18 & 0.32 & 0.24 & 0.18 & 0.32 & 0.24 & 0.06 \\
3 & 0.49 & 0.39 & 0.30 & 0.59 & 0.51 & 0.18 & 0.35 & 0.26 & 0.19 & 0.31 & 0.24 & 0.06 \\
5 & \cellcolor{winnerblue}0.49 & \cellcolor{winnerblue}0.40 & \cellcolor{winnerblue}0.30
  & \cellcolor{winnerblue}0.60 & \cellcolor{winnerblue}0.52 & 0.18
  & 0.36 & 0.27 & 0.18
  & 0.35 & 0.28 & 0.06 \\
7 & 0.49 & 0.39 & 0.30 & 0.60 & 0.52 & 0.18 & 0.36 & 0.27 & 0.19 & 0.37 & 0.30 & 0.06 \\
9 & 0.48 & 0.39 & 0.30 & 0.60 & 0.53 & \cellcolor{winnerblue}0.19 & \cellcolor{winnerblue}0.37 & 0.27 & 0.19 & \cellcolor{winnerblue}0.39 & \cellcolor{winnerblue}0.31 & 0.06 \\
\bottomrule
\end{tabular}
}
\label{tab:bfs_C_ablation_app}\label{tab:bfs_C_ablation}
\end{table*}

\paragraph{MRR@$K$ Gains Depend on Graph Neighborhoods}
The MRR@$K$ metric benefits from \ours{} primarily when the first relevant ground-truth item within the top-$K$ results is a neighbor of a center node. In cases where the first relevant item is a non-center node, particularly in TypeScript as shown in Tab.~\ref{tab:swepoly_combined}—\ours{} may rank neighbors of center nodes ahead, slightly reducing MRR@$K$. Since \ours{} is not explicitly designed to optimize the precise ranking order within the top-$K$ results, improvements in MRR@$K$ over baseline \der{} methods are modest but still observable.

\subsection{Edge-Type Ablation: Statistical Significance}\label{sec:edge_significance_app}

\cref{tab:edge_ablation} in the main text reports point estimates for each exploration edge type. Because \ours{} and \der{} are evaluated on identical instances and their per-instance scores are highly correlated, we assess significance with paired per-instance tests rather than by comparing the marginal confidence intervals of \cref{tab:swepolybench_ci_compact}, which have substantially less power. For each language and edge configuration we form the per-instance differences $\Delta_i$ between the \ours{} and \der{} score on instance $i$, and report (i) a paired-bootstrap $p$-value, $\Pr(\bar{\Delta} \le 0)$ over $B=10{,}000$ resamples of $\{\Delta_i\}$, and (ii) a two-sided paired-permutation (sign-flip) test over $10{,}000$ permutations. \cref{tab:paired_significance_app} reports both, for Recall@20 and Acc@20.

The two tests agree in every cell. All structural-edge gains on Recall@20 are significant at $p<0.01$ in all four languages, and at $p<0.001$ for Java and TypeScript under \textit{invokes} or \textit{contains}$+$\textit{invokes}. On Acc@20 the only non-significant structural cell is Java with \textit{contains} alone, which becomes significant once \textit{invokes} edges are added (permutation $p=0.0224$); this is consistent with the locality analysis of \cref{tab:locality_app}, where Java combines the highest invoke-connectivity with the lowest dense top-$100$ coverage. \textit{imports} and \textit{inherits} are uniformly non-significant, and Proposition~\ref{prop:recall_condition} predicts these two nulls for structurally different reasons. For \textit{inherits}, no edge in our schema is incident to a function node in any language, so the explored shell contains no functions at all, $n_h = 0$, and no exchange is offered. The configuration returns exactly the \der{} set, which is what \cref{tab:edge_ablation} reports. For \textit{imports}, edges do reach function nodes but their prevalence sits at the background rate outside Python, so $\eta$ falls below the frontier $\tau$ and the condition genuinely fails. A framework that could not separate these two cases would be vacuous, whereas here each predicts a distinct observable, an identical-to-\der{} result in one case and a non-significant difference in the other.

\begin{table*}[!ht]
\centering
\caption{\textbf{Paired significance of \ours{} over \der{} on \swepoly{}}, by exploration edge type, with \texttt{SweRankEmbed-Small} at $K=20$, $N=500$, $C=5$. $\Delta$ is the mean per-instance difference; ``bootP'' is the paired-bootstrap $\Pr(\bar{\Delta} \le 0)$ over $B=10{,}000$ resamples; ``perm'' is the two-sided paired-permutation $p$-value over $10{,}000$ sign-flips. $^{***}p<0.001$, $^{**}p<0.01$, $^{*}p<0.05$; unmarked entries are not significant. $\Delta$ is computed on unrounded per-instance scores and may therefore differ by up to $0.01$ from the difference of the rounded means in \cref{tab:edge_ablation}. \textit{inherits} is marked N/A outside Python because the configuration is degenerate at function granularity, for the reason given in \cref{tab:edge_ablation}.}
\label{tab:paired_significance_app}
\resizebox{\textwidth}{!}{%
\begin{tabular}{l|ccc|ccc|ccc|ccc}
\toprule
\multirow{2}{*}{\textbf{Exploration edges / depth}}
& \multicolumn{3}{c|}{\textbf{Python}}
& \multicolumn{3}{c|}{\textbf{Java}}
& \multicolumn{3}{c|}{\textbf{JavaScript}}
& \multicolumn{3}{c}{\textbf{TypeScript}} \\
& $\Delta$ & bootP & perm
& $\Delta$ & bootP & perm
& $\Delta$ & bootP & perm
& $\Delta$ & bootP & perm \\
\midrule
\multicolumn{13}{l}{\textit{Recall@20}} \\
\textit{contains} / $d{=}4$
 & $+0.07$ & $<0.0001^{***}$ & $0.0008^{***}$
 & $+0.05$ & $0.0001^{***}$ & $0.0060^{**}$
 & $+0.08$ & $<0.0001^{***}$ & $<0.0001^{***}$
 & $+0.05$ & $<0.0001^{***}$ & $0.0006^{***}$ \\
\textit{invokes} / $d{=}2$
 & $+0.15$ & $<0.0001^{***}$ & $<0.0001^{***}$
 & $+0.11$ & $<0.0001^{***}$ & $<0.0001^{***}$
 & $+0.13$ & $<0.0001^{***}$ & $<0.0001^{***}$
 & $+0.17$ & $<0.0001^{***}$ & $<0.0001^{***}$ \\
\textit{imports} / $d{=}2$
 & $+0.01$ & $0.1318$ & $0.4979$
 & $+0.00$ & $1.0000$ & $1.0000$
 & $+0.00$ & $1.0000$ & $1.0000$
 & $+0.00$ & $1.0000$ & $1.0000$ \\
\textit{inherits} / $d{=}2$
 & $+0.00$ & $1.0000$ & $1.0000$
 & N/A & N/A & N/A
 & N/A & N/A & N/A
 & N/A & N/A & N/A \\
\textit{contains} / $d{=}4$ + \textit{invokes} / $d{=}2$
 & $+0.17$ & $<0.0001^{***}$ & $<0.0001^{***}$
 & $+0.11$ & $<0.0001^{***}$ & $<0.0001^{***}$
 & $+0.15$ & $<0.0001^{***}$ & $<0.0001^{***}$
 & $+0.17$ & $<0.0001^{***}$ & $<0.0001^{***}$ \\
\midrule
\multicolumn{13}{l}{\textit{Acc@20}} \\
\textit{contains} / $d{=}4$
 & $+0.05$ & $0.0026^{**}$ & $0.0304^{*}$
 & $+0.03$ & $0.3690$ & $1.0000$
 & $+0.07$ & $<0.0001^{***}$ & $0.0001^{***}$
 & $+0.07$ & $0.0005^{***}$ & $0.0080^{**}$ \\
\textit{invokes} / $d{=}2$
 & $+0.14$ & $<0.0001^{***}$ & $0.0001^{***}$
 & $+0.07$ & $0.0038^{**}$ & $0.0224^{*}$
 & $+0.12$ & $<0.0001^{***}$ & $<0.0001^{***}$
 & $+0.17$ & $<0.0001^{***}$ & $<0.0001^{***}$ \\
\textit{imports} / $d{=}2$
 & $+0.01$ & $0.3648$ & $1.0000$
 & $+0.00$ & $1.0000$ & $1.0000$
 & $+0.00$ & $1.0000$ & $1.0000$
 & $+0.00$ & $1.0000$ & $1.0000$ \\
\textit{inherits} / $d{=}2$
 & $+0.00$ & $1.0000$ & $1.0000$
 & N/A & N/A & N/A
 & N/A & N/A & N/A
 & N/A & N/A & N/A \\
\textit{contains} / $d{=}4$ + \textit{invokes} / $d{=}2$
 & $+0.15$ & $<0.0001^{***}$ & $0.0003^{***}$
 & $+0.06$ & $0.0082^{**}$ & $0.0392^{*}$
 & $+0.14$ & $<0.0001^{***}$ & $<0.0001^{***}$
 & $+0.17$ & $<0.0001^{***}$ & $<0.0001^{***}$ \\
\bottomrule
\end{tabular}}
\end{table*}

\paragraph{What the Stage-2 selector contributes.}
Disabling Stage 2 does not yield a comparable operating point: every node in $\Gamma_d(\mathcal{C}_Q) \cap \mathcal{S}_N(Q)$ is promoted, the retrieved set grows to hundreds of candidates, and the fixed budget $K$ is abandoned. The resulting recall is therefore an \emph{upper bound} on what any selection rule over the same neighborhood could achieve, and the useful question is how much of that ceiling \ours{} captures while still returning only $K$ functions. \cref{tab:stage2_ceiling_app} reports both. \ours{} recovers $79$--$89\%$ of the unfiltered neighborhood's reachable recall at $K=20$, so Stage 2 discards the large majority of surfaced nodes while preserving nearly all of the recoverable ground truth; without it, the same recall is attainable only by giving up the fixed budget. The residual gap also quantifies the headroom available to a better Stage-2 selector.

\begin{table}[!ht]
\centering
\caption{\textbf{Stage-2 selector against the unfiltered neighborhood ceiling} on \swepoly{} with \texttt{SweRankEmbed-Small}, $N=500$, $C=5$. The ceiling is Recall of the full Stage-1 neighborhood with Stage 2 disabled, which returns hundreds of candidates rather than $K$; the \ours{} column is Recall@20 under \textit{invokes} expansion from \cref{tab:edge_ablation}, and the last column is its share of the corresponding ceiling.}
\label{tab:stage2_ceiling_app}
\resizebox{\columnwidth}{!}{%
\begin{tabular}{lcccc}
\toprule
\multirow{2}{*}{\textbf{Language}}
& \multicolumn{2}{c}{\textbf{Ceiling (Stage 2 disabled)}}
& \textbf{\ours{}} & \textbf{\% of} \\
& \textit{contains}/$4$ & \textit{invokes}/$2$ & \textbf{Recall@20} & \textbf{ceiling} \\
\midrule
Python      & 0.517 & 0.626 & 0.56 & 89 \\
Java        & 0.376 & 0.533 & 0.42 & 79 \\
JavaScript  & 0.667 & 0.730 & 0.65 & 89 \\
TypeScript  & 0.400 & 0.593 & 0.47 & 79 \\
\bottomrule
\end{tabular}}
\end{table}

\subsection{Comparison with Lexical and Agentic Retrieval}\label{sec:lexical_agentic_app}

Coding agents in practice often locate code with lexical \texttt{grep} search and iterative tool calls rather than with a dense retriever. We therefore compare \ours{} against both, on \swepoly{} at $K=20$.

\paragraph{Baselines.} The \emph{lexical} baseline operates over the same function nodes as the dense retriever: an LLM extracts up to $25$ search terms from the issue, every function in the code graph is ranked by keyword-hit count, and the top-$K$ are returned. The \emph{agentic} baseline is \texttt{OpenHands} \citep{wang2025openhands}, a general-purpose software agent that explores the repository with its own tool suite and is not restricted to a fixed retrieval budget.

\begin{table*}[!ht]
\centering
\caption{\textbf{Lexical and agentic retrieval on \swepoly{}} at $K=20$. \ours{} rows use \texttt{SweRankEmbed-Small} with $N=500$, $C=5$; \der{} is carried over from \cref{tab:swepoly_combined} for reference. ``Tok'' is total (input $+$ output) LLM tokens per instance. The upper block operates under a \emph{fixed} retrieval budget of $K$ functions, and the best entry per language and metric within that block is in light blue. \texttt{OpenHands} is listed separately because it is an iterative agent that is not budget-constrained: it explores until it decides to stop, so its recall is not obtained at a comparable operating point.}
\label{tab:lexical_agentic_app}
\resizebox{\textwidth}{!}{%
\begin{tabular}{l|ccc|ccc|ccc|ccc}
\toprule
\multirow{2}{*}{\textbf{Method}}
& \multicolumn{3}{c|}{\textbf{Python}}
& \multicolumn{3}{c|}{\textbf{Java}}
& \multicolumn{3}{c|}{\textbf{JavaScript}}
& \multicolumn{3}{c}{\textbf{TypeScript}} \\
& Recall@20 & Acc@20 & Tok
& Recall@20 & Acc@20 & Tok
& Recall@20 & Acc@20 & Tok
& Recall@20 & Acc@20 & Tok \\
\midrule
\multicolumn{13}{l}{\textit{Fixed retrieval budget ($K=20$)}} \\
\der{}
 & 0.42 & 0.35 & 0 & 0.31 & 0.24 & 0 & 0.52 & 0.45 & 0 & 0.29 & 0.21 & 0 \\
\texttt{grep} $+$ LLM keywords
 & 0.46 & 0.40 & 1.1K
 & 0.29 & 0.22 & 0.7K
 & \cellcolor{winnerblue}0.69 & \cellcolor{winnerblue}0.61 & 0.7K
 & 0.28 & 0.24 & 0.7K \\
\ours{} (\textit{contains}/$4$)
 & 0.49 & 0.40 & 12.9K
 & 0.36 & 0.27 & 7.8K
 & 0.60 & 0.52 & 14.9K
 & 0.35 & 0.28 & 8.3K \\
\ours{} (\textit{invokes}/$2$)
 & \cellcolor{winnerblue}0.56 & \cellcolor{winnerblue}0.50 & 109.9K
 & \cellcolor{winnerblue}0.42 & \cellcolor{winnerblue}0.30 & 87.4K
 & 0.65 & 0.58 & 55.5K
 & \cellcolor{winnerblue}0.47 & \cellcolor{winnerblue}0.41 & 92.7K \\
\hdashline
\multicolumn{13}{l}{\textit{Iterative agent (no fixed budget)}} \\
\texttt{OpenHands}
 & 0.84 & 0.77 & 273.2K
 & 0.71 & 0.59 & 371.4K
 & 0.93 & 0.89 & 223.4K
 & 0.66 & 0.62 & 599.7K \\
\bottomrule
\end{tabular}}
\end{table*}

\paragraph{Findings.} \cref{tab:lexical_agentic_app} supports three observations. First, \ours{} outperforms the lexical baseline on Python, Java and TypeScript. Second, JavaScript is an exception where keyword matching wins, and this is the outcome our own locality analysis predicts: \cref{tab:locality_app} identifies JavaScript as the dense-ranker-dominated split, with $53.6\%$ of instances already fully covered in the top-$100$ and only $21.8\%$ having any ground truth beyond the top-$500$, leaving little headroom for a graph mechanism, and it is also the split where lexical matching is strongest. Third, \texttt{OpenHands} attains the highest recall in every language, but by iterative exploration costing $223$K--$600$K tokens per instance, between two and forty times the cost of \ours{} depending on the edge configuration, and without the fixed-budget guarantee that makes the retrieved set directly consumable by a downstream generator under a known context cost. The comparisons are therefore complementary rather than competing: \ours{} is agnostic to its first stage and improves both dense and sparse \bm{} retrievers (\cref{tab:swepoly_combined}), so the same expansion could be applied on top of a lexical first stage in the regime where that stage is strongest.

\paragraph{Deployability.} The two regimes also differ in how they can be served. \ours{}'s cost is spread over $C$ independent LLM calls that are issued concurrently and carry no context between them, so each call sees a small, bounded context regardless of repository size; \cref{tab:spider_results} shows an open $30$B coder model filling the filter role to within $0.01$--$0.05$ Recall@20 of \texttt{Claude Sonnet~4}, and the bounded per-call context is what makes serving or fine-tuning a smaller model for this step practical. An iterative agent's context instead grows with its tool-call history, so its budget is neither fixed in advance nor parallelizable in the same way, and the same substitution is correspondingly harder. This matters for the developer-in-the-loop setting \ours{} targets, where retrieval runs on an active repository at session start under a cost ceiling the developer chooses.

\subsection{Per-Language Hyperparameter Sensitivity}\label{sec:sensitivity_app}

$C$, $d$ and $N$ were grid-searched on the Python split of \swepoly{} and then held fixed for every language and benchmark. The per-language sweeps already reported let us check directly whether that transfer costs anything. \cref{tab:bfs_C_ablation_app} shows Recall@20 saturating at $C=5$ for Python, JavaScript and Java, with only TypeScript still improving marginally up to $C=9$ ($+0.04$ Recall@20). \cref{tab:bfs_d_ablation_app} shows monotone improvement in $d$ for all four languages, so no language prefers a smaller depth. \cref{tab:bfs_N_ablation_app} shows saturation at $N \approx 300$--$500$ in all four.

The Python-tuned configuration $(C{=}5, d{=}4, N{=}500)$ is therefore at or near the per-language optimum everywhere we measure, and the single deviation, TypeScript preferring a larger $C$, would \emph{increase} \ours{}'s reported gains rather than reduce them. The cross-language conclusions in \cref{tab:swepoly_combined} are consequently not an artifact of tuning on Python, and the fixed configuration is, if anything, conservative for TypeScript.

\subsection{Qualitative Analysis}\label{sec:case_study_app}

\paragraph{Which issues benefit from graph expansion?} Those where the issue names or points at one function, but the code that must change is a structural neighbor of it; dense retrieval ranks the named function highly and stops there. In \texttt{huggingface/transformers-5749} (Python, \textit{contains}), the issue observes that \texttt{BasicTokenizer.\_clean\_text} appears never to be called. Dense retrieval ranked \texttt{\_clean\_text} at the top but never surfaced the ground truth \texttt{BasicTokenizer.tokenize}, the function that must be edited to call it; \ours{} recovered \texttt{tokenize} as a same-class \textit{contains} neighbor of the \texttt{\_clean\_text} center. In \texttt{apache/dubbo-3479} (Java, \textit{invokes}), the issue concerns URL keys in simplify mode. Dense retrieval missed both ground-truth functions, \texttt{ServiceConfig.doExport} and \texttt{ServiceConfig.doExportUrlsFor1Protocol}, which live in a different file from any top-ranked function; \ours{} recovered both as \textit{invokes} neighbors of the center \texttt{ZKTools.pathToKey}. These are cross-file targets that \textit{contains} expansion cannot structurally reach. Quantitatively, \textit{invokes} expansion changes the retrieved set on $23\%$ of instances (macro-averaged over languages) against $10\%$ for \textit{contains}, and improves recall in the large majority of those cases, with losses of $1$--$2\%$.

\paragraph{When does the Stage-2 filter judge incorrectly?} In two opposite ways. \emph{Over-restrictive:} the prompt (\cref{sec:prompt_app}) instructs the selector to keep only high-confidence candidates and to return an empty set when in doubt. When the ground truth is reached from a center that is itself only tangentially related to the issue, the selector sees no issue-relevance and discards it. In \texttt{keras-team/keras-18975}, the ground truth \texttt{CompileLoss.build} was surfaced as a neighbor of \texttt{Trainer.compile} but rejected by the filter, so the instance remained a miss. Across languages the filter retains only $50$--$70\%$ of the ground-truth functions that graph expansion successfully surfaces, which is the same conclusion \cref{tab:stage2_ceiling_app} reaches from the opposite direction: \emph{the selector, not the graph, is the current bottleneck.} \emph{Over-permissive:} dense call graphs can surface hundreds of neighbors per center, and the selector then admits tangential callers and test functions which, under a fixed budget, can evict a correctly retrieved ground-truth function. In \texttt{prettier/prettier-14788} (JavaScript), dense retrieval already had the sole ground truth \texttt{handleError} inside the top-$K$ and neighbor flooding displaced it. This is why we cap the \textit{invokes} neighborhood at the $100$ highest-ranked neighbors per query (\cref{tab:edge_ablation}).

\paragraph{Operating boundaries.} \ours{} helps when the target is both (i) within $d$ hops of a top-$C$ center and (ii) inside the top-$N$ pool. Targets that are simultaneously low-ranked and graph-distant remain unreachable. The best-fit cases are call-chain bugs and localized refactors (\textit{invokes}) and same-file or same-class fixes (\textit{contains}); the worst-fit cases are large multi-function refactors and fixes that are semantically distant from every highly ranked function. Taken together with the filter's $50$--$70\%$ retention, this points at issue-aware neighbor ranking before filtering, and per-center neighbor caps on dense \textit{invokes} graphs, as the most promising directions for improving \ours{}.

\subsection{Evaluating Open Source Models}\label{sec:open_source_models_app}

We examine whether \ours{} can be run end-to-end with open-source language models in place of \texttt{Claude Sonnet 4} for the LLM neighborhood-filtering step. We substitute three Qwen-Coder variants of increasing scale and re-run the dense retrieval pipeline on \swepoly{}. \cref{tab:spider_results} reports function-level Recall and Accuracy across all four languages.

\begin{table*}[t]
\centering
\caption{\ours{} pipeline results on \swepoly{} (function-level localization, dense retrieval, $K{=}20$, $N{=}500$, $C{=}5$, $d{=}4$, with a single non-iterative Stage-2 call per seed center). Each cell reports macro-averaged recall and accuracy over the language's evaluation subset. The top block reports two references with \texttt{SweRankEmbed-Small} taken from \cref{tab:swepoly_combined} (retrieval-only): \texttt{DER} is the dense embedding baseline without \ours{}, and \texttt{Claude Sonnet 4} is the closed-source LLM filter used inside \ours{}. \textbf{Bold} marks the best score in each column across all rows (\texttt{DER}, \texttt{Claude Sonnet 4}, and the open-source Qwen variants); \underline{underline} marks the best score in each column among the open-source models only.}
\label{tab:spider_results}
\setlength{\tabcolsep}{5pt}
\renewcommand{\arraystretch}{1.1}
\resizebox{0.7\textwidth}{!}{
\begin{tabular}{lcccccccc}
\toprule
 & \multicolumn{2}{c}{\textbf{Java}} & \multicolumn{2}{c}{\textbf{JavaScript}} & \multicolumn{2}{c}{\textbf{Python}} & \multicolumn{2}{c}{\textbf{TypeScript}} \\
\cmidrule(lr){2-3} \cmidrule(lr){4-5} \cmidrule(lr){6-7} \cmidrule(lr){8-9}
\textbf{Model} & Rec. & Acc. & Rec. & Acc. & Rec. & Acc. & Rec. & Acc. \\
\midrule
DER (no \ours{})             & 0.31             & 0.24             & 0.52             & 0.45             & 0.42             & 0.35             & 0.29             & 0.21             \\
Claude Sonnet 4              & \textbf{0.36}    & \textbf{0.27}    & \textbf{0.60}    & \textbf{0.52}    & \textbf{0.49}    & \textbf{0.40}    & \textbf{0.35}    & \textbf{0.28}    \\
\hdashline%
\multicolumn{9}{l}{\textit{Open-source}} \\
Qwen2.5-Coder-7B-Instruct    & 0.32             & \underline{0.24} & 0.51             & 0.44             & 0.43             & 0.35             & 0.31             & 0.23             \\
Qwen2.5-Coder-14B-Instruct   & 0.33             & 0.23             & 0.53             & 0.45             & 0.44             & 0.35             & 0.31             & 0.23             \\
Qwen3-Coder-30B-A3B-Instruct & \underline{0.34} & \underline{0.24} & \underline{0.55} & \underline{0.47} & \underline{0.48} & \underline{0.38} & \underline{0.33} & \underline{0.25} \\
\bottomrule
\end{tabular}
}
\end{table*}

Among the open-source backbones, \texttt{Qwen3-Coder-30B-A3B-Instruct} obtains the best Recall and Accuracy in every language, with the largest gains over the 7B and 14B variants on Python and JavaScript.

The top block of \cref{tab:spider_results} carries over two references from \cref{tab:swepoly_combined}: the \texttt{DER} baseline (no \ours{}) and the closed-source \texttt{Claude Sonnet 4} used as the LLM filter inside \ours{}. \texttt{Qwen3-Coder-30B-A3B-Instruct} improves over the \texttt{DER} baseline on every language, while the smaller 7B and 14B variants show smaller and less consistent gains over \texttt{DER}. Relative to \texttt{Claude Sonnet 4}, \texttt{Qwen3-Coder-30B-A3B-Instruct} closes the Recall@20 gap to roughly 0.02 on Java, 0.05 on JavaScript, 0.01 on Python, and 0.02 on TypeScript; the Accuracy@20 gaps follow a similar pattern. This indicates that a sufficiently large open-source coder model is a practical drop-in substitute for \ours{}'s LLM filter on \swepoly{}.

\subsection{Frequently Asked Questions}\label{sec:faq_app}
This section collects clarifications on points that recur in discussion of this work.

\paragraph{Is the headline ``$\ge 13\%$'' gain absolute or relative?}
Relative. The minimum is taken over all language $\times$ dataset cells in \cref{tab:swepoly_combined} comparing \ours{} to \der{} with \texttt{SweRankEmbed-Small} at $K=20$. The corresponding absolute Recall@20 gains range from $+0.05$ to $+0.12$ points. For example, Python on \swepoly{} goes from $0.42$ to $0.49$, a $+0.07$ absolute / $+16.7\%$ relative gain.

\paragraph{Why is \textit{contains} the default exploration edge?}
It does not: \cref{tab:edge_ablation} evaluates all four edge types, and \textit{invokes} expansion outperforms \textit{contains}-only expansion in every language. \textit{contains} is the \emph{default} because it is the cheapest reliable signal. It captures hierarchical structure (function $\subseteq$ class $\subseteq$ file $\subseteq$ directory), is identified reliably by syntax-based parsers across all four languages, and costs roughly an order of magnitude fewer tokens per instance than \textit{invokes}, whose accurate call-graph resolution is also more brittle in dynamic languages (e.g., method dispatch through Python decorators, JavaScript prototype chains, dynamic imports). Where the token budget permits, \textit{invokes} or \textit{contains}$+$\textit{invokes} is the stronger choice, and the margin is largest exactly where \cref{tab:locality_app} predicts: Java, with the highest invoke-connectivity ($0.63$) and the lowest dense top-$100$ coverage ($29\%$).

\paragraph{Is the \ours{}-over-\der{} improvement statistically significant?}
We report bootstrapped $95\%$ confidence intervals in \cref{tab:swepolybench_ci_compact} and KDE plots in \cref{fig:kde_plots_recall_swepolybench,fig:kde_plots_acc_swepolybench}. The comparison is paired per-instance, so the paired test has substantially more power than comparing marginal CIs. \cref{tab:paired_significance_app} reports paired-bootstrap and two-sided paired-permutation results for every language and edge configuration; the two tests agree in every cell, and all structural-edge Recall@20 gains are significant at $p<0.01$ in all four languages. \cref{sec:edge_significance_app} discusses the one non-significant structural cell.

\paragraph{Why is \locagent{} evaluated only on Python?}
\locagent{} is an iterative LLM-driven agent that issues many tool calls per instance and took over $48$ hours on \sweverified{} and roughly $24$ hours on the Python split of \swepoly{} (see \cref{tab:locagent_performance}). Running it at the same scale on Java, JavaScript, and TypeScript was not feasible within our compute budget. The comparison in \cref{tab:locagent_vs_spider} is therefore illustrative for Python only; we do not extrapolate cross-method conclusions to other languages.

\paragraph{Why is the TypeScript function-edit percentage in \cref{tab:data_stats} low?}
Many TypeScript repositories in \swepoly{} and \multiswe{} are linter- or configuration-heavy, and patches more often touch type definitions, build configurations, or other non-function regions than they do in Python or Java. Our function-level evaluation excludes these instances, which is why only $29.6\%$ (\swepoly{}) and $27.7\%$ (\multiswe{}) of TypeScript instances have function-level ground truth. The class-level breakdown in \cref{tab:swepolybench_class} is correspondingly small and is not reported for TypeScript.

\paragraph{When does \ours{} reduce MRR@$K$?}
\ours{} promotes graph neighbors of top-ranked seeds, which can occasionally rerank a neighbor above a high-ranked non-center ground-truth node. The MRR@$K$ drop is most visible in TypeScript with \swerankszs{} (\cref{tab:swepoly_combined_app}) and on \multiswe{} TypeScript. \ours{} is Recall-oriented and not designed to optimize ranking order within the top-$K$; Recall@$K$ and Acc@$K$, the relevant metrics for downstream patch generation under a fixed budget, are not affected. A fuller discussion is in App.~\ref{sec:additional_exps_app}.

\paragraph{What is the wall-clock and token-cost overhead of \ours{} relative to \der{} and \bm{}?}
\ours{} adds one Stage-2 LLM round-trip per seed center (default $C{=}5$), bounding per-query overhead by the input-token budget reported in \cref{tab:bfs_d_ablation_tokens_app,tab:bfs_N_ablation}. At the default $C{=}5$, $d{=}4$, $N{=}500$ setting, the per-query input is $\sim 18$--$32$K tokens (per language), with output $\sim 150$--$300$ tokens; calls are issued concurrently across seeds. Empirically, this corresponds to a per-query latency of roughly $5$--$10$ seconds with \texttt{Claude Sonnet~4}, compared with $<1$ second for \der{} alone (dominated by the encoder forward pass and a top-$K$ scan) and a few milliseconds for \bm{}. \locagent{} required $24$--$48$ hours on the same datasets due to its iterative tool-call architecture (\cref{tab:locagent_performance}), so \ours{} sits roughly two orders of magnitude below LLM-agent localization while remaining only one LLM round-trip slower than embedding retrieval.

\paragraph{Does graph expansion alone (without the LLM filter) already provide most of the gain?}
No. The Stage-2 LLM filter is the component that makes graph expansion safe at fixed budget $K$. Without it, every node in $\Gamma_d(\mathcal{C}_Q) \cap \mathcal{S}_N(Q)$ would be promoted into the top-$K$ set, which scales super-linearly with $d$ and quickly displaces high-confidence semantic candidates with noise; equivalently, Stage~1 alone admits too many false-positive neighbors per seed for $d \ge 4$. \cref{tab:stage2_ceiling_app} measures this directly: the unfiltered neighborhood's recall is an upper bound attained only by abandoning the fixed budget, and \ours{} captures $79$--$89\%$ of it while returning just $K$ functions.

\paragraph{Why are GraphCodeBERT, RepoGraph, CoRet, CoRNStack, RepoDeepSearch and RepoNavigator not included as baselines?}
GraphCodeBERT \citep{guo2021graphcodebertpretrainingcoderepresentations} is a code encoder rather than an issue-localization system; the role of a strong embedding-based baseline is filled by \codesageszs{} and \codesagespeft{} in \cref{tab:swepoly_combined}, which are stronger code encoders for issue-driven retrieval. CoRNStack \citep{suresh2025cornstack} is a training procedure whose representative model is the \codesage{} family we evaluate. CoRet \citep{fehr-etal-2025-coret} has released neither embeddings nor code, preventing direct comparison. RepoGraph \citep{ouyang2025repograph} is an agentic localization pipeline built on line-of-code-level graphs, constructed with the Python \texttt{ast} module, so it operates at a different granularity than function-level retrieval at fixed $K$ and is not directly usable on \oursbench{} or on existing non-Python benchmarks. RepoDeepSearch \citep{ma2025repodeepsearch} and RepoNavigator \citep{zhang2026reponavigator} are recent agentic localization systems, but, like \locagent{}, both lean heavily on Python-specific tooling such as \texttt{ast} and Python-oriented parsers; running them on the Java, JavaScript and TypeScript splits of \oursbench{} would require substantial pipeline modification, which is itself part of the gap \oursbench{} is intended to expose. RepoNavigator additionally has no public implementation. The agentic comparators within reach are therefore \locagent{} (\cref{tab:locagent_performance,tab:locagent_vs_spider}) and \texttt{OpenHands} (\cref{tab:lexical_agentic_app}), which is language-agnostic.

\paragraph{How is \oursbench{} constructed, and what are the per-language sizes?}
For each instance from \sweverified{}, \swepoly{}, and \multiswe{}, we (i) clone the repository at the issue's base commit, (ii) parse all primary-language files into syntax trees (Python \texttt{ast}; Tree-sitter for Java/JS/TS), (iii) extract function, class, file, and directory nodes plus their \textit{contains}, \textit{invokes}, \textit{imports}, and \textit{inherits} edges following the schema of \citet{chen2025locagent}, and (iv) map each patch's modified function bodies back to graph nodes to obtain $\mathcal{V}^*$. After excluding instances without function-level ground truth, the retained per-language counts are: Python $182$ from \swepoly{} and $469$ from \sweverified{}; Java $158$ from \swepoly{}; JavaScript $573$ from \swepoly{}; TypeScript subsets across \swepoly{} and \multiswe{}; full per-dataset numbers are in \cref{tab:data_stats,tab:locality_app}. Construction is deterministic given the repository state and uses no manual annotation; ground-truth labels are inherited from the upstream benchmarks.

\paragraph{How sensitive are the results to the initial dense ranker?}
The initial ranker bounds \ours{} in two ways. First, \cref{tab:locality_app} shows that $22\%$--$62\%$ of instances (depending on language) contain at least one ground-truth function outside the dense top-$500$ candidate pool $\mathcal{S}_N(Q)$, and \ours{} cannot recover these targets regardless of graph structure. Second, the relative gain ordering across languages tracks the per-language top-$100$ coverage of the dense baseline: JavaScript benefits less in relative terms because \der{} already covers most of its ground truth at $K{=}20$, while Python and Java see larger relative gains because their dense top-$100$ coverage is lower. A stronger initial ranker would lift the ceiling proportionally and is the most direct lever for closing the residual gap.

\end{document}